\documentclass{iopbk2e}

\usepackage{amssymb}
\usepackage{amsmath,amsthm,amsfonts,amstext,amssymb,latexsym}
\usepackage{iopams}
\usepackage[dvips]{graphicx}

\makeindex

\begin{document}
\title{Gravitational Waves, Sources and Detectors}
\author{\textbf{Bernard F Schutz} \\
Max Planck Institute for Gravitational Physics\\ (Albert Einstein Institute)
\\Golm bei Potsdam, Germany \\ and \\ Department of Physics and Astronomy \\
Cardiff University, Wales \\ and \\[1cm] \textbf{Franco Ricci} \\ University ``La Sapienza'' \\ Rome, Italy}
\date{}
\maketitle
\pagenumbering{roman}
\setcounter{page}{5}
\tableofcontents
\newcommand{\lecture}{Lecture}
\newcommand{\supr}{{}^}
\newcommand{\sub}{{}_}
\newcommand{\ibar}{\mbox{\rlap{$I$}--}}
\newcommand{\correction}[2]{#2}

\chapter*{Preface}

Gravitational waves and their detection are becoming more and more important
both for the theoretical physicist and the astrophysicist. In fact, 
technological developments have enabled the construction such sensitive
detectors (bars and interferometers) that the detection of gravitational radiation
could become a reality during the next few years. In these lectures we give a brief overview of  this interesting and challenging field of
modern physics.

The topics to be covered are divided into 6 \lecture s.  We begin by
describing gravitational waves in linearized general relativity, where 
one can examine most of the
basic properties of gravitational radiation itself: propagation, gauge invariance, and
interactions with matter (and in particular with detectors).

The second \lecture\ deals with gravitational waves detectors: how they operate, 
what their most important sources of noise are, and what mechanisms are used to overcome noise. We report here  on the most important detectors planned or under
construction (both ground-based ones and space-based), their likely
sensitivity, and their prospects for making detections. Other
speakers will go into much
more detail on specific detectors, such as LISA.

The third \lecture\ deals with the astrophysics of likely sources of gravitational waves: binary systems, neutron stars, pulsars, X-ray sources,
supernovae/hypernovae, $\gamma $-ray bursts and the big bang. We estimate there the expected wave amplitude $h$ and the suitability of specific detectors for seeing waves from each source.

The fourth \lecture\ is much more theoretical. Here we develop the mathematical
theory of gravitational waves in general, their effective stress-energy tensor, the energy carried by
gravitational waves, and the energy in a random wave field (gravitational
background generated by the big bang).

The fifth \lecture\ carries the theory further and examines the generation of gravitational radiation in linearized theory. We show in some detail how both mass-quadrupole and current-quadrupole radiation is generated, including how characteristics of the radiation like its polarization are related to the motion of the source. 
Current-quadrupole radiation has become important very recently and may indeed be one of the first forms of gravitational radiation to be detected. We attempt to give a physical description of the way it is generated.

The final \lecture\ will explore applications of the 
theory we have developed to some sources. We calculate the quadrupole moment of a binary system, the energy radiated in the Newtonian approximation and the back-reaction on the orbit. We conclude with a brief introduction the current-quadrupole-driven instability in the  $r$-modes of neutron stars. 

Chapters one and four are followed by a few exercises to assist students. We presume the reader has some background in general relativity and its mathematical tools in differential geometry, at the level of the introductory chapters of Schutz (1985). See the list of references at the end of these lectures for sources suitable for further and background reading.

\bigskip
\begin{flushright}
\textbf {B.F.\ Schutz, F.\ Ricci}\\
May 2000
\end{flushright}

\newpage

\pagenumbering{arabic}
\setcounter{page}{1}

\chapter{Elements of gravitational waves}

General relativity is a theory of gravity that is consistent with 
special relativity in many respects, and in particular with the principle 
that nothing travels faster than light. This means that changes in the 
gravitational field cannot be felt everywhere instantaneously: they must 
propagate. In general relativity they propagate at exactly the same speed 
as vacuum electromagnetic waves: the speed of light. These propagating 
changes are called gravitational waves. 

However, general relativity is a non-linear theory and there is, in general, no
sharp  distinction between the part of the metic that represents the waves and rest of the metric. Only in certain approximations can we clearly define 
gravitational radiation. Three interesting approximations in which it is possible to make
this distinction are:

\begin{itemize}
\item  {linearized theory;}

\item  {small perturbations of a smooth, time-in\-de\-pen\-dent back\-ground metric;}

\item  {post-newtonian theory.}
\end{itemize}

The simplest starting point for our discussion is certainly linearized theory, which is a
weak-field approximation to general relativity, where the equations are
written and solved in a nearly flat space-time. The static and wave parts of 
the field cleanly separate. We idealize
gravitational waves as a ``ripple'' propagating through a flat and empty
universe.

This picture is a simple case of the more general ``short-wave approximation'', in which
waves appear as small perturbations of a smooth background that is time-dependent 
and whose radius of curvature is much larger than the wavelength of the waves.
We will describe this in detail in \lecture~4. This approximation describes wave propagation well, but it is inadequate for wave generation. The most 
useful approximation for sources is the post-Newtonian approximation, where 
waves arise at a high order in corrections that carry general relativity away from its Newtonian limit; we treat these in \lecture s~5 and 6.  

For now we concentrate our attention
on linearized theory. We follow the notation and conventions of Misner, et al, (1973) and Schutz (1985). In particular we choose units in which $c=G=1$; Greek indices run from 0 to 3; Latin indices run from 1 to 3; repeated indices are summed; commas in subscripts or superscripts denote partial derivatives; and semicolons denote covariant derivatives. The metric has positive signature. See these two textbooks or others referred to at the end of these lectures for more details on the theory that we sketch here. For an even simpler introduction, based on a scalar analogy to general relativity, 
see reference \cite{bern}.

\section{Mathematics of linearized theory}

Consider a perturbed flat space-time. Its metric tensor can be written as 
\begin{equation}\label{eq:linear}
g_{\alpha \beta }=\eta _{\alpha \beta }+h_{\alpha \beta }\correction{}{,}\hspace{2cm}\left|
h_{\alpha \beta }\right| \ll 1\hspace{2cm}\alpha ,\beta =0,...,3
\end{equation}
where $\eta _{\alpha \beta }$ is the Minkowski metric (-1,1,1,1) and $%
h_{\alpha \beta }$ is a very small perturbation of the flat space-time metric. Linearized theory is an approximation to general relativity that is correct to first order in the size of this perturbation. Since the size of tensor components depends on coordinates, one must be careful with such a definition. What we require for linearized theory to be valid is that there should exist a coordinate system in which \Eref{eq:linear} holds in a suitably large region of space-time. Even though $%
\eta _{\alpha \beta }$ is not the true metric tensor,  we are free to 
\textit{define} raising and
lowering indices of the perturbation with $\eta _{\alpha \beta }$, as if it wree a tensor on flat space-time. We  write 
\[
h^{\alpha \beta }\correction{}{:}=\eta ^{\alpha \gamma }\eta ^{\beta \delta }h_{\gamma
\delta } .
\]
This leads to the following equation for the inverse metric, correct to first order (all we want in linearized theory):
\begin{equation}
g^{\alpha \beta }=\eta ^{\alpha \beta }-h^{\alpha \beta }.
\end{equation}

The mathematics is simpler if we define the \textit{trace-reversed} metric perturbation: 
\begin{equation}
\bar{h}_{\alpha \beta }\correction{}{:}=h_{\alpha \beta }-\frac{1}{2}\eta _{\alpha \beta }h,
\end{equation}
where $h\correction{}{:}=\eta _{\alpha \beta }h^{\alpha \beta }$. There is considerable coordinate freedom in the components $h_{\alpha \beta }$, since we can wiggle and 
stretch the coordinate system with a comparable amplitude and change the components. This coordinate freedom is called \correction{gauge freedom,}{\emph{gauge freedom,}} by analogy with electromagnetism. We use this freedom to enforce the \emph{Lorentz (or
Hilbert)} gauge:
\begin{equation}
\bar{h}\supr{\alpha \beta }\sub{,\beta}=0
\end{equation}

In this gauge the Einstein field equations (neglecting the quadratic and
higher terms in $h^{\alpha \beta }$) are just a set of decoupled linear
wave equations: 
\begin{equation}
\left( -\frac{\partial ^{2}}{\partial t^{2}}+\nabla ^{2}\right) \bar{h}%
_{\quad }^{\alpha \beta \;}=-16\pi T^{\alpha \beta }\correction{}{.}
\end{equation}
To understand wave propagation we look for the easiest solution of the vacuum
gravitational field equations: 
\begin{equation}
\Box \bar{h}^{\alpha \beta }\equiv \left( -\frac{\partial ^{2}}{\partial
t^{2}}+\nabla ^{2}\right) \bar{h}^{\alpha \beta }=0
\end{equation}
\emph{Plane wave\correction{s}{ solutions}} have the form: 
\begin{equation}
\bar{h}_{\alpha \beta }=\mathcal{A}\mathbf{e}_{\alpha \beta }\exp
(ik_{\gamma }x^{\gamma })\correction{}{,}
\end{equation}
where the amplitude $\mathcal{A}$, polarization tensor $\mathbf{e}^{\alpha \beta }$
and wave vector $k^{\gamma }$ are all constants. (As usual one has to take the real part of this expression.)

The Einstein equations imply
that the wave vector is ``light-like'', $k^{\gamma }k_{\gamma }=0$, and the gauge
condition implies that the amplitude and the wave vector are orthogonal: $\mathbf{e}^{\alpha \beta }k_{\beta }=0$.

Linearized theory describes a classical gravitational field whose 
quantum description would be a massless 
spin 2 field that propagates at the speed of light. We expect from this 
that such a field will have only \textbf{2} independent degrees of freedom 
(helicities in quantum language, polarizations in classical terms). To show this classically we remember that $h_{\alpha \beta }$ is
symmetric, so it has 10 independent components, and that Lorentz gauge applies 4 independent
conditions to these, reducing the freedom to 6. However, Lorentz gauge doe snot fully fix the coordinates. 
In fact if we perform another infinitesimal coordinate transformation ($x^{\mu }\rightarrow x^{\mu }+\xi ^{\mu }$, \correction{}{with }$\xi ^{\mu }\correction{}{\sub{,\nu}}=O(h)$) and
impose $\Box \xi ^{\mu }=0$, we remain in Lorentz gauge. We can use
this freedom to demand: 
\begin{eqnarray}\label{eq:transverse}
\mathbf{e}^{0\alpha } &=&0\Longrightarrow \mathbf{e}^{ij}k_{j}=0\quad \text{%
\emph{(transverse wave)}}\correction{}{,} \\ \label{eq:traceless}
\mathbf{e}\supr{i}\sub{i} &=&0\correction{\Longrightarrow}{} \text{\emph{(traceless wave)}}\correction{}{.}
\end{eqnarray}
These conditions can only be applied outside a sphere surrounding the
source. Together they put metric into the \textbf{transverse-traceless} (TT)
gauge. We will explicitly construct this gauge in chapter 5.

\section{Using TT gauge to understand gravitational waves}

The TT gauge leaves only \textbf{two independent polarizations} out of the
original ten,  and it ensures that $\overline{h}_{\alpha \beta }= h_{\alpha \beta }$. In order to understand the polarization degrees of freedom, let us 
take the wave to move in z-direction, so that $k_{z}=\omega $, $%
k^{0}=\omega $, $k_{x}=0$, $k_{y}=0$; the TT gauge conditions in \Eref{eq:transverse} and~(\ref{eq:traceless}) lead to $\mathbf{e}^{0\alpha }=\mathbf{e}^{z\alpha }=0$ and $\mathbf{e}^{xx}=-%
\mathbf{e}^{yy}$. This leaves only 2 independent components of
the polarization tensor, say $\mathbf{e}^{xx}$ and $\mathbf{e}^{xy}$ (\correction{}{which we denote by the symbols }$\mathbf{\oplus, \otimes }$).

A wave for which $\mathbf{e}^{xy}=0$ (pure $\oplus $ polarization) produces a
metric of the form: 
\begin{equation}\label{eq:oplus}
ds^{2}=-dt^{2}+(1+h_{+})dx^{2}+(1-h_{+})dy^{2}+dz^{2}\correction{}{,}
\end{equation}
where $h_{+}=\mathcal{A}\mathbf{e}^{xx}\exp [-i\omega (t-z)]$. Such a metric
produce\correction{}{s} opposite effects on proper distance on the two transverse axes, contracting one
while expanding the other. 
\begin{figure}[t]
\begin{center}
\caption{Illustration of two linear polarizations and the associated wave 
amplitude.}\label{fig:pol}
\includegraphics[width=8cm]{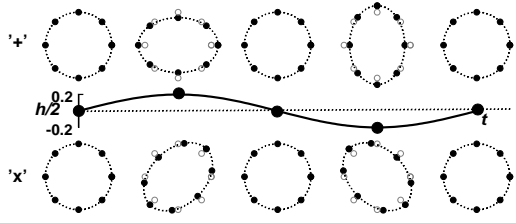}
\end{center}
\end{figure}
If $e^{xx}=0$ we have pure $\mathbf{\otimes }$
polarization $h_{\times }$ which can be obtained from the previous case by a
simple 45$^{{{}^{\circ }}}$ rotation, as in \Fref{fig:pol}. Since the wave equation and TT conditions are linear, a general wave will be a linear combination of these two
polarization tensors. A circular polarization basis would be: 
\begin{equation}
\mathbf{e}_{R}=\frac{1}{\sqrt{2}}(\mathbf{e}_{+}+i\mathbf{e}_{\times
}),\qquad \mathbf{e}_{L}=\frac{1}{\sqrt{2}}(\mathbf{e}_{+}-i\mathbf{e}%
_{\times })\correction{}{,}
\end{equation}
where $\mathbf{e}_{+}$, $\mathbf{e}_{\times}$ are the two linear polarization tensors and $\mathbf{e}_{R}$ and $\mathbf{e}_{L}$ are polarizations that rotate in the righthanded and lefthanded directions respectively.  
It is important to understand that, for circular polarization, the polarization pattern rotates around the central position, but  test particles themselves rotate only in small circles relative to the central position.

Now we compute the effects of a wave in the TT gauge on a particle at rest in the
flat background metric $\eta _{\alpha \beta \text{ }}$ before the passage of the 
gravitational wave. The geodesic equation 
\[\frac{d^{2}x^{\mu}}{d\tau ^{2}}+\Gamma\supr{\mu}\sub{\alpha\beta}\frac{dx^\alpha}{d\tau}\frac{dx^\beta}{d\tau} = 0 \]
implies in this case: 
\begin{equation}
\frac{d^{2}x^{i}}{d\tau ^{2}}=-\Gamma\supr{i}\sub{00}=-\frac{1}{2}%
(2h_{i0,0}-h_{00,i})=0,
\end{equation}
so that the particle \textit{does not move}. \textbf{The TT gauge, to first order
in }$h_{\alpha \beta }$\textbf{, represents a coordinate system that is
comoving with freely-falling particles. Because }$h_{0\alpha }=0$, \textbf{%
TT-time is proper time on the clock of freely-falling particles at rest.}

Tidal forces show the action of the wave independently of the coordinates. Let us consider the equation of geodesic deviation, which governs the separation of two neighboring freely falling test particles A and B. If the particles are initially at rest, then as the wave
passes it produces an oscillating curvature tensor, and the separation $\xi$ of the two particles is: 
\begin{equation}\label{eq:geoddev}
\frac{d^{2}\xi ^{i}}{dt^{2}}=R\supr{i}\sub{0j0}\xi ^{j}.
\end{equation}
To calculate the component $R\supr{i}\sub{0j0}$ of Riemann tensor in
\Eref{eq:geoddev}, we can use the metric in the TT gauge, because the 
Riemann tensor is gauge-invariant at linear order (see
exercise~4 at the end of this \lecture).
So we can replace $R\supr{i}\sub{0j0}$ by $R\correction{^{\text{TT}}}{}\supr{i}\sub{0j0}=\frac{1}{2}\correction{h_{j,00}^{i}}{h\supr{TTi}\sub{j,00}}$ and write: 
\begin{equation}\label{eq:stretch}
\frac{d^{2}\xi ^{i}}{dt^{2}}=\frac{1}{2}h^{\text{TT}}\supr{i}\sub{j,00}\xi ^{j}\correction{}{.}
\end{equation}
This equation, with an initial condition $\xi _{(0)}^{j}=$ \emph{constant},
describes the oscillations of B's location as measured in the proper reference frame
of A. The validity of \Eref{eq:stretch} is the
same as that of the geodesic deviation equation: geodesics have to be close to one another, in a
neighborhood where the change in curvature is small. In this approximation a gravitational wave is like an extra force, called a \textit{tidal force}, perturbing the proper distance between two test particles. If there are other forces on the particles, so that 
they are not free, then as long as the gravitational field is weak, one can 
just add the tidal forces to the other forces and work as if the particle were 
in special relativity.

\section{Interaction of gravitational waves with detectors}

We have shown above that the TT gauge is a 
particular coordinate system in which
the polarization tensor of a plane gravitational wave assumes a very simple
form. This gauge is comoving for freely-falling particles and so
it is not the locally Minkowskian coordinate system that would be used by an
experimenter to analyze an experiment. In general relativity one must 
always to be aware of how one's 
coordinate\correction{s}{} system is defined.

We shall analyze two typical situations:
\begin{itemize}
\item  the detector is small compared to the wavelength of the 
gravitational waves it is measuring; and 
\item  \correction{detectors}{the detector} is comparable to or larger than that wavelength.
\end{itemize}

In the first case we can use the geodesic deviation equation above to 
represent the wave as a simple extra
force on the equipment. Bars detectors can always be analyzed in this way.
Laser interferometers on the Earth can be treated this way too. In these
cases a gravitational wave simply produces a force to be measured. There is 
no more to say from the relativity point of view. The rest of the detection 
story is the physics of the detectors. Sadly, this 
is not as simple as the gravitational wave physics!

In the second case, the geodesic deviation
equation is not useful because we have to abandon the ``local mathematics'' of 
geodesic deviation and return to the 
``global mathematics'' of the TT gauge and metric components $%
h^{\text{TT}}\sub{\alpha \beta }$. Space-based interferometers like LISA, accurate ranging to solar-system spacecraft, and pulsar timing are all in this class. \correction{They are all}{Together with ground interferometers, these are} \textbf{beam detectors}: they use light (or radio waves) to register the waves.

To study these detectors, it is easiest to remain in the TT gauge and to calculate the effect of the waves on the (coordinate) speed of light. Let consider, for example, the $\oplus $ metric from \Eref{eq:oplus} 
and examine a null geodesic moving in the $x$-direction. The speed along
this curve is: 
\begin{equation}\label{eq:coordspeed}
\left( \frac{dx}{dt}\right) ^{2}=\frac{1}{1+h_{+}}.
\end{equation}
This is only a \emph{coordinate speed}, not a contradiction to special relativity.

To analyze the way in which detectors work, suppose one arm of an
interferometer lies along the $x$-direction and the wave, for simplicity, is
moving in the $z$-direction with a $\oplus $ polarization of \emph{any}
waveform $h_{+}(t)$ along this axis. (It is a plane wave, so its \correction{wave form}{waveform}
does not depend on $x$.) Then a photon emitted at time $t$ from the origin
reaches the other end, at a fixed coordinate position $x=L$, at the
coordinate time 
\begin{equation}
t_{far}=t+\int_{0}^{L}\sqrt{1+h_{+}(t(x))}dx\correction{}{,}
\end{equation}
where the argument $t(x)$ denotes the fact that one must know the time to
reach position $x$ in order to calculate the wave field. This implicit
equation can be solved in linearized theory by using the fact that $h_{+}$
is small, so we can use the first-order solution of \Eref{eq:coordspeed} above to
calculate $h_{+}(t)$ to sufficient accuracy.

To do this we expand the square-root in powers of $h_{+}$, and consider as a zero-order solution a photon
travelling at speed of light in the $x$-direction of a flat space-time. We can set $t(x)=t+x$%
. The result is: 
\begin{equation}
t_{out}=t+L+\frac{1}{2}\int_{0}^{L}h_{+}(t+x)dx\correction{}{.}
\end{equation}
In an interferometer, the light is reflected back, so the return trip takes
\begin{equation}
t_{return}=t+L+\frac{1}{2}\left[
\int_{0}^{L}h_{+}(t+x)dx+\int_{0}^{L}h_{+}(t+x+L)dx\right]\correction{}{.}
\end{equation}
What one monitors is changes in the time taken by a 
return trip as a function of
time at the origin. If there were no gravitational wave $t_{return}$
would be
 constant because $L$ is fixed, so changes indicate a gravitational wave. 

The rate of variation of the return time as a 
function of the start time $t$ is
\begin{equation}
\frac{dt_{return}}{dt}=1+\frac{1}{2}\left[ h_{+}(t+2L)-h_{+}(t)\right]\correction{}{.}
\end{equation}
This depends only on the wave amplitude when the beam leaves and when it
returns. 

Let us consider now a more realistic geometry than the previous one, and in
particular suppose that the wave travels 
at an angle $\theta $ to the $z$-axis in
the $x$-$z$ plane. If we re-do this calculation, allowing the phase of the
wave to depend on $x$ in an appropriate way, and taking into account the fact
that $h^{\text{TT}}_{+}\supr{xx}$ is reduced if the wave is not moving in a direction perpendicular to $x$, we find (see exercise~1 at the end of this chapter
for the details of the calculation) 
\begin{eqnarray}
\frac{dt_{return}}{dt} &=&\frac{1}{2}\left\{ \left( 1-\sin \theta \right)
h_{+}^{xx}\left( t+2L\right) -\left( 1+\sin \theta \right)
h_{+}^{xx}(t)\right.  \nonumber \\
&&\left. +2\sin \theta h_{+}^{xx}\left[ t+L\left( 1-\sin \theta \right)
\right] \right\}\correction{}{. }\label{eq:threeterm}
\end{eqnarray}
This three-term relation is the starting point for analyzing the response of
all beam detectors. This is directly what happens in radar 
ranging or in transponding to spacecraft, where a beam in only one 
direction is used. In long-baseline interferometry, one must analyze 
the second beam as well. We shall discuss these cases in turn.

\section{Analysis of beam detectors}

\subsection{Ranging to spacecraft}
Both NASA and ESA perform experiments in which they monitor the return time
of communication signals with inter-planetary spacecraft for the
characteristic effect of gravitational waves.  For missions to Jupiter and
Saturn, the return times are of order $2$-$4\times 10^{3}$~s. Any
gravitational wave event shorter than this will leave an imprint on the 
delay time 3 times: 
once when the wave passes the Earth-based transmitter, once when
it passes the spacecraft, and once when it passes the Earth-based receiver.
Searches use a form of pattern matching to look for this characteristic 
imprint. There are two dominant sources of noise: \correction{transmission}{propagation}-time 
irregularities caused by fluctuations in the solar wind plasma, and 
timing noise in the clocks used to measure the signals. The plasma 
delays depend on the radio-wave frequency, so by 
using two
transmission frequencies one can model \correction{}{and subtract }the plasma noise. Then if one 
uses the most stable atomic clocks, it is possible to
achieve sensitivities for $h$ of order $10^{-13}$. In the future, 
using higher radio frequencies, such experiments may reach $10^{-15}$. 
No positive detections have yet been made, but the chances are not 
zero. For example, if a small black hole fell into 
the massive black hole in the center of the Galaxy, it would produce 
a signal with a frequency of about 10~mHz and an amplitude significantly 
bigger than $10^{-15}$. Rare as this might be, it would be 
a dramatic event to observe.

\subsection{Pulsar timing}
Many pulsars, particular old millisecond pulsars, are extraordinarily
regular clocks, whose random timing irregularities are 
too small for even the best
atomic clocks to measure. Other pulsars have weak but observable 
irregularities. Measurements of or even upper limits on any of 
these timing irregularities for single
pulsars can be used to set \textit{upper limits} on any 
background gravitational wave field with periods comparable to or shorter 
than the observing time.  Here
the 3-term formula is replaced by a simpler two-term expression (see  
exercise~2 at the end of this chapter), because we only have a one-way
transmission from the pulsar to Earth. 
Moreover, the transit time of a signal to the Earth from the
pulsar may be thousands of years, so we cannot look for correlations between
the two terms in a given signal. Instead, the delay time is a combination of
the effects of uncorrelated waves at the 
pulsar when the signal was emitted and at
the Earth when it is received.

If one simultaneously observes two or more pulsars, the Earth-based part of
the delay is correlated between them, and this offers a means of actually detecting
long-period gravitational waves. Observations require timescale of several years in order to achieve the long-period stability of pulse arrival times, so this method is suited to looking for strong gravitational waves with periods of  several years.

\subsection{Interferometry}
An interferometer essentially measures changes in the difference in the return times
along two different arms. It does this by looking for changes in the 
interference pattern formed when the returning light beams are superimposed on 
one another.  The response of each arm will follow the three-term formula in 
\Eref{eq:threeterm}, 
but with a different value of $\theta $ for each arm, depending in a complicate way on
the orientation of the arms relative to the direction of travel and the 
polarization of wave. Ground-based interferometers are small enough to use
the small-$L$ formulas we derived earlier. But LISA, the space-based interferometer that will be described by Bender at this meeting, is
larger than a wavelength of gravitational waves for frequencies above 10
mHz, so a detailed analysis of its sensitivity requires the full three-term formula.

\section*{Exercises for \lecture~1}
Suggested solutions for these exercises are at the end of the lectures.
\begin{enumerate}
\item[\textbf{1}]  \emph{(a) Derive the full three-term return equation,
reproduced here: } 
\begin{eqnarray}
\frac{dt_{return}}{dt} &=&\frac{1}{2}\left\{ \left( 1-\sin \theta \right)
h_{+}^{xx}\left( t+2L\right) -\left( 1+\sin \theta \right)
h_{+}^{xx}(t)\right.   \nonumber \\
&&\left. +2\sin \theta h_{+}^{xx}\left[ t+L\left( 1-\sin \theta \right)
\right] \right\} 
\end{eqnarray}
\emph{(b) Show that, in the limit where }$L$\emph{\ is small compared to \correction{}{the }
wavelength of the gravitational wave, the derivative of the return time is
the derivative of the excess proper distance }$\delta L=Lh_{+}^{xx}(t)\cos
^{2}\theta $\emph{\ for small }$L$\emph{. Make sure you know how to
interpret the factor of }$\cos ^{2}\theta $\emph{.}

\emph{(c) Examine the limit of the three-term formula when the gravitational
wave is travelling along the }$x$\emph{-axis too (}$\theta =\pm \frac{\pi }{2%
}$\emph{): what happens to light going parallel to a gravitational
waves\bigskip }

\item[\textbf{2}]  \emph{Derive the two-term formula governing the delays
induced by gravitational waves on a signal transmitted only one-way, for example
from a pulsar to Earth.\bigskip }

\item[\textbf{3}]  \emph{A frequently asked question is: if gravitational
waves alter the speed of light, as we seem to have used here, and if they
move the ends of interferometer closer and further apart, might these
effects not cancel, so that there would be no measurable effects on light?
Answer this question. You may want to examine the calculation above: did we
make use of the changing distance between the ends, and why or why not?\bigskip }

\item[\textbf{4}]  \emph{Show that Riemann tensor is gauge-invariant in
linearized theory.} 
\end{enumerate}

\chapter{Gravitational wave detectors}

Gravitational radiation is a central prediction of general
relativity and \correction{their}{its} detection is a key test of the
integrity of the theoretical structure of Einstein's work. But in 
the long run, \correction{their}{its} importance as a tool for observational 
astronomy is likely to be even more important. We have excellent 
observational evidence from the Hulse-Taylor binary pulsar system 
(described in \lecture~3) that the predictions of general relativity 
concerning gravitational radiation are quantitatively correct. But 
we have very incomplete information from astronomy today about the 
likely sources of detectable radiation.

The gravitational wave spectrum is completely
unexplored, and whenever a new electromagnetic waveband has been opened 
to astronomy, \correction{we}{astronomers} have discovered completely unexpected phenomena. This 
seems just as likely to me to happen again with gravitational waves, 
especially because gravitational waves carry some kinds of information
that electromagnetic radiation cannot convey. Gravitational 
waves are generated by bulk motions of masses, and they encode the 
mass distributions and speeds. They are coherent and their low 
frequencies reflect the dynamical timescales of their sources. 

By contrast, electromagnetic waves 
come from individual electrons executing complex and partly random 
motions inside their sources. They are incoherent, and individual 
photons must be interpreted as samples of the large statistical ensemble 
of photons being emitted. Their frequencies are determined by microphysics 
on length scales much smaller than the structure of the astronomical 
system emitting them. From electromagnetic \correction{}{observations }we can make inferences about 
this structure only through careful modelling of the source. Gravitational 
waves, by contrast, carry information whose connection to the source 
structure and motion is fairly direct.

A good example is that of massive black holes in galactic nuclei. From 
observations that span the electromagnetic spectrum from radio waves to 
X-rays, astrophysicists have inferred that black holes of masses up 
to $10^9M_\odot$ are responsible for quasar emissions and control the 
jets that power the giant radio emission regions. The evidence for the 
black hole is very strong but indirect: no other known object can 
contain so much mass in such a small volume. Gravitational wave 
observations will tell us about the dynamics of the holes themselves, 
providing unique signatures from which they can be identified, 
measuring their masses and spins directly from their vibrational frequencies.
The interplay of electromagnetic and gravitational observations will 
enrich many branches of astronomy.

The history of gravitational wave detection started in the 1960's with J.
Weber at the University of Maryland. He built the first \textit{bar detector:} it was a
massive cylinder of aluminium ($\sim $2$\times 10^{3}$~kg) operating at room
temperature (300~K) with a resonant frequency of about 1600 Hz. This early prototype had a modest 
sensitivity, around 10$^{-13}$ or 10$^{-14}$.

Despite this poor sensitivity, in the late 1960's 
Weber announced the detection of a population of 
coincident events between two similar bars at a rate far higher than 
expected from instrumental noise. This news stimulated a number of other groups (at Glasgow, Munich,\correction{}{Paris, }
Rome, Bell Labs, Stanford, Rochester, LSU, MIT, Beijing, Tokyo) to build and
develop bar detectors to check Weber's results.
Unfortunately for Weber and for the idea that gravitational waves were easy
to detect, none of these other detectors found anything, even at times 
when Weber continued to find coincidences. Weber's 
observations remain unexplained even today. 
However the failure to confirm Weber was in a real sense a
confirmation of general relativity, because theoretical calculations 
had never predicted that reasonable signals would be strong enough 
to be seen by Weber's bars.

Weber's announcements have had a mixed effect on gravitational wave 
research. On the one hand, they have created a cloud under which 
the field has labored hard to re-establish its respectability in 
the eyes of many physicists. The legacy of this even today is 
an extreme cautiousness among the major projects, a conservatism
that will ensure that the next claim of a detection will be 
ironclad. On the other hand, the stimulus that Weber gave to 
other groups to build detectors has directly led to the present 
advanced state of detector development.

From 1980 to 1994 groups developed detectors in two different directions:
\begin{itemize}
\item  \textbf{Cryogenic bar detectors,} developed primarily 
at Rome/Frascati,  Stanford, LSU, and
Perth (Australia). The best of these detectors reach 
below $10^{-19}$. They are the only detectors operating continuously 
today and and they have performed a number of joint coincidence 
searches, leading to upper limits but no detections.
\item  \textbf{Interferometers,} developed at MIT, Garching (where the Munich
group moved), Glasgow, 
Caltech, and Tokyo. The typical sensitivity of these prototypes 
was $10^{-18}$. The first long coincidence
observation with interferometers was  the Glasgow/Garching 100-hour experiment  in 1989.\correction{}{\cite{100hrs}}
\end{itemize}

In fact, interferometers had apparently been considered by
Weber, but at that time the technology was not good enough for this kind of detectors.
Only 10-15 years later, technology had progressed. Lasers, mirror \correction{coatings 
mirror}{coating and} polishing techniques, and 
materials science had advanced far enough 
to allow the first practical interferometers, and it was 
clear that further progress would continue unabated. 
Soon thereafter several major collaborations were formed to 
build large-scale interferometric detectors:
\begin{itemize}
\item  LIGO: Caltech \& MIT (NSF)
\item  VIRGO: France (CNRS) \& Italy (INFN)
\item  GEO600: Germany (Max Planck) \& UK (PPARC)
\end{itemize}
Later other collaborations were formed in Australia (AIGO) and Japan 
(TAMA and JGWO).

At present there is still considerable effort in building successors to
Weber's original resonant-mass detector: ultra-cryogenic bars are in operation 
in Frascati and Padova, and they are expected to reach below $10^{-20}$.
Further, there are proposals for a new generation
of spherical or icosahedral solid-mass detectors from the USA (LSU), Brasil,
the Netherlands, and Italy.  Arrays of smaller bars have been 
proposed for observing the highest frequencies, where neutron star normal 
modes lie.

However, the real goal for the 
near future is to break through the $10^{-21}$ level, 
which is where theory predicts that it is not unreasonable to expect gravitational waves of the order of once per year. (See the discussion in \lecture~3 below.\correction{}{)}  The
first detectors to reach this level will 
be the large-scale interferometers that are
now under construction. They have very long
arms: LIGO, Hanford (WA) and Livingstone (LA), 4 km; VIRGO: Pisa, 3 km;
GEO600: Hannover, 600 m; TAMA300: Tokyo, 300 m.

The most spectacular detector in the near future 
is the space-based detector LISA, which has been  adopted by ESA 
(European Space Agency) as a Cornerstone mission
for the 21$\supr{st}$ century. 
The project is now gaining a considerable amount of
momentum in the USA,  and a collaboration between ESA and NASA seems likely.
This mission could be launched around 2010.

\section{Gravitational wave observables}
We have described earlier how different gravitational wave observables 
are from electromagneric observables. Here are the 
things that we want to measure when we detect gravitational waves:
\begin{itemize}
\item  $h_{+}(t) $, $h_{\times }(t) $, $phase(t) $: the amplitude and polarization of the wave, and the phase of polarization,  as functions of
time.  These contain most of the information about
gravitational wave.
\item  $\theta $, $\phi $: the direction on the sky of the source 
(except for observations of a stochastic background).
\end{itemize}

From this it is clear that gravitational wave detection is not the same
as electromagnetic radiation detection. In electromagnetic 
astronomy one almost always 
rectifies the electromagnetic wave, while we can follow the oscillations of
the gravitational wave. Essentially in electromagnetism one detects the power
in the radiation, while for gravitational radiation, 
as we have said before, one detects
the wave coherently.

Let us consider now what we can infer from a detection. 
If the gravitational wave has a short duration, of the order of the sampling 
time of the signal stream, then each detector will usually give
just a single number, which is the amplitude of the wave 
projected on the detector (a projection of the two
polarizations $h_{+}$ and $h_{\times }$). If the wave lasts more than 
one sampling time, then this information is a
function of time.

If the signal lasts for a sufficiently long time, then 
both the amplitude and the phase of the wave 
can be affected by the motion of the detector, which 
moves and turns with the motion of Earth. This produces 
an amplitude and phase modulation which is
not intrinsic to signal. If the signal's intrinsic form is understood, 
then this modulation can be used to determine the location of the source.
\correction{There are therefore two }{

We distinguish three }distinct kinds of signals, from the point of 
view of observations.

\textbf{Bursts} have a duration so short that modulation due 
to detector motion is not observable. During the detection, 
the detector is effectively
stationary. In this case we need at least 3, and preferably 4, 
\correction{detectors}{interferometers} to triangulate the
positions of bursts on the sky and to find the two polarizations $h_{+}$ and $%
h_{\times }$. (See discussions in Schutz 1989.) A network of detectors is essential to extract all the 
information in this case.

\textbf{Continuous waves} by definition last long enough for 
the motion of the detector to induce amplitude and
phase modulation. In this case, assuming a simple model for the intrinsic signal, 
we can use the information imprinted on the signal (the amplitude
modulation and phase modulation) to infer the position and polarization
amplitude of the source on the sky. A single detector, effectively, performs
aperture synthesis, finding tne position of the 
source and the amplitude of the wave entirely by itself. However, in 
order to be sure that the signal is not an artefact, it will be important
that the signal is seen by a second or third detector.

\textbf{Stochastic backgrounds }can be detected just like noise in a single
detector. If the detector noise is well understood, this excess noise may be detected as a stochastic background. This is closely analogous to the way 
the original microwave background detection was discovered.

A more reliable method for detecting stochastic radiation is the cross-correlation between two detectors, which
experience the same cosmological noise but have a different intrinsic noise.
Coherent cross-correlation between two detectors eliminates much detector noise and works best when detectors are closer than a wavelength.

In general, detection of
gravitational waves requires joint observing by a network of detectors,
both to increase the confidence of the detection and to provide accurate
information on other physical observables (direction, amplitude and so on).
Networks can be assembled from interferometers,  bars, or both.

\section{The physics of interferometers}

Interferometric gravitational wave detectors are the most sensitive 
instruments, and among the most complex, that have ever been constructed. 
They are remarkable for the range of physics that is important for 
their construction. Interferometer groups work at the forefront of development 
in lasers, mirror polishing and coating, quantum measurement, 
materials science, mechanical isolation, optical system design, 
and thermal science.  In this section 
we shall only be able to take a fairly superficial look at 
one of the most fascinating instrumentation stories of our age. A good 
introduction to interferometer design is Saulson (1994).

\begin{figure}[t]
\begin{center}
\caption{Five steps to a gravitational wave interferometer. (a)  
The simple Michelson. Notice that there are two return beams: one goes 
toward the photodetector and the other toward the laser. (b) Delay line: a 
Michelson with multiple bounces in each arm to enhance the signal. (c) Power 
recycling. The extra mirror recycles the light that goes towards the laser, 
which would otherwise be wasted. (d) Signal recycling. The mirror in front 
of the photodetector recycles only the signal sidebands, provided that 
in the absence of a signal no light goes to the photodetector. (e) Fabry-Perot 
interferometer. The delay lines are converted to cavities with 
partially silvered interior mirrors.  }\label{fig:interferometer}
\includegraphics[clip=true,width=0.45\textwidth]{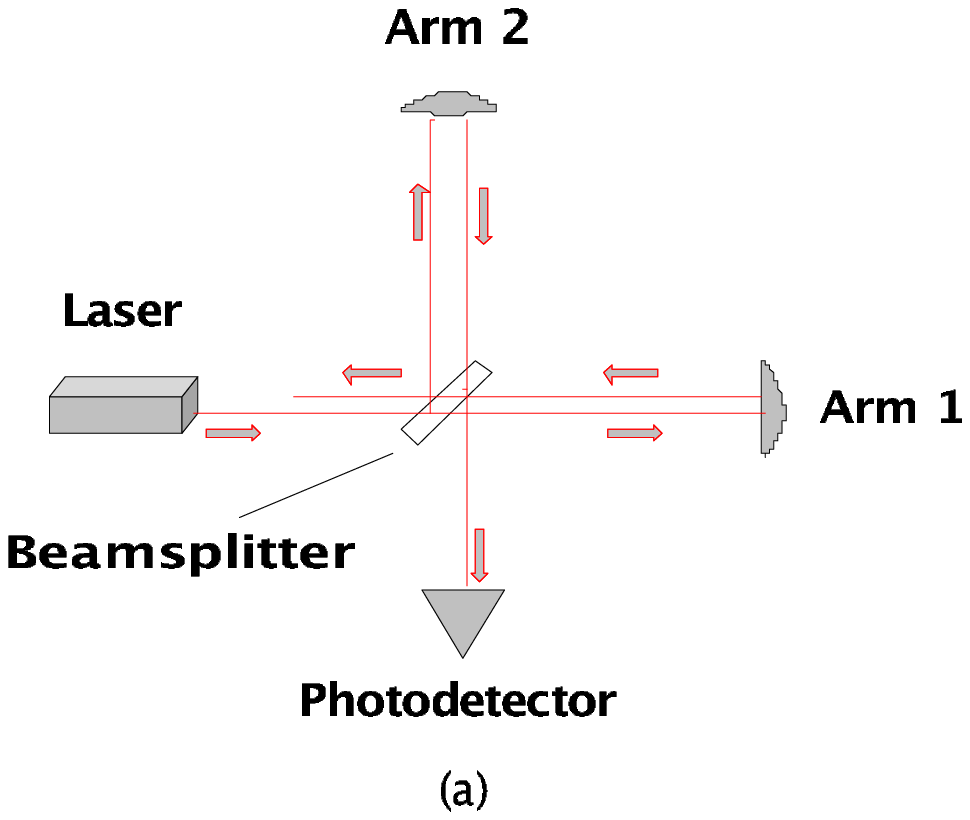}\hfill
\includegraphics[clip=true,width=0.45\textwidth]{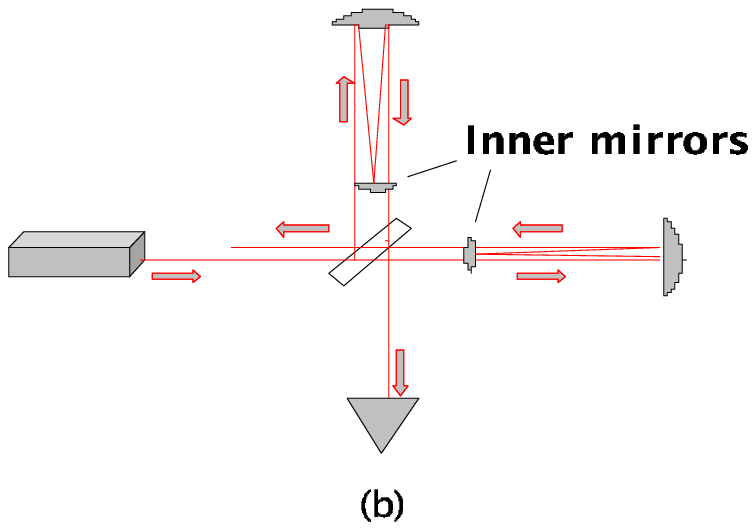}
\includegraphics[clip=true,width=0.45\textwidth]{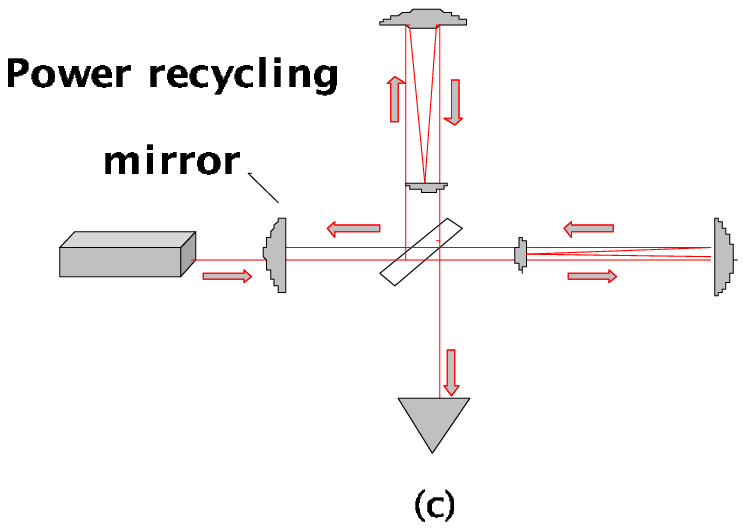}\hfill
\includegraphics[clip=true,width=0.45\textwidth]{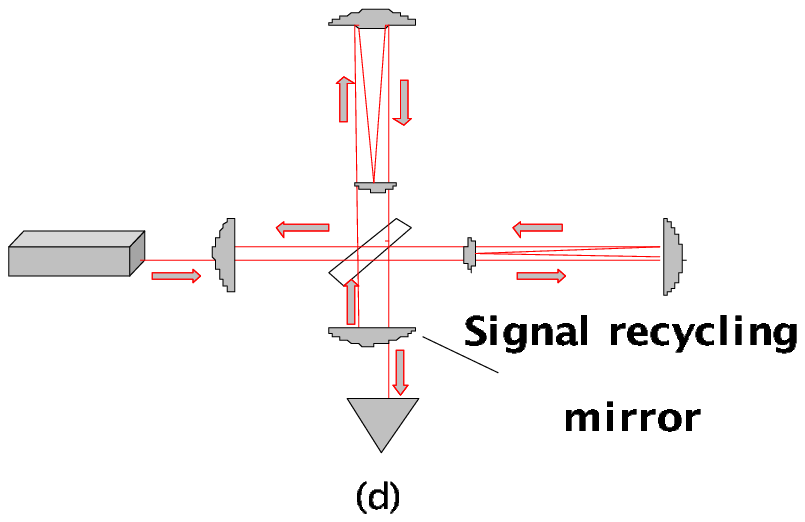}
\includegraphics[clip=true,width=0.45\textwidth]{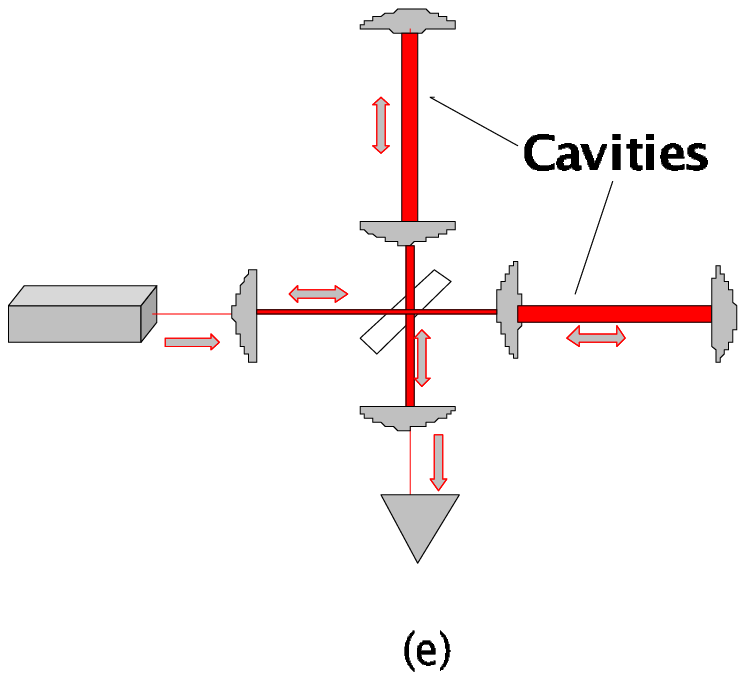}
\end{center}
\end{figure}

Interferometers use laser light to compare the lengths of two perpendicular
arms. The simplest design, originated by Michelson for his famous experiment 
on the velocity of light, uses light that passes up and down each arm 
once, as in the first panel in \Fref{fig:interferometer}. 
Imagine such an instrument with identical arms defined 
by mirrors that hang from supports, so they are 
free to move horizontally in response to a gravitational wave. 
If there is no wave, 
the arms have the same length, and the light from one arms returns 
exactly in phase with that from the other. 
When the wave arrives, the two arms typically 
respond differently. The arms are no longer the same length, and so 
the light that arrives back at the center from one arm will no longer be 
in phase with that arriving back from the other arm. This will produce 
a shift in the interference fringes between the two beams. This is the 
principle of detection.

Real detectors are designed to store the light in each arm for longer 
than just one reflection. (See part (b) of \Fref{fig:interferometer}.) 
It is optimum to store the light for one-half 
of the period of the gravitational wave, so that on each reflection the 
light gains an added phase shift. Michelson-type \correction{}{\emph{delay-line} }interferometers store 
the light by arranging multiple reflections. \textit{Fabry-Perot} interferometers 
store the light in cavities in each arm, allowing only a small fraction 
to escape for the interference measurement (part (e) of \Fref{fig:interferometer}).

An advantage of interferometers as detectors  is
that the gravitational-wave-induced phase shift of the light 
can be made larger simply by making the arm-length larger, since 
gravitational waves act by tidal forces.  
A detector with an arm length $l=4$~km responds to a
gravitational wave with an amplitude of $10^{-21}$ with 
\begin{equation}\label{eq:dlgw}
\delta l_{gw}\sim \frac{1}{2}hl\sim 2\times 10^{-18}\text{ m} 
\end{equation}
where $\delta l_{gw}$ is the change in the length of one arm. If 
the orientation of the interferometer is optimum, then the 
other arm will change by the same amount in the opposite direction, 
so that the interference fringe will shift by twice this length. 

If the light path is folded or resonated, as in Panels (b) or (d) of 
\Fref{fig:interferometer}, then the effective number of bounces can 
be traded off against overall length to achieve a given desired 
total path length, or storage time. Shorter interferometers with 
many bounces have a disadvantage, however: even though they can 
achieve the same response as a longer interferometer, the extra 
bounces introduce noise from the mirrors, as discussed below. There 
is therefore a big advantage to long-arm interferometers. 

There are three main sources of noise in interferometers: thermal, shot, 
and vibrational. To understand 
the way they are controlled, it is important to think in frequency space. 
Observations with ground-based detectors will be made in a range from 
perhaps 10~Hz up to 10~kHz, and initial detectors will have a much 
smaller observing bandwidth within this. Disturbances by noise that 
occur at frequencies outside the observation band can simply be filtered 
out. The goal of noise control is to reduce disturbances in the observation 
band.
\begin{itemize}
\item  \textbf{Thermal noise. }Interferometers work at room temperature, and
vibrations of the mirrors and of the suspending pendulum can mask
gravitational waves. To control this noise, scientists take advantage of 
the fact that thermal noise has its maximum amplitude at the 
frequency of the vibrational mode, and if the resonance of the mode 
is narrow (a high quality factor $Q$) then the amplitude at other 
frequencies is small. Therefore pendulum suspensions are designed with 
the pendulum frequency at about 1~Hz, well below the observing window, 
and mirror masses are designed to have principal vibration modes above 
1~kHz, well above the optimum observing frequency for initial 
interferometers. These systems are 
constructed with high values of $Q$ ($10^6$ or more) to reduce the noise 
in the observing band. Even so, thermal noise is typically a dominant 
noise below 100 or 200~Hz. 
\item  \textbf{Shot noise. }This is the principal limitation to sensitivity
at higher frequencies, above 200 to 300~Hz. It arises from the quantization 
of photons. When photons form interference fringes, 
they arrive at random times and make
random fluctuations in the light intensity that can look like a
gravitational wave signal; the more photons one uses, the smoother will be
the interference fringe. We can easily calculate this intrinsic noise. If $N$
is the number of photons emitted by the laser during our measurement, then 
as a random process the
fluctuation number $\delta N$ is proportional to the square root of $N$. If
we are using light with a wavelength $\lambda $ (for example infrared light
with $\lambda \sim 1$ $\mu $m) one can expect to measure lengths to an accuracy of
\[
\delta l_{shot}\sim \frac{\lambda }{2\pi \sqrt{N}}
\]
To measure a gravitational wave at a frequency $f$, one has to make at least 
$2f$ measurements per second, so one can accumulate photons for a time $%
\frac{1}{2f}$. If $P$ is the light power, one has 
\[
N=\frac{P}{\frac{hc}{\lambda }\cdot \frac{1}{2f}}
\]
It is easy to work out from this that, for $\delta l_{shot}$ to be equal to $\delta l_{gw}$ in \Eref{eq:dlgw}, one needs
light power of about 600 kW. No continuous laser could provide this much light 
to an interferometer.

The key to reaching such power levels inside the arms of a detector is a 
technique called power recycling (see Saulson 1994), 
first proposed by Drever and independently by Schilling. Normally, 
interferometers work on a ``dark fringe'', that is they are arranged so 
that the light reaching the photodetector is zero if there is no 
gravitational wave. Then, as shown in part (a) of \Fref{fig:interferometer}, 
the 
whole of the input light must emerge from the interferometer travelling 
towards the laser. If one places another mirror, 
correct positioned, between the laser and the beam splitter (part (c) of the figure), it 
will reflect this wasted light back into 
the interferometer in such a way that it adds coherently in phase with light
emerging from laser. In this way, light can be recycled and the required 
power levels in the arms achieved.

Of course, there will be a maximum recycling gain, which is set by 
mirror losses. Light power builds up until the laser merely resupplies 
the losses at the mirrors, due to scattering and absorption. 
The maximum power gain is 
\[
P=\frac{1}{1-R^{2}}
\]
where $1-R^{2}$ is the total loss summed over all the optical surfaces. 
For the very high-quality mirrors used in these projects, $%
1-R^{2}\sim 10^{-5}$. This reduces the power requirement 
for the laser by the same factor, down to about 6 W. 
This is attainable with modern laser technology.
\item  \textbf{Ground vibration }and \textbf{mechanical vibrations} are
another source of noise that must be screened out. Typical seismic 
vibration spectra fall sharply with frequency, so this is a problem 
primarily below 100~Hz. Pendulum suspensions are excellent mechanical 
filters above the pendulum frequency: 
it is a familiar elementary-physics demonstration that one can wiggle 
the suspension point of a pendulum vigorously at a high frequency 
and the pendulum itself remains undisturbed. Suspension designs 
typically involve multiple pendula, each with a frequency around 1~Hz. 
These provide very \correction{faat}{fast} roll-off of the noise above 1~Hz. Interferometer
spectra normally show a steep low-frequency noise ``wall'': this is
the expected vibrational noise amplitude.
\end{itemize}

In addition, there are noise sources that are not dominant in the 
present interferometers but will become important as sensitivity
increases. 
\begin{itemize}
\item  \textbf{Quantum effects: uncertainty principle noise. }Shot noise is a quantum noise, but in
addition there are other effects similar to those that bar detectors face, 
as described below: zero-point vibrations of suspensions and 
mirror surfaces, and back-action of light pressure fluctuations on the mirrors. These are small compared to present operating
limits of detectors, but they may become important in 5 years or so.
Practical schemes to reduced this noise have already been demonstrated in
principle, but they need to be improved considerably. This is the 
subject of considerable theoretical work at the moment.
\item \textbf{Gravity gradient noise. }Gravitational wave detectors respond 
to any changes in the gradients (tidal forces) of the 
local gravitational field, not just those carried 
by waves. The environment always contains changes in the Newtonian 
fields of nearby objects. Besides obvious ones, like people, there are 
changes caused by density waves in ground vibrations, atmospheric 
pressure changes, and many other disturbances. Below about 1~Hz, these
gravity gradient changes will be stronger than waves expected from 
astronomical objects, and they make it impossible to do observing 
at low frequencies from Earth. This is the reason that scientists 
have proposed the LISA mission, discussed below. Above 1~Hz, this 
noise does not affect the sensitivity of present detectors, but in 
10 years this could become a limiting factor.
\end{itemize}

Besides these noise sources, which are predictable and therefore 
can be controlled by detector design, it is possible that there
will be unexpected or unpredicted noise sources. 
Interferometers will be instrumented with 
many kinds of environmental monitors, but there may  
occasionally be noise that is 
impossible to identify. For this reason, short bursts of 
gravitational radiation must be identified at two or more 
separated facilities.  Even if detector noise is not at all 
understood, it is relatively easy to estimate from the observed 
noise profile of the individual detectors what the chances are of 
a coincident noise event between two detectors.

\subsection{New Interferometers and their capabilities}

Interferometers work over a broad bandwidth and they do not have any natural
resonance in their observing band. They are ideal for detecting 
\emph{bursts}, since one can perform pattern-matching over the 
whole bandwidth and detect such signals optimally. They are also 
ideal for searching for unknown \emph{continuous signals}, such 
as surveying the sky for neutron stars. And in observations 
of \emph{stochastic signals} by cross-correlating two detectors, they 
can give information about the spectrum of the signal. 

If an interferometer wants to study a signal with a known frequency, 
such as a known pulsars, then there is another optical technique 
available to enhance its sensitivity in a narrow bandwidth, at the 
expense of sensitivity outside that band. This is called \emph{signal 
recycling}\cite{meers}. In this technique, a further mirror is placed in front of 
the photo-detector, where the signal emerges from the interferometer 
(see part (d) of \Fref{fig:interferometer}).
If the mirror is chosen correctly, it will build up the signal, 
but only in a certain bandwidth. This modifies the shot noise 
in the detector, but not other noise sources. So it can improve 
sensitivity only at the higher frequencies where shot noise is 
the limiting factor.

\begin{figure}[t] 
\begin{center} 
\caption{TAMA300 sensitivity as a function of frequency. The vertical 
axis is the $1\sigma$ noise level, measured in strain per root Hz. 
To get a limit on the gravitational wave amplitude $h$, one must 
multiply the height of the curve by the square root of the 
bandwidth of the signal. This takes account of the fact that the 
noise power at different frequencies is independent, so the power
is proportional to bandwidth. The noise amplitude is therefore 
proportional to the square root of the bandwidth.}\label{fig:tama} 
\includegraphics[width=9cm]{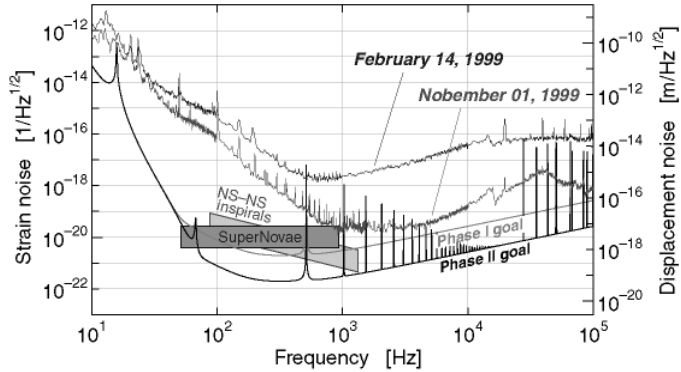} 
\end{center} 
\end{figure} 

Four major interferometer projects are now under construction,
and they could begin acquiring good data in the period between 2000 and
2003. They will all operate initially with a sensitivity approaching 10$%
^{-21}$ over a bandwidth between 50 and 1000 Hz. Early detections are by no means
certain, but recent work has made prospects look better for an early
detection than when these detectors were funded.

\textbf{TAMA300}\cite{tama} (Japan) is located in Tokyo, and its
arm length is 300~m. It began taking data without power recycling in 1999,
but its sensitivity is not yet near $10^{-21}$. Following improvements, 
especially power recycling, it should get to within a factor of 10 of 
this goal. But it is not planned as an observing instrument: it is a 
prototype for a kilometer-scale interferometer in Japan, currently called 
JGWO. By 2005 this may be operating, possibly with cryogenically cooled 
mirrors.

\textbf{GEO600}\cite{geo} (Germany \& Britain) is
located near Hannover (Germany). Its arm length is 600~m and the target date 
for first good data is now the end of 2001. Unlike TAMA, GEO600 is designed 
as a leading-edge-technology detector, where high-performance suspensions 
and optical tricks like signal recycling can be developed and applied. 
Although it has a short baseline, it will have a similar sensitivity 
to the larger LIGO and VIRGO detectors at first. At a later stage, LIGO 
and VIRGO will incorporate the advanced methods developed in GEO, and 
at that point they will advance in sensitivity, leaving GEO behind.

\begin{figure}[t]
\begin{center}
\caption{GEO600 noise curves. As for the TAMA curve, these 
are calibrated in strain per root Hz. The figure on the left shows 
GEO's wideband configuration; that on the right shows a 
possible narrowband operating mode.}\label{fig:geo}
\includegraphics[width=0.45\textwidth]{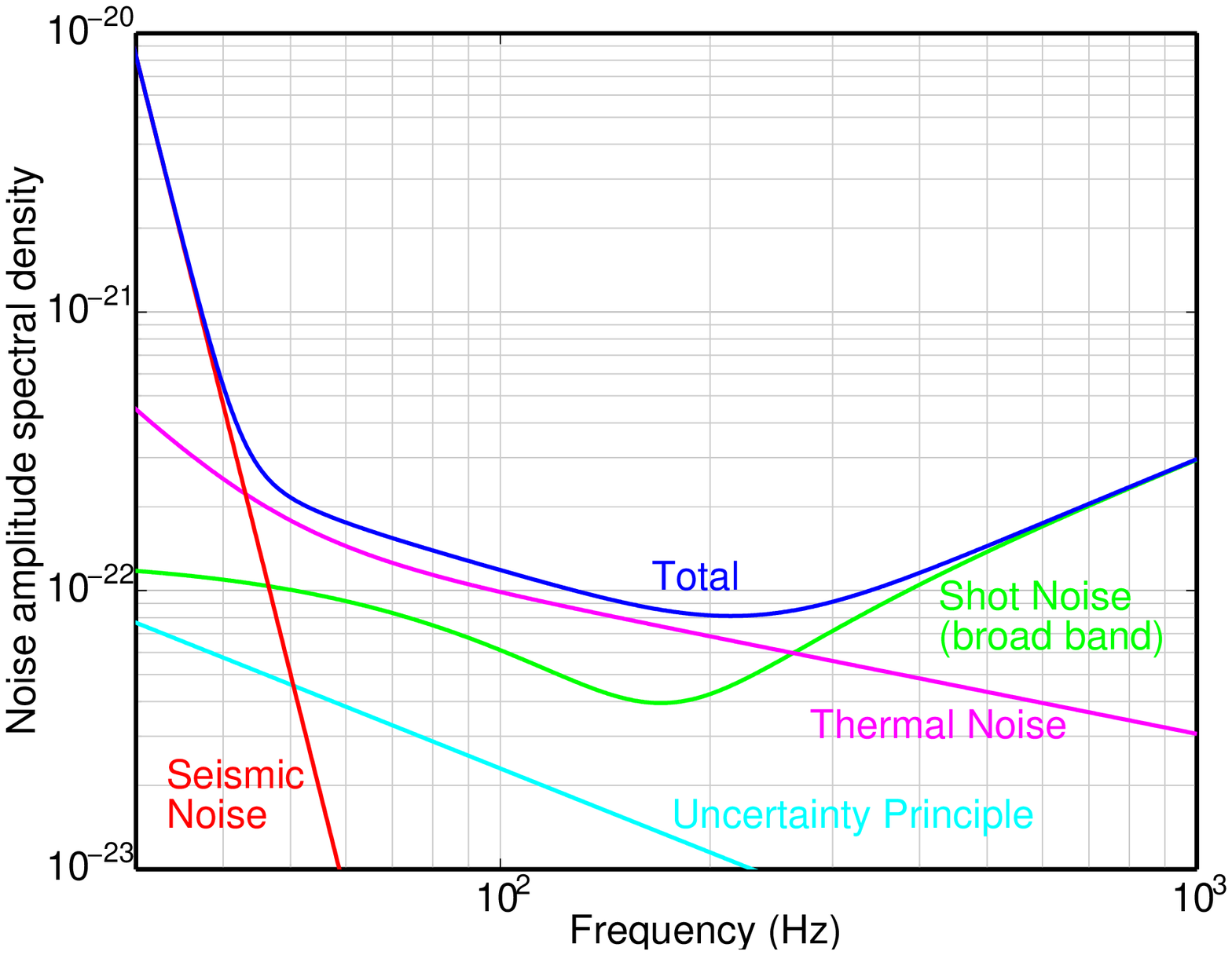}\hfill
\includegraphics[width=0.45\textwidth]{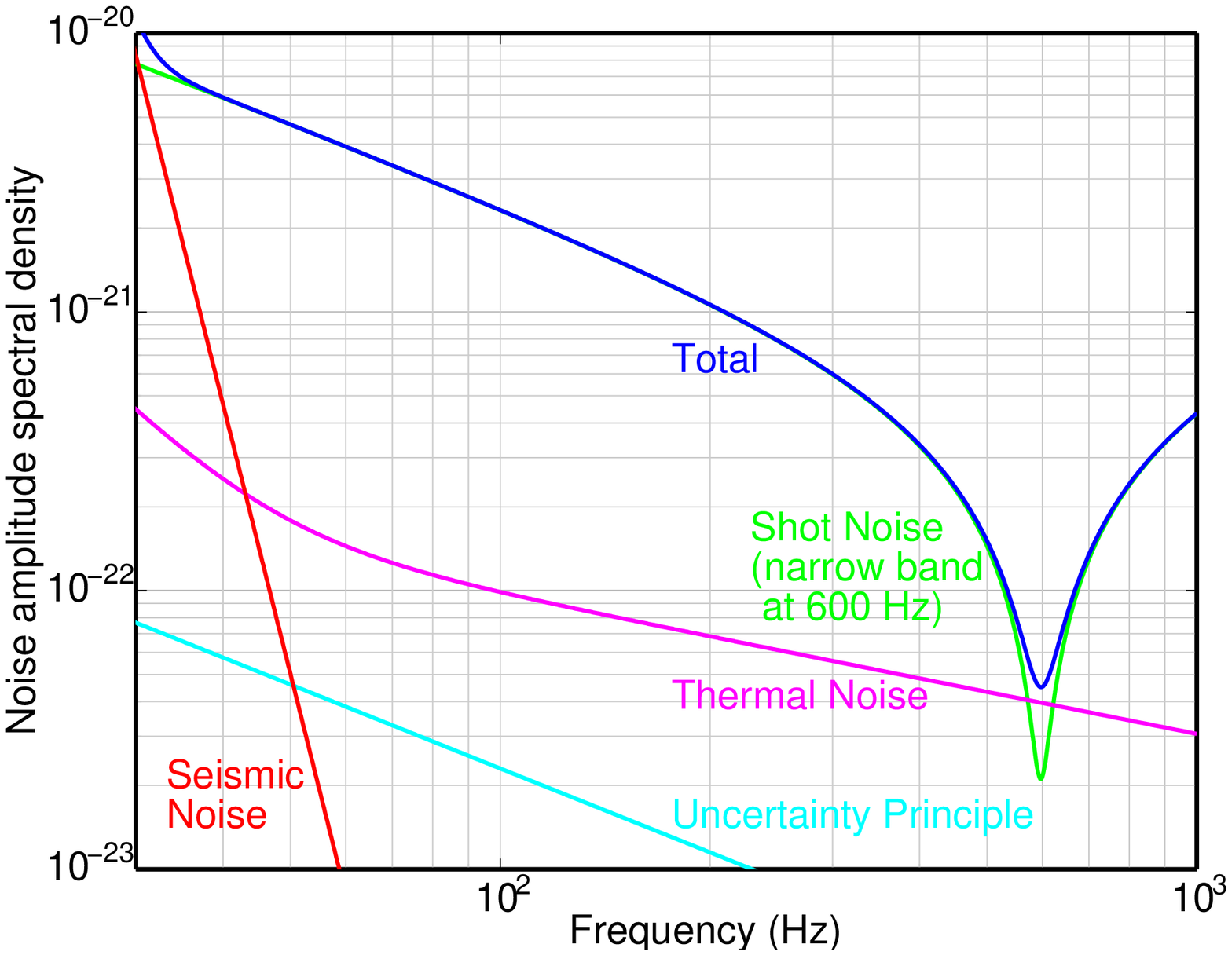}
\end{center}
\end{figure}
As we can see from \Fref{fig:geo} that the sensitivity of GEO600 depends on its
bandwidth, which in its turn depends on the signal recycling factor. GEO600 
can change its observing bandwidth in response to observing goals. 
By choosing low or high reflectivity for 
the signal recycling mirror, scientists can make GEO600 wide-band 
or narrow-band, respectively. The center 
frequency of the observing band (in the \correction{figure}{right panel of \Fref{fig:geo}} it is 
$\sim $ 600 Hz) can be tuned to any desired frequency by shifting
the position of the signal recycling mirror, 
thus changing the resonance frequency of the
signal recycling cavity. This feature could be useful when 
interferometers work with bars or when performing wide-band surveys.

\begin{figure}[t] 
\begin{center} 
\caption{Noise curves of the initial LIGO (left) and VIRGO (right) detectors. 
The VIRGO curve is in strain per root Hz, as the GEO curves earlier. The 
LIGO curve is calibrated in meters per 
root Hz, so to convert to a limit on $h$ one multiplies by the 
square root of the bandwidth and divides by the length of 
the detector arm, 4000~m.}\label{fig:ligovirgo}
\includegraphics[width=.45\textwidth]{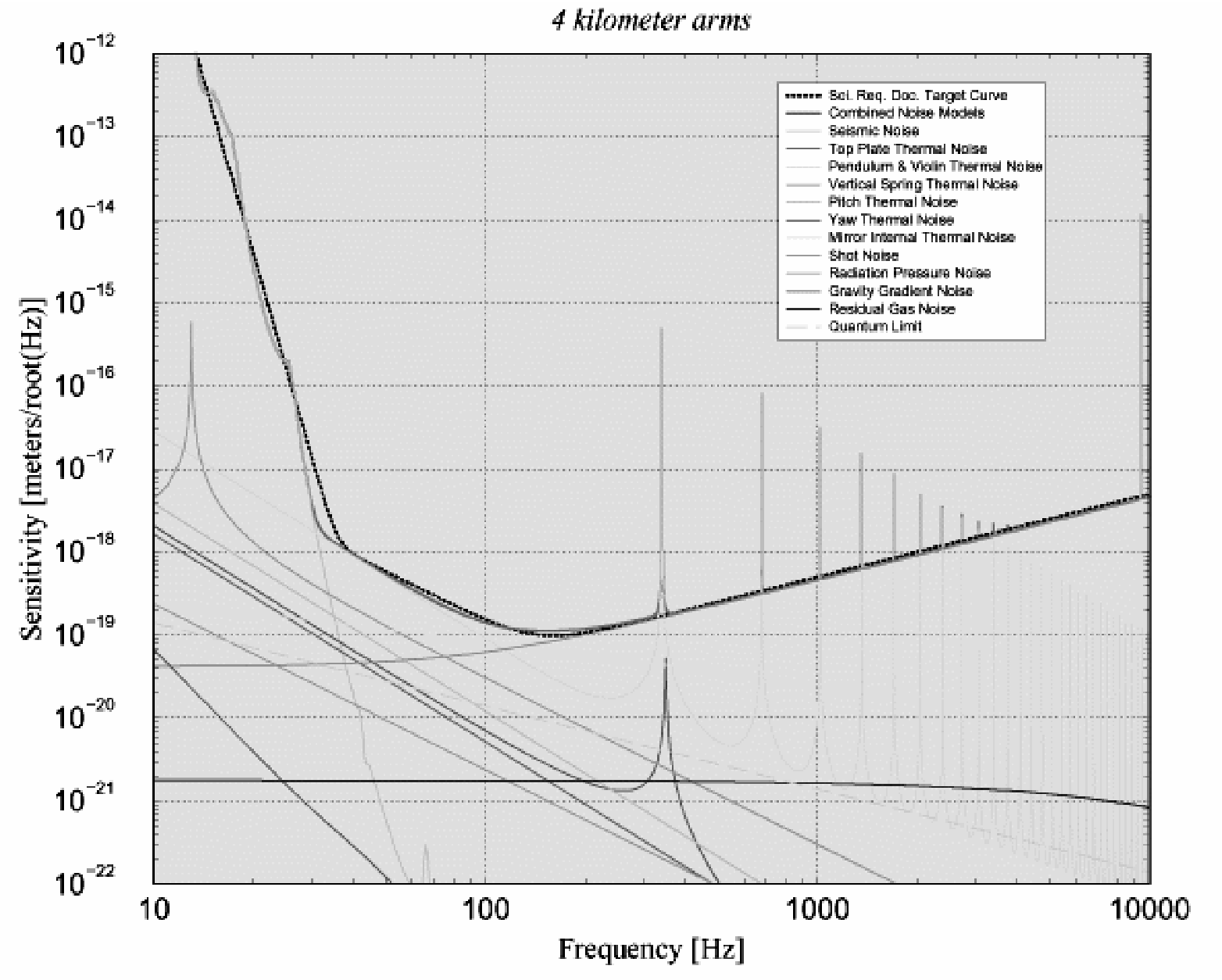} \hfill
\includegraphics[width=.45\textwidth]{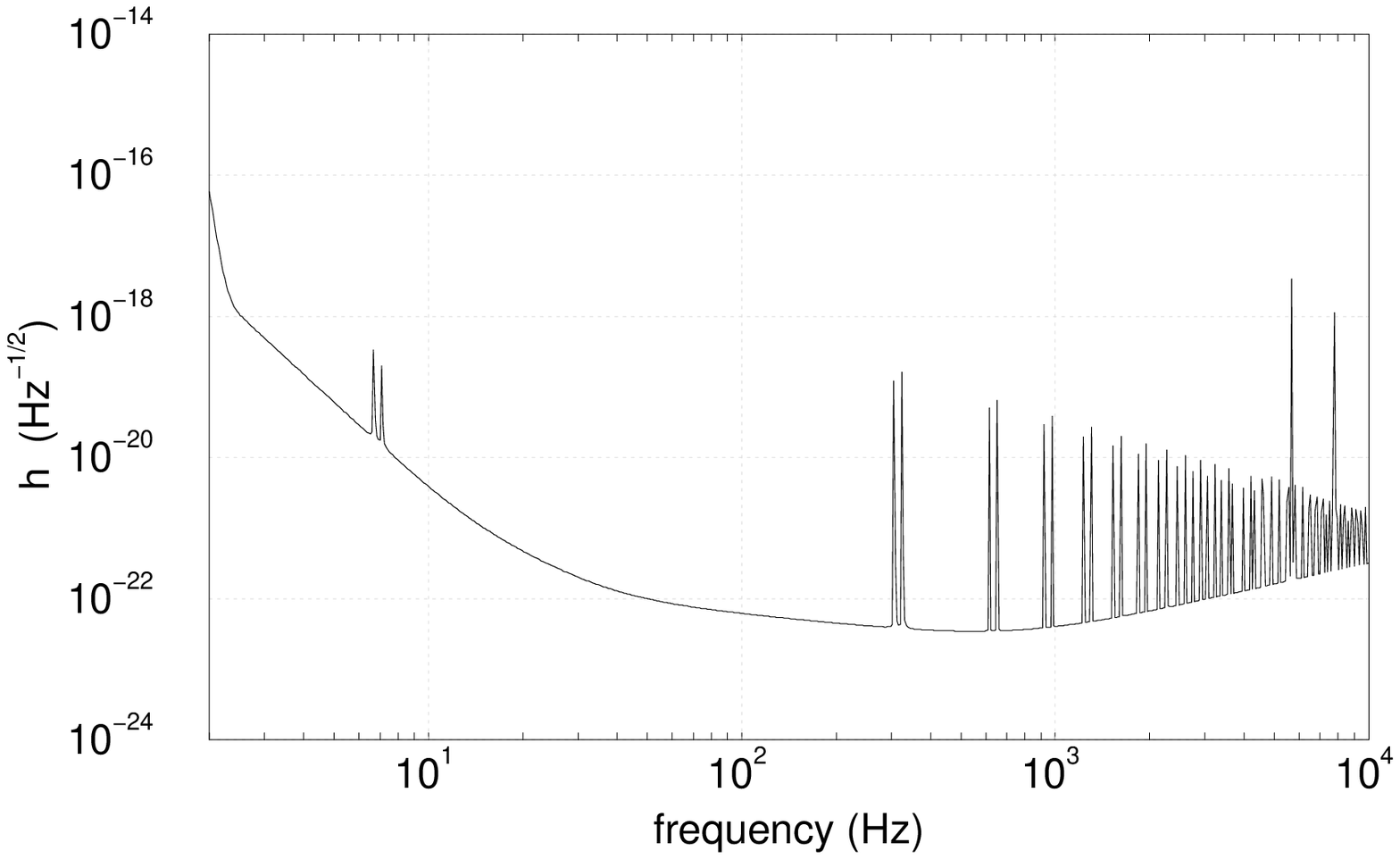} 
\end{center} 
\end{figure}

\textbf{LIGO}\cite{ligo} (USA) is building two detectors of arm length 
4~km. One is located in Hanford WA and the other in  Livingstone LA.
The target date for observing is mid-2002. The two detectors are 
placed so that their antenna patterns overlap as much as possible
and yet they are far enough apart that there will be a measurable 
time-delay in most coincident bursts of gravitational radiation. This 
delay will give some directional information. The Hanford detector 
also contains a half-length interferometer to assist in coincidence
searches. The two LIGO detectors are the best placed for doing 
cross-correlation for a random background of gravitational waves. 
LIGO's expected initial noise curve is shown in \Fref{fig:ligovirgo}.
These detectors have been constructed to have a long lifetime. 
With such long arms they can benefit from upgrades in laser 
power and mirror quality. LIGO has defined an upgrade goal called 
LIGO II, which it hopes to reach by 2007, which will observe 
at $10^{-22}$ or better over a bandwidth from 10~Hz up to 1~kHz.

\textbf{VIRGO}\cite{virgo} (Italy \& France) is building a 3~km detector
near Pisa. Its target date for good data is 2003.  Its expected 
initial noise curve is  shown in \Fref{fig:ligovirgo}. Like LIGO, it can 
eventually be pushed to much higher
sensitivities with more powerful lasers and other optical enhancements. 
VIRGO specializes in sophisticated suspensions, and the control of 
vibrational noise. Its goal is to observe at the lowest possible 
frequencies from the ground, at least partly to be able to examine 
as many pulsars and other neutron stars as possible.

\section{The physics of Resonant Mass Detectors}

The principle of operation of bar detectors 
is to use the gravitational tidal force
of the wave to stretch a massive cylinder along its axis, and then to
measure the elastic \correction{energy of vibrations gained by the cylinder}{vibrations of the cylinder}.

Let us suppose we have a typical bar with length $L\sim 1$~m. (In 
the future, spheres may go up to 3~m.) Depending on the length 
of the bar and its material, the resonant frequency will be $%
f\sim 500$~Hz to 1.5~kHz and mass $M\sim 1000$~kg. A short burst
gravitational wave $h$ will make the bar vibrate with an amplitude 
\[
\delta l_{gw}\sim hl\sim 10^{-21}\;\text{m.} 
\]
Unlike the interferometers, whose response is simply given by this 
equation, the bars respond in a complicated way depending on all their 
internal forces. But if the duration of the wave is short, 
the amplitude will be of the same order as that 
given here. If the wave has long duration and is near the bar's 
resonant frequency, then the signal can build up to much larger
amplitudes. Normally, bar detector searches have been targeted at  
short-duration signals.

The main sources of noise that compete with this very small amplitude are:
\begin{itemize}
\item  \textbf{Thermal noise. }This is the most serious source of 
noise. Interferometers can live with room-temperature thermal noise 
because their larger size makes their response to a gravitational 
wave larger, and because they observe at frequencies far from the 
resonant frequency, where the noise amplitude is largest.  
But bars observe at the resonant frequency and have a very short 
length, so they must reduce thermal noise by going to low 
temperatures. The best ultra-cryogenic 
bars today operate at about $T=100$ mK, 
where the r.m.s.\ amplitude of vibration is found by setting the 
kinetic energy of the normal mode, $M(\delta\dot{l})^2/2$, equal 
to $kT/2$, the equipartition thermal energy of a single degree
of freedom. This gives then
\[
\left\langle \delta l^{2}\right\rangle _{th}^{\frac{1}{2}}=\left( \frac{kT}{%
4\pi ^{2}Mf^{2}}\right) ^{\frac{1}{2}}\sim 6\cdot 10^{-18}\ \text{m}
\]
This is far larger than the gravitational wave amplitude. 
In order to
detect gravitational waves against this noise, bars are constructed to
have a very high $Q$, of order $10^{6}$ or better. 

The reason that 
bars need a high $Q$ is different from the reason that interferometers 
also strive for high-$Q$ systems. To see how $Q$ helps bars, recall
that  $Q$ is defined 
as $Q=f\correction{/}{\cdot} \tau $ where $f$ is the resonant frequency of the mode and 
$\tau $ is the decay time of the oscillations. If $Q$ is large, then 
the decay time is long. If the decay time is long, then  
the amplitude of oscillation changes very slowly in thermal 
equilibrium. Essentially, the bar's mode of vibration 
changes its amplitude by a random walk with very small steps, 
taking a time $\frac{Q}{f}%
\sim 1000$ s to change by the full amount. On the other hand, 
a gravitational wave burst will
cause an amplitude change in a time of order  1~ms, during which
the thermal noise will have random walked to an
expected amplitude change that is 
$Q^{\frac{1}{2}}=\left( \frac{1000\text{ s}}{1\ 
\text{ms}}\right) ^{\frac{1}{2}}$ times smaller. In  this case 
\[
\left\langle \delta l^{2}\right\rangle _{th:\ 1\ \text{ms}}^{\frac{1}{2}%
}=\left( \frac{kT}{4\pi ^{2}Mf^{2}Q}\right) ^{\frac{1}{2}}\sim 6\cdot
10^{-21}\ \text{m}
\]
Thus, thermal noise only affects a measurement to the extent that 
it changes the amplitude of vibration during the time of the 
gravitational wave burst. This change is similar to that produced
by a gravitational wave of amplitude $6\times10^{-21}$. It 
follows that, if thermal noise were the only noise source, 
bars would be operating at around $10^{-20}$ today. Bar groups expect in fact 
to reach this level during the next few years, as they 
reduce the other competing sources of noise. Notice that 
the effect of thermal noise has nothing to do with the frequency 
of the disturbance, so it is not the reason \correction{bar's}{that bars} observe near
their resonant frequency. In fact, both thermal impulses and 
gravitational wave forces
are mechanical forces on the bar, 
and the ratio of their induced vibrations is the
same at all frequencies for a given applied impulsive force.

\item \textbf{Sensor noise. }Because the oscillations of the bar
are very small, bars require a \emph{transducer }to 
convert the mechanical energy of vibration 
into electrical energy, and an \emph{amplifier} that increases the electrical
signal to a level where it can be recorded. If the 
amplifier were perfect, then the detector would in fact 
be broad-band: it would amplify the smaller off-resonant responses
just as well as the on-resonance ones. Conversely, real bars are 
narrow-bad because of sensor noise, not because of their mechanical resonance.

Unfortunately sensing is not perfect: amplifiers introduce noise and this
makes small amplitudes harder to measure. The amplitudes of vibration are
largest in the resonance band near the resonant frequency $f_0$, 
so amplifier noise limits the
detector sensitivity to frequencies near $f_0$. Now, the  signal (a typical
gravitational wave burst) has a duration time $\tau _{w}\sim 1$~ms, so the 
amplifier's bandwidth should be at least $1\diagup \tau _{w}$ in order 
for it to be able to record a signal
every $\tau _{w}$. In other words, bars require amplifiers with very small noise
in a large bandwidth ($\sim $ 1000 Hz) near $f_0$ (note that this band is much
larger than $f\diagup Q$). Today typical bandwidths of realizable 
amplifiers are 1~Hz, but in the very near 
future it is hoped to extend these to 10~Hz, and eventually to 100~Hz.
\item  \textbf{Quantum limit.} According to the Heisenberg uncertainty 
principle, the zero-point vibrations of a
bar with a frequency of 1~kHz have r.m.s.\ amplitude
\[
\left\langle \delta l^{2}\right\rangle _{quant}^{\frac{1}{2}}=\left( \frac{%
\hslash }{2\pi Mf}\right) ^{\frac{1}{2}}\sim 4\cdot 10^{-21}\ \text{m}
\]
This is bigger than the expected signal, and comparable to the thermal limit
over 1 ms. It represents the accuracy with which one can measure the 
amplitude of vibration of the bar. 
So as soon as current detectors improve their thermal limits,
they will run into the quantum limit, which must be overcome before a signal at 10$%
^{-21}$ can be seen with such a detector. One way to overcome this limit 
is  by increasing the size of the detector
and even by making it spherical. This increases its mass dramatically, 
pushing the quantum limit down below 10$^{-21}$.

Another way around the quantum limit is to avoid measuring $\delta l$, but 
instead to measure other
observables. After all, the goal is to infer the gravitational wave
amplitude, not to measure the state of vibration of the bar. It is 
possible to define a pair of conjugate observables that have the 
property that one of them can be measured arbitrarily accurately 
repeatedly, so that the resulting inaccuracy of knowing the conjugate
variable's value does not disturb the first variable's value. Then,
if the first variable responds to the gravitational wave, the 
gravitational wave may be measured accurately, even though the 
full state of the bar is poorly known. 
This method is called \emph{``back reaction evasion''.} The theory
was developed in a classic paper by
Caves et al\cite{caves}.  
However, no viable schemes to do 
this have been demonstrated for bar detectors so far. 
\end{itemize}

\subsection{New Bar Detectors and their capabilities}
Resonant-mass detectors are limited by properties of materials and, as we
have just explained, they have their best sensitivity in a narrow band around
their resonant frequency. However they can usefully explore higher
frequencies (above 500~Hz), where the interferometer noise curves
are rising (see earlier figures).
 
From the beginning, bars were designed to detect \emph{bursts.} If 
the burst radiation carries significant energy in the bar's
bandwidth, then the bar can do well. Standard assumptions about gravitational
collapse suggest a signal with a broad spectrum to 1 kHz or more, 
so that most of the sensitive bars today would be suited to observe 
such a signal. Binary
coalescence has a spectrum that peaks at low frequencies, so bars are not 
partiularly well-suited for such signals. On the other hand, 
neutron-star and stellar-mass black-hole normal 
modes range in frequency from about 1~kHz up to 10~kHz, so suitably 
designed bars could in principle go after these interesting signals.

A bar gets all of its sensitivity in a relatively  narrow bandwidth, 
so if a bar and an interferometer can both barely detect a burst of amplitude 
$10^{-20}$, then the bar has much greater sensitivity than the interferometer
in its narrow band, and much worse at other frequencies. This has led
recently to interest in bars as detectors of  
\emph{continuous signals. } If 
the signal frequency is in the observing band of the bar, it can do very well 
compared to interferometers. Signals from millisecond pulsars and possible 
signals from X-ray binaries 
are suitable if they have the right frequency. But most known 
pulsars will radiate at frequencies rather low compared to the 
operating frequencies of present-day bars. 

The excellent sensitivity of bars in their narrow bandwidth also suits
them to detecting    \emph{stochastic signals.} 
Cross-correlations of two bars or of bars with interferometers can 
be better than searches with first-generation interferometers.\cite{astone}
One gets no spectral information, of course, and in the long run 
expected improvements in interferometers will overtake bars in 
this regard.

Today's best bar detectors are orders of magnitude more sensitive 
than the original Weber bar.  Two \textbf{%
ultra-cryogenic} bars have been built and are operating at thermodynamic 
temperatures below 100~mK: \textbf{NAUTILUS}\cite{nautilus} at Frascati, 
near Rome, and \textbf{AURIGA}\correction{}{\cite{auriga}} in Legnaro. With a mass of several tons, 
these may be the coldest
massive objects ever seen anywhere 
in the universe. These are expected soon to reach a sensitivity of 
$10^{-20}$ near 1~kHz. Already they are performing coincidence 
experiments with bars at around 4~K at Perth, Australia, and at 
LSU.

Proposals exist in the Netherlands, Brazil, Italy, and the 
USA for \textbf{spherical or
icosahedral detectors}. (See links from reference \cite{nautilus}.) 
These detectors
have more mass, so they could reach $10^{-21}$ near 1~kHz. Because 
of their shape, they have omni-directional antenna patterns; if 
they are instrumented so that all 5 independent fundamental quadrupolar modes
 of vibration can be monitored, they can do all-sky observing and determine
directions as well as verify detections using  
coincidences between modes of the same antenna.

\section{A detector in space}

As we have noted earlier, gravitational waves from astronomical 
objects at frequencies below 1~Hz are obscured by Earth-based gravity-gradient 
noise. \correction{Ddetectors}{Detectors} must go into space to observe in this very 
interesting frequency range.

The \textbf{LISA}\cite{lisa} mission is likely to be the first 
such mission to fly. LISA will be a triangular array of spacecraft, 
with arm lengths of $5\times 10^6$~km, orbiting the Sun in the 
Earth's orbit, about $20^o$ behind the Earth. The spacecraft will 
be in a plane inclined to the ecliptic by $60^o$. The three arms 
can be combined in different ways to 
form two independent interferometers. During the mission the 
configuration of spacecraft rotates in its plane, and the plane 
rotates as well, so that LISA's antenna pattern sweeps the sky.
 
LISA has been named a Cornerstone mission 
of the European Space Agency (ESA), and NASA has recently formed its 
own team to study the same mission, with a view toward a collaboration
with ESA. LISA  will be sensitive in a range
from 0.3 mHz to about 0.1 Hz, and it will be able to  
detect known binary star systems in the Galaxy and 
binary coalescneces of supermassive black holes anywhere they occur 
in the Universe. A joint ESA-NASA project looks very likely,
aiming at a launch around 2010. A technology demonstration mission might 
be launched in 2005 or 2006.

\begin{figure}[t]
\begin{center}
\caption{LISA sensitivity to binary systems in the Galaxy (top) 
and to massive black hole coalescences (bottom). The top figure 
is calibrated in the intrinsic amplitude of the signal, and the 
noise curve shows the detection threshold ($5\sigma$) for a 1-year 
observation. It also shows the confusion limit due to unresolved 
binary systems. The bottom panel shows the effective amplitude 
of signals from coalescences of massive black holes. Since 
some such events last less than 1 year, what is shown is the 
expected signal-to-noise ratio of the observation.}\label{fig:lisa}
\includegraphics[width=0.75\textwidth,clip=true]{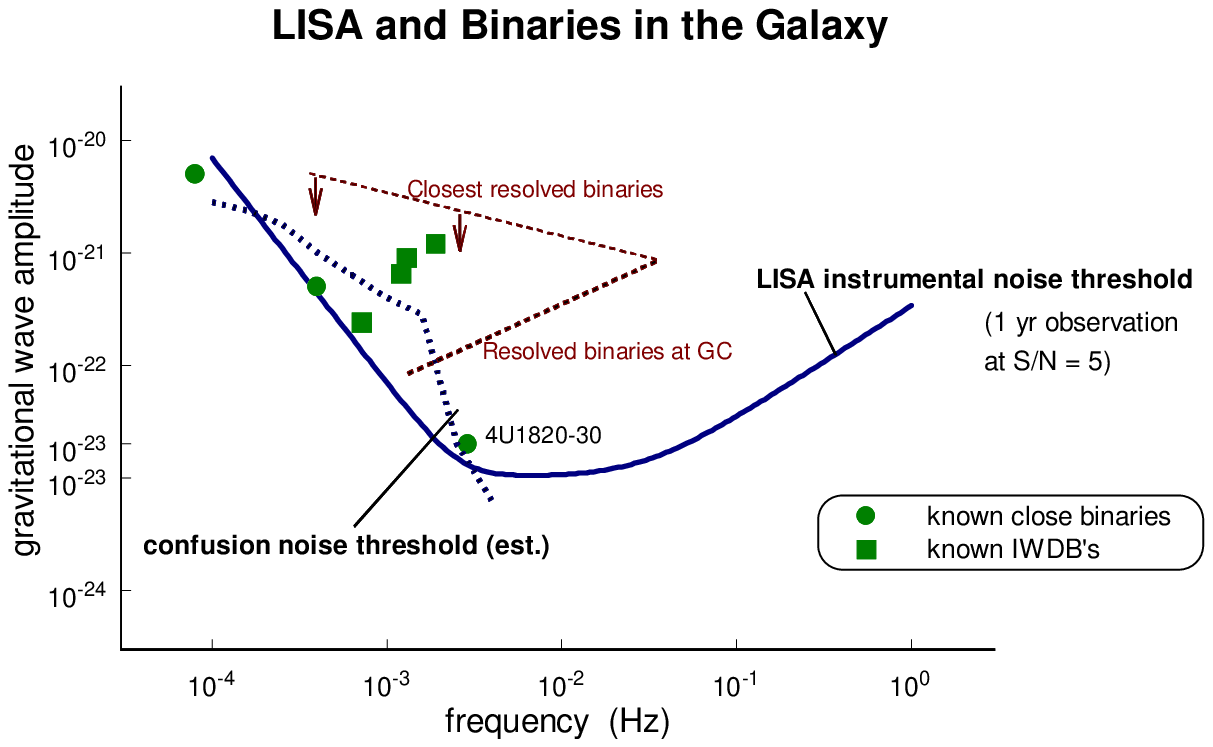}
\includegraphics[width=0.75\textwidth,clip=true]{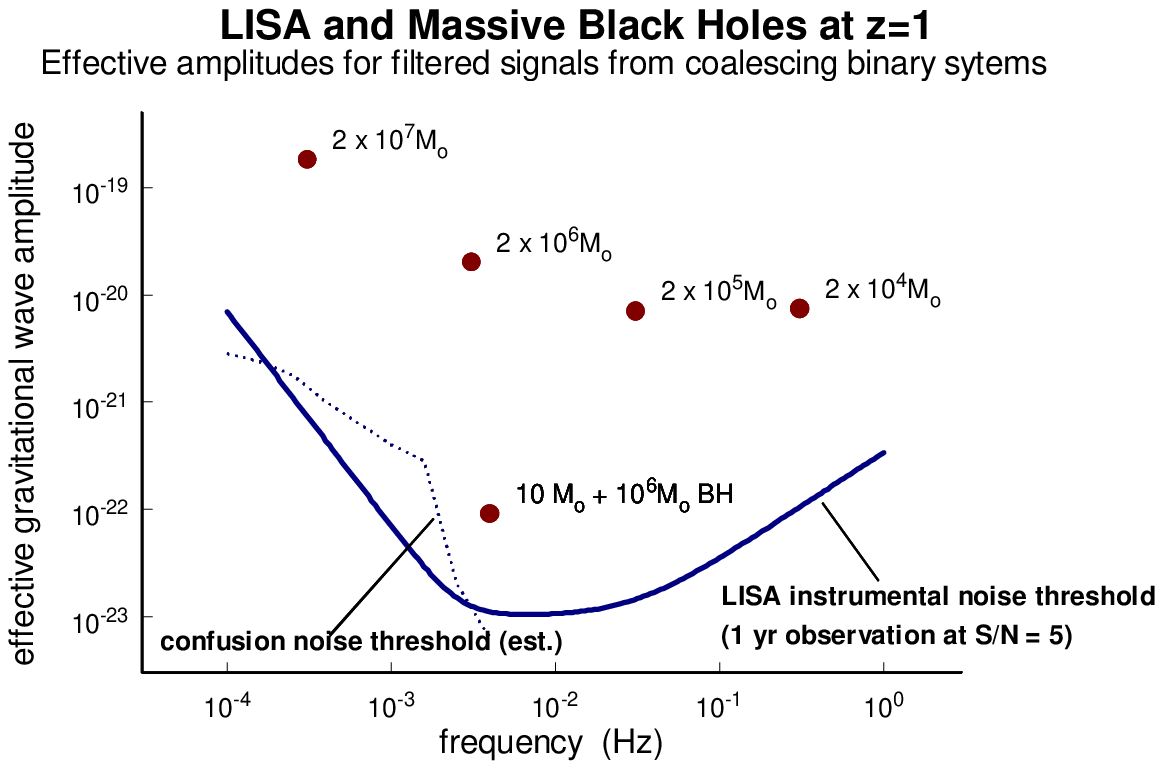}
\end{center}
\end{figure}
LISA's technology is fascinating. We can only allude to the most 
interesting parts of the mission here. A full description can be found 
in the pre-Phase A study document.\cite{prephasea} The most innovative 
aspect of the mission is \emph{drag-free control.} In order to guarantee 
that the interferometry is not disturbed by external forces, such as 
fluctuations in solar radiation pressure, the mirror that is the 
reference point for the interferometry is on a free mass inside 
the spacecraft. The spacecraft acts as an active shield, sensing
the position of the free mass, firing jets to counteract external 
forces on itself and ensure that it does not disturb the free mass. The 
jets themselves are remarkable, in that they must be very weak 
compared to most spacecraft\correction{s'}{'s} control jets, and they must be capable 
of very precise control. They will work by expelling streams of 
ions, accelerated and controlled by a high-voltage electric field. 
Fuel for these jets is not a problem: one gram will be enough for 
a mission lifetime of 10 years! 

LISA interferometry is not done with reflection from mirrors. When 
a laser beam reaches one spacecraft from the other, it is too weak 
to reflect: the sending spacecraft would only get the occasional photon!
Instead, the incoming light is sensed, and an on-board laser is 
slaved to it, returning an amplified beam with the same phase 
and frequency as the incoming one. No space mission has yet implemented
this kind of laser-transponding. The LISA team had to ensure that 
there was enough information in all the signals to compensate 
for inevitable frequency fluctuations among all six on-board lasers. 

A further serious problem that the LISA 
team had to solve was how to compensate for the relative motions 
of the spacecraft. The laser signals converging on a single spacecraft 
from the other two corners will be Doppler shifted so that their 
fringes change at MHz frequencies. This has to be sensed on board 
and removed from the signal that is sent back to Earth, which can 
only be sampled a few tens of times per second. 

When LISA flies it will, on a technical as well as a scientific level, 
be a worthy counterpart to its Earth-based interferometer cousins!

\subsection{LISA's capabilities}
In the low-frequency LISA window, most sources will be relatively 
long-lived, at least a few months. During an observation, LISA 
will rotate and change its velocity by a significant amount. This 
will induce Doppler shifts into the signals, and modulate their 
amplitudes, so that LISA should be able to infer the position, 
polarization, and amplitude of sources entirely from its own 
observations. Below about 1~mHz, this information weakens, because
the wavelength of the radiation becomes comparable to or greater
than the radius of LISA's orbit. The amplitude modulation is the 
only directional information in this frequency range. 

\section{Gravitational and electromagnetic waves compared and contrasted}

To conclude this \lecture\ is useful to discuss the most important 
differences and similarities between gravitational waves and 
electromagnetic ones. We do this in the form of a table.

\begin{center}
\begin{tabular}{p{5.25cm}p{5.25cm}}
\br
\large \emph{Electromagnetism} & \large \emph{General Relativity} \\
\mr
Two signs of charges - large bodies usually neutral - waves usually emitted by single particles, often incoherently - waves carry ``local'' information. & One sign of mass - gravity accumulates - waves emitted more strongly by larger body - waves carry ``global'' information. \\
\end{tabular}
\end{center}

\begin{center}
\begin{tabular}{p{5.25cm}p{5.25cm}}
\mr
A genuine physical force, acting differently on different bodies. Detected by measuring accelerations. & Equivalence principle: gravity affects all bodies in the same way. Represented as a space-time curvature rather than a force. Detected only by tidal forces - differential accelerations. \\
\mr
\end{tabular}
\end{center}

\begin{center}
\begin{tabular}{p{5.25cm}p{5.25cm}}
\mr
Maxwell's equations are \emph{linear}. Physical field is $F_{\mu \nu }$ (%
\textbf{E} and \textbf{B}). Gauge field is vector potential $A$. & Einstein's equations are \emph{nonlinear}. Physical field is Riemann
curvature tensor $R_{\mu \nu \alpha \beta }$. Gauge fields are metric $g_{\mu
\nu }$ and connection $\Gamma _{\mu \nu }^{\alpha }$. Gauge transformations
are coordinate transformations. \\

\end{tabular}
\end{center}

\begin{center}
\begin{tabular}{p{5.25cm}p{5.25cm}}
\mr
Source is charge-current density $J_{\mu }$. Charge creates electric field,
current magnetic field. & Source is stress-energy tensor $T_{\mu \nu }$. Mass creates a Newtonian-like field, momentum as gravito-magnetic effects. Stress creates field too. \\
\end{tabular}
\end{center}

\begin{center}
\begin{tabular}{p{5.25cm}p{5.25cm}}
\mr
Moderately strong force on the atomic scale: $\frac{e^{2}/4\pi \epsilon _{0}%
}{Gm_{p}^{2}}=10^{39}$. & Weaker than ``weak'' interaction. \\
\end{tabular}
\end{center}

\begin{center}
\begin{tabular}{p{5.25cm}p{5.25cm}}
\mr
Wave generation for $A_{\mu }$: $\partial
^{\beta }\partial _{\beta }A_{\mu }=4\pi \epsilon _{0}J_{\mu }$ in a
convenient gauge (Lorentz gauge). & Wave generation for $h_{\mu \nu }=g_{\mu \nu }-\eta _{\mu \nu }$: $\partial
^{\beta }\partial _{\beta }\left( h_{\mu \nu }-\frac{1}{2}\eta _{\mu \nu }h_{%
\hspace{0.12cm}\alpha }^{\alpha }\right) =8\pi T_{\mu \nu }$ in a
convenient gauge. \\
\end{tabular}
\end{center}

\begin{center}
\begin{tabular}{p{5.25cm}p{5.25cm}}
\mr
Propagate at the speed of light, amplitude falls as $1/r$. & Propagate at the speed of light, amplitude falls as $1/r$. \\
\end{tabular}
\end{center}

\begin{center}
\begin{tabular}{p{5.25cm}p{5.25cm}}
\mr
Conservation of charge $\Rightarrow $ radiation by low-velocity charges is
dominated by dipole component. & Conservation of mass and momentum $\Rightarrow $ radiation by low-velocity masses is dominated by quadrupole component. \\
\end{tabular}
\end{center}

\begin{center}
\begin{tabular}{p{5.25cm}p{5.25cm}}
\mr
Simple detector: oscillating charge. Action is along a line, transverse to
the directions of propagation. Spin $s=1$ and two states of linear polarisation that are inclined each other at an angle of 90${{}^\circ}$. & Simple detector: distorted ring of masses. Action is elliptic in a plane transverse to the direction of propagation. Spin $s=2$ and two states of linear polarisation that are inclined each other at an angle of 45${{}^\circ}$. Equivalence principle $\Rightarrow $ action depends only on $h_{\mu \nu }$, which is dimensionless. \\
\end{tabular}
\end{center}

\begin{center}
\begin{tabular}{p{5.25cm}p{5.25cm}}
\mr
Strength of force $\Rightarrow $ waves scatter and refract easily. & Weakness of gravity $\Rightarrow $ waves propagate almost undisturbed and
transfer energy very weakly. Dimensionless amplitude $h$ is small. \\
\end{tabular}
\end{center}

\begin{center}
\begin{tabular}{p{5.25cm}p{5.25cm}}
\mr
Local energy and flux well-defined: Poynting vector etc. & Equivalence principle $\Rightarrow $ local energy density cannot be defined
exactly. Only \emph{global} energy balance is exact. \\
\mr
\end{tabular}
\end{center}

\begin{center}
\begin{tabular}{p{5.25cm}p{5.25cm}}
\mr
Multipole expansion in slow-motion limit is straightforward, radiation
reaction well-defined. & Multipole expansion different if fields are weak or strong. For
quasi-Newtonian case fields are weak, and the resulting post-Newtonian
expansion is delicate. Radiation reaction is still not fully understood. \\

\end{tabular}
\end{center}

\begin{center}
\begin{tabular}{p{5.25cm}p{5.25cm}}
\mr
Exact solutions, containing waves, are available and can guide the
construction of approximation methods for more complicated situations. & Fully realistic exact solutions for dynamical situations of physical
interest are not available. Extensive reliance on approximation methods. \\
\br
\end{tabular}
\end{center}

\chapter{Astrophysics of gravitational wave sources}

There are a large number of possible gravitational wave sources in 
the observable wave band,  which spans 8
orders of magnitude in frequency: from $10^{-4}$~Hz (lower bound of current space-based
detector designs) to $10^{4}$~Hz (frequency limit of likely ground-based detectors). Some of these
sources are highly relativistic and not too massive, especially above 10~Hz: a black hole of
mass 1000 $M_{^{\odot }}$ has a characteristic frequency of 10~Hz, and
larger holes have lower frequencies in inverse proportion to the mass. 
Neutron stars have even higher characteristic frequencies. Other 
systems are well-described by Newtonian dynamics, such as binary orbits.

For nearly-Newtonian sources the post-Newtonian approximation (see \correction{\lecture\~5}{\lecture~5})
provides a good framework for calculating gravitational waves. More 
relativistic systems, and unusual sources like the early universe, require
more sophisticated approaches (see \correction{\lecture\~6}{\lecture~6}). 

\section{Sources detectable from ground and from space}

\subsection{Supernovae and gravitational collapse}

The longest-expected and still probably the least-understood source, 
gravitational collapse is one of the most violent
events know to astronomy. Yet, because we have little direct 
information about the deep interior, we cannot make reliable 
predictions about the gravitational radiation from it.

Supernovae are triggered by the gravitational collapse of the interior
degenerate core of an evolved star. According to current theory the result
should be a neutron star or black hole. The collapse  releases an
enormous amount of energy, about $0.15M_{\odot }c^{2}$, most of which is carried
away by neutrinos; an uncertain fraction is converted into 
gravitational waves. One mechanism for producing this radiation could be 
dynamical instabilities in the rapidly rotating core before it becomes a neutron
star. Another likely source of radiation is the $r$-mode instability (see \lecture~6). This could release $\sim0.1M_\odot c^2$ in 
radiation  every time a neutron star is formed.

However, both kinds of mechanisms are difficult to model. The problem 
with gravitational collapse is that perfectly spherical 
motions do not emit gravitational waves, and it
is still not possible to estimate in a reliable way the amount of asymmetry in
gravitational collapse. Even modern computers are not able to perform
realistic simulations of gravitational collapse in 3D, including all the
important nuclear reactions and neutrino-\ and photon-transport. Similarly, 
it is hard to model the $r$-mode instability because its evolution 
depends on nonlinear hydrodynamics and on poorly known physics, such as 
the cooling and viscosity of neutron stars. 

An alternative 
approach is to use general energy considerations. 
If for example we assume that 1\% of the available energy is converted
into gravitational radiation, \correction{}{then, from formulas we 
will derive in the next \lecture, }the amplitude $h$ 
would be large enough to be detected
by the first ground-based in\-ter\-fero\-meters (LIGO/GEO600/VIRGO) at the
distance of Virgo Cluster (18 Mpc) if the emission centers at 300~Hz. 
Moreover, bar  and
spherical-mass detectors with an effective sensitivity of $10^{-21}$ and 
the right resonant
frequency could see these signals as well.

The uncertainties in our predictions have a positive aspect: it is 
clear that if we can detect radiation from supernovae,  
we will learn much that we don't know about 
the end stages of stellar evolution and about neutron-star physics.

\subsection{Binary stars}

Binary systems have given us our best proof of the reliability of general relativity for
gravitational waves. The most famous example of such systems is the binary
pulsar PSR1916+16, discovered by Hulse and Taylor in 1974; they were 
awarded the Nobel Prize for this discovery in 1993. From the observations 
of the modulation of the pulse period as the stars move in their
orbits, one knows many
important parameters of this system 
(orbital period, eccentricity, masses of the
two stars, etc), and the data also show directly the decrease
of the orbital period due to the emission of gravitational radiation. The 
observed value is $2.4\cdot 10^{-12}$ s/s. Post-Newtonian theory allows 
one to predict this from the other measured parameters of the system, 
without any free parameters (see \lecture~6); the prediction is 
$2.38\cdot 10^{-12}$, in agreement within the measurement errors. 

Unfortunately the radiation from the Hulse-Taylor system 
will be too weak and of too low frequency 
to be detectable by LISA.

\subsection{Chirping binary systems}

If a binary gives off enough energy for  its orbit to shrink by an observable
amount during an observation, it is said to \emph{chirp}: as 
the orbit shrinks, the frequency and amplitude go up. LISA will see a
few chirping binaries. If a binary system 
is compact enough to radiate above $10^{-3}$~Hz, 
it will always chirp within one year, provided its components have 
a mass above about $1\;M_\odot$. If they are above about $10^3\;M_\odot$, 
the binary will go all the way to coalescence within the one-year 
observation.

Chirping binary systems are  more easily detectable than
gravitational collapse events because one can model with great accuracy 
the gravitational wave-form during the inspiral phase. There will 
be radiation, possibly with considerable energy, during the poorly understood 
\emph{plunge} phase (when the objects reach the last 
stable orbit and fall rapidly
towards one another) and during the merger event, but 
the detectability of such systems rests on tracking their orbital
emissions.

The major uncertainty about this kind of source is the event rate. Current
pulsar observations suggest that there will be $\sim 1$ coalescence per 
year of a
Hulse-Taylor binary out to about 200~Mpc. This is a \emph{lower limit }
on the event rate, since it comes from systems we actually observe. 
It is possible that 
there are other kinds of binaries that we have no 
direct knowledge of, which 
will boost the event rate.

Theoretical modeling of binary \correction{systems}{populations} gives a wide spectrum of 
mutually inconsistent predictions. Some authors\cite{lipunov}
suggest that there may be a large population that escapes
pulsar surveys but brings the nearest neutron star coalescence in one year
as far as 30 Mpc, only slightly further than the Virgo cluster; but 
other models\cite{yungelson} put the rate near to the observational limit.

The most exciting motivation  for detecting coalescing binaries is that they
could be associated with gamma-ray bursts. The event rates are consistent, 
and neutron stars are able to provide the required energy. If gamma-bursts
are associated with neutron-star coalescence, then observations of 
coalescence radiation should be followed within a second or so by 
a strong gamma-ray burst. 

LISA will see a few chirping binaries in the Galaxy, 
but the sensitivity of the first
generation of ground based detectors is likely to be \emph{too poor} to see
many such events (see \Tref{tab:cb}).

\begin{table}
\caption{The range for detecting a $2\times 1.4$ M$_{\odot }$ NS binary\correction{.}{ coalescence.} The threshold for detection is taken to be $5\sigma$. The binary and detector orientations are assumed optimum. The average S/N ratio for randomly oriented systems is reduced from the optimum by $1/\sqrt{5}$.}\label{tab:cb}
\begin{center}
\begin{tabular}{p{1.5cm}ccccc}
\br
Detector: & \textbf{TAMA300} & \textbf{GEO 600} & \textbf{LIGO I} & \textbf{VIRGO} & \textbf{LIGO II} \\
\mr
Range (S/N=5) & \textbf{3 Mpc} & \textbf{14 Mpc} & \textbf{30 Mpc} & \textbf{36 Mpc} & \textbf{500 Mpc}  \\
\br
\end{tabular}
\end{center}
\end{table}

A certain fraction of such system could contain
black holes instead of neutron stars. In fact black holes should be
over-represented in binary systems (relative to their birth rate) 
because their formation is much less likely to
disrupt a binary system (there is much less mass lost) than the formation of
a neutron star would be. Pulsar observations have not yet turned up 
a black-hole/neutron-star system, and of course one does not 
expect to see binary black holes electromagnetically. So we can 
only make theoretical estimates, and there are big uncertainties.

Some evolution calculations\cite{lipunov} suggest that the coalescence rate of BH-BH
systems may be of the same order as  the NS-NS rate. Other models\cite{yungelson} suggest it
could even be zero, because stellar-wind mass loss (significant in very massive stars) could drive the stars far apart before
the second BH forms, leading to coalescence times longer than the
age of the Universe. A recent proposal identifies globular clusters as 
``factories'' for binary black holes, forming binaries by 3-body collisions
and then expelling them.\cite{zwart} Gamma-ray bursts may also come 
from black-hole/neutron-star coalescences. If the more optimistic event rates 
are correct, then black-hole coalescences may be among the 
first sources detected
by ground-based detectors (\Tref{tab:bhbin}).

\begin{table}
\caption{The range for detecting a $10$ M$_{\odot }$ black-hole binary\correction{.}{ coalescence.} Conventions as in Table~\protect\ref{tab:cb}.}\label{tab:bhbin}
\begin{center}
\begin{tabular}{lcccc}
\br
Detector: & \textbf{GEO 600} & \textbf{LIGO I} & \textbf{VIRGO} & \textbf{LIGO II} \\
\mr
Range (S/N=5) & \textbf{75 Mpc} & \textbf{160 Mpc} & \textbf{190 Mpc} & \textbf{2.6 Gpc} \\
\br
\end{tabular}
\end{center}
\end{table}

\subsection{Pulsars and other spinning neutron stars}

There are a number of ways in which a spinning neutron star may give off
a continuous stream of gravitational waves. They will be weak, so 
they will require long continuous observation times, up to many months. Here 
are some possible emission mechanisms for neutron stars.

\textbf{The r-modes.} Neutron stars are born hot and probably rapidly 
rotating. Before they cool (during their first year) they  have a family of 
\emph{unstable normal modes, the $r$-modes.} These modes 
are excited to instability by the emission of gravitational radiation, 
as predicted originally by Andersson \correction{(1998)}{\cite{and}}.
They are particularly interesting theoretically because the 
radiation is gravito-magnetic, generated by mass currents rather 
than mass asymmetries. We will study the theory of this radiation in 
\lecture~5. In \lecture~6 we will 
discuss how the emission of this radiation excites the 
instability (the CFS instability mechanism). 

Being unstable, young neutron stars will presumably radiate away enough angular
momentum to reduce their spin and become stable. This could lower the spin
of a neutron star to $\sim 100$ Hz within one year 
after its formation \correction{(Lindblom
et al. 1998)}{\cite{lind}}. The energy emitted in this way should be a good fraction of
the star's binding energy, so in principle this radiation 
could be detected from the
Virgo Cluster by LIGO II, provided matched filtering can be used effectively.

We discuss a possible stochastic background of gravitational waves 
from the $r$-modes below.

\begin{figure}
\begin{center}\label{fig:accrete}
\caption{Accreting neutron star in a Low-Mass X-Ray binary system}
\includegraphics[width=8cm]{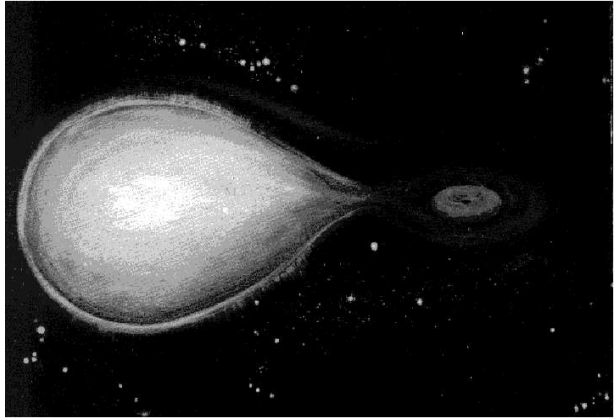}
\end{center}
\end{figure}

\textbf{Accreting neutron stars} (\Fref{fig:accrete}) 
are the central objects of most of the binary X-ray
sources in the Galaxy. Astronomers divide them into two distinct groups: the
low-mass and high-mass binaries, according to the mass of the companion
star. In these systems mass is pulled from the low or high-mass giant by the tidal
forces exerted by its neutron star companion. In low-mass X-ray binaries (LMXBs)
the accretion lasts long enough to spin the neutron star up 
to the rotation rates of millisecond
pulsars. Astronomers have therefore supposed for some time that 
the neutron stars in LMXBs would have a range of spins, from near zero 
(young systems) to near 500 or 600~Hz (at the end of the accretion 
phase). Until the launch of the Rossi X-ray Timing Explorer (RXTE), 
there was no observational evidence for the neutron star spins. But 
in the last two years there has been an accumulation of evidence 
that most, if not all, of these stars have angular velocities in 
a narrow range around 300~Hz.\cite{klis} It is not known yet what 
mechanism regulates this spin,
but a strong candidate is the emission of gravitational radiation.

A novel proposal by Bildsten\cite{bildsten} suggests that the temperature gradient
across a neutron star that is accreting preferentially at its magnetic poles should lead
to a composition and hence a density gradient in the deep crust. Spinning at
300~Hz, such a star could radiate as much as it accretes. It would 
then be a steady source for as long as accretion lasts, which could be 
millions of years.

In this model the gravitational wave energy flux is proportional to the observed X-ray
energy flux. The strongest source in this model is Sco X-1, which could be
detected by GEO600 in a two-year-long narrow-band mode if the 
appropriate matched filtering can be done. LIGO II would have no difficulty
in detecting this source.

Older stars may also be \emph{lumpy}. For known pulsars, this is constrained by
the rate of spin down: the energy radiated in gravitational waves can not exceed the
total energy loss. In most cases, this limit is rather weak, and 
stars would have to sustain strains in their crust of order $10^{-3}$ or 
more. It is unlikely that crusts could sustain this kind of strain, 
so the observational limits are probably significant overestimates for 
most pulsars. However, millisecond pulsars have much slower spindown 
rates, and it would  be easier to account for the strain in their 
crusts, for example as a remnant Bildsten asymmetry. Such stars 
could in principle be radiating more energy in gravitational waves than electromagnetically.

Observations of individual neutron stars would be rich with information about
astrophysics and fundamental nuclear physics. So little is known about 
the physics of these complex objects that the incentive to observe 
their radiation is great.

However, making such observations presents challenges for data analysis, since 
the motion of the Earth puts a strong phase modulation on the 
signal, which means that even if its rest-frame frequency is constant 
it cannot be found by simple Fourier analysis. More sophisticated 
pattern-matching (matched-filtering) techniques are needed, which 
track and match the signal's phase to within one cycle over the 
entire period of measurement. This is not difficult if the source's 
location and frequency are known, but the problem of doing a wide-area 
search for unknown objects is very challenging.\cite{brady}  Moreover, 
if the physics of the source is poorly known, such as for LMXBs or 
$r$-mode spindown, the job of building an accurate family of templates
is a difficult one. These questions are the subject of much research today, 
but they will need much more in the future.

\subsection{Random backgrounds}

The big bang was the most violent event of all, and it may have created a
significant amount of gravitational radiation. Other events in the early
universe may also have created radiation, and there may be backgrounds from
more recent epochs. We have seen earlier, for example, that compact 
binary systems in the Galaxy will merge into a confusion-limited noise 
background in LISA observations below about 1~mHz.

Let us consider the $r$-modes as another important example. 
This process may have occurred in a good fraction of all neutron stars formed 
since the beginning of star formation. The sum of all of their $r$-mode 
radiation will be a stochastic background, with a spectrum that 
extends from a lower cutoff of about 200~Hz in the rest frame of the 
emitter to an upper limit that depends
on the initial angular velocity of stars. If significant star formation 
started at, say, a redshift of 5, then this background should extend 
down to about 25~Hz. If $10^{-3}$ of the mass of the Galaxy is in 
neutron stars, and each of them radiates 10\% of its mass in this 
radiation, then the gravitational wave background should have a density 
equal to $10^{-4}$ of the mean cosmological density of visible stars. 
Expressed as a fraction $\Omega _{gw}$ of the closure density 
of the Universe, per logarithmic frequency interval, this converts to 
\[
\Omega _{gw}^{\text{r-modes}}\left( 25-1000\text{Hz}\right) \approx
10^{-8}-10^{-7}
\]
This background 
would be easily detectable by LIGO II.

There may also be a cosmological background from either topological defects
(e.g. cosmic strings) or from inflation (which amplifies initial quantum 
gravitational fluctuations
as it does the scalar ones that lead to galaxy formation). Limits 
from COBE observations suggest that standard inflation could not 
produce a background stronger than $\Omega _{gw}^{\text{inflation}}\sim 10^{-14}$ today. This is too weak for any of the planned detectors to reach, 
but it remains an important long-range goal for the field. But there could 
also be a component of background radiation that depends on what happened 
\emph{before} inflation: string
cosmological models, for example, predict spectra growing 
with frequency.\cite{Veneziano} 

First-generation interferometers are not likely to detect these
backgrounds: they may not be able to go below the upper limit 
set by the requirement that gravitational waves should not 
disturb cosmological nucleosynthesis, which is 
$\Omega _{gw}=10^{-5}.$ (This limit does not apply to backgrounds
generated after nucleosynthesis, like the r-mode background.) Bar detectors
may do as well or better than the first generation of interferometers for a
broad-spectrum primordial background: as we have noted earlier, their noise levels within their
resonance bands are very low. But their frequencies are not right for the
r-mode background.

Second-generation interferometers may be able to reach to $10^{-11}$ of closure or
even lower, by cross-correlation of the output of the two detectors. But 
they are unlikely to get to the inflation target of $10^{-14}$.
LISA may be able to go as low as 10$^{-10}$ (if we have a confident
understanding of the instrumental noise), but it is likely to 
detect only the confusion background of binaries, which is expected 
to be much stronger than a cosmological background in the LISA band.

\subsection{The unexpected}

At some level, we are bound to see things we did not expect. LISA, with its
high signal-to-noise ratios for predicted sources, is particularly well placed to do this. Most of the Universe
is composed of dark matter whose existence we can infer only from its
gravitational effects. It would not be particularly surprising if 
a component of this dark
matter produced gravitational radiation in unexpected ways, such as from
binaries of small exotic compact objects of stellar mass. We will have to
wait to see!

\chapter{Waves and Energy}

Here we discuss wave-like perturbations $h_{\mu \nu }$ of a general
background metric $g_{\mu \nu }$. The mathematics is similar to that of
linearized theory: $h_{\mu \nu }$ is a tensor with respect to background
coordinate transformations (as it was for Lorentz transformations in
linearized theory) and it undergoes a gauge transformation when one makes an
infinitesimal coordinate transformation. As in linearized theory, 
we will assume that the amplitude of the waves is small. Moreover, 
the waves must have a wavelength that is short compared to the 
radius of curvature of the background metric. These two assumptions 
allow us to visualize the waves  as small ripples running through a curved and
slowly changing space-time.

\section{Variational principle for general relativity}

We start our analysis of small perturbation $h_{\mu \nu }$ by 
introducing the standard Hilbert variational principle for Einstein's equations. The field equations of general relativity can be derived from an action
principle using 
the Ricci scalar curvature as the Lagrangian density. The Ricci scalar (second
contraction of the Riemann tensor) is an invariant quantity which contains in
addition to $g_{\mu \nu }$ and its first derivatives also the second
derivatives of $g_{\mu \nu }$, so our action can be written symbolically as: 
\begin{equation}
I[g_{\mu \nu }]=\frac{1}{16\pi }\int R(g_{\mu \nu }\text{, }g_{\mu \nu
,\alpha }\text{, }g_{\mu \nu },_{\alpha \beta })\sqrt{-g}\;d^{4}x\text{ }
\end{equation}
where $\sqrt{-g}$ is the square-root of the determinant of the metric
tensor. As usual in  variational principles, the metric tensor
components are varied $g_{\mu \nu }\rightarrow g_{\mu \nu }+h_{\mu \nu }$,
and one demands that the resulting change in the action should vanish to
first order in any small variation $h_{\mu \nu }$ of compact support: 
\begin{eqnarray}
\delta I &=&I[g_{\mu \mu }+h_{\mu \nu }]-I[g_{\mu \nu }] \nonumber \\
&=&\frac{1}{16\pi }\int \frac{\delta \left( R\sqrt{-g}\right) }{\delta
g_{\mu \nu }}h_{\mu \nu }d^{4}x+O\left( 2\right)\label{eq:var} \\
&=&-\frac{1}{16\pi }\int G^{\mu \nu }h_{\mu \nu }\sqrt{-g}d^{4}x+O(2) 
\end{eqnarray}
where ``$O(2)$'' denotes terms quadratic and higher in $h_{\mu \nu }$. 
All the divergences obtained in the intermediate steps of this 
calculation integrate to zero since $h_{\mu \nu }$
is of compact support. This variational principle therefore yields the vacuum
Einstein equations: $G^{\mu \nu }=0$.

Let us consider how this changes if we include  matter. This will 
help us to see how we can treat gravitational waves as a new 
kind of ``matter'' field on spacetime.

Suppose we have a matter field, described by a variable $\Phi $ (which may represents a
vector, a tensor or a set of tensors). It will have a Lagrangian density $%
L_{m}=L_{m}\left( \Phi ,\Phi _{,\alpha },\ldots ,g_{\mu \nu }\right) $ that
depends on the field and also on the metric. 
Normally derivatives of the metric tensor do not appear in $%
L_{m}$, since by the equivalence principle, matter fields should behave locally as if they
were in flat space-time, where of course  there are no metric derivatives. Variations
of $L_{m}$ with respect to $\Phi $ will produce the field equation(s) for the
matter system , but here we are more 
interested in variations with respect to $g_{\mu \nu }$, which is how 
we will find the matter field's contribution to the gravitational field 
equations.
The total action has the form: 
\begin{equation}\label{eq:total}
I=\int \left( R+16\pi L_{m}\right) \sqrt{-g}\;d^{4}x,
\end{equation}
whose variation is 
\begin{equation}
\delta I=\int \frac{\delta \left( R\sqrt{-g}\right) }{\delta g_{\mu \nu }}%
h_{\mu \nu }d^{4}x+\int 16\pi \frac{\partial \left( L_{m}\sqrt{-g}\right) }{%
\partial g_{\mu \nu }}h_{\mu \nu }d^{4}x.
\end{equation}
This variation must yield the full Einstein equations, so we must have the
following result for the stress-energy tensor of the matter: 
\begin{equation}
T^{\mu \nu }\sqrt{-g}=2\frac{\partial L_{m}\sqrt{-g}}{\partial g_{\mu \nu }},
\end{equation}
leading to 
\begin{equation}
\fbox{$G^{\mu \nu }=8\pi T^{\mu \nu }$}.
\end{equation}
This way of deriving the stress-energy tensor of the matter field has deep
connections to the conservation laws of general relativity, to the way of
constructing conserved quantities when the metric has symmetries and to the 
so-called pseudo-tensorial definitions of gravitational wave 
energy (see Landau and Lifshitz 1962).\cite{sork}. 
We shall use it in the latter sense.

\section{Variational principles and the energy in gravitational waves}

Before we introduce the mathematics of gravitational waves, it is important
to understand which geometries we are going to examine.
We have said that these geometries consist of a slowly and smoothly changing 
\emph{background metric} which is altered by\emph{\ perturbations }of small
amplitude and high frequency. If $L$ and $\lambda $ are the characteristic lengths over which
the background and ``ripple'' metrics change significantly, we  assume
that the ratio $\frac{\lambda }{L}$ will be very much smaller than unity and that
$\left| h_{\mu \nu }\right| $ is of the same order of smallness as $\frac{\lambda }{L}$.
In this way the total metric remains slowly changing on a macroscopic scale,
while the high-frequency wave, when averaged over several wavelengths, will 
be the principal source of the curvature of the background metric. This is the 
``short-wave'' approximation\cite{isa}. Obviously this is a direct generalization of the treatment in \lecture~1.

\subsection{Gauge transformation and invariance}

Consider an infinitesimal coordinate transformation generated by a 
vector field $\xi ^{\alpha }$,
\begin{equation}
x^{\alpha }\rightarrow x^{\alpha }+\xi ^{\alpha }.
\end{equation}
In the new
coordinate system, neglecting quadratic and higher terms in $h^{\alpha \beta
}$, it is not hard to show that the general gauge transformation of the metric is
\begin{equation}
h_{\mu \nu }\rightarrow h_{\mu \nu }-\xi _{\mu ;\nu }-\xi _{\nu ;\mu },
\end{equation}
where a semicolon denotes the covariant derivative. We assume that the 
derivatives of the coordinate displacement field are of the same order 
as the metric perturbation: $\left| \xi
^{\alpha,\beta }\right| \sim \left| h^{\alpha \beta }\right| $. 

Isaacson\cite{isa} showed that the gauge transformation of the 
Ricci and Riemann curvature tensors has the property
\begin{eqnarray}
\bar{R}_{\mu \nu }^{{(1)}}-R_{\mu \nu }^{{ (1)}}
&\approx &\left( \frac{\lambda }{L}\right) ^{2} \\
\bar{R}_{\alpha \mu \beta \nu }^{{ (1)}}-R_{\alpha \mu \beta
\nu }^{{ (1)}} &\approx &\left( \frac{\lambda }{L}\right) ^{2} 
\nonumber
\end{eqnarray}
where $R_{\mu \nu }^{(1)}$ and $R_{\alpha \mu \beta \nu }^{(1)}$ are the
first order  Ricci and Riemann tensors \correction{9}{(}in powers of
perturbation $h_{\mu \nu }$) and an over-bar denotes their values after 
the gauge transformation. In our high-frequency limit, therefore, 
these tensors are gauge-invariant
to linear order, just as in linearized theory. 

\subsection{Graviational wave action}

Let us suppose that we are in vacuum so we have only the metric, no matter
fields, but we work in the high-frequency approximation. The full metric
is $g_{\mu \nu }$ (smooth \emph{background }metric)\emph{\ }$+ h_{\mu \nu }$
(high-frequency perturbation). Our purpose is to show that the wave 
field can be treated as a  ``matter''
field, with a Lagrangian and its own stress-energy tensor. 
To do this we have to
expand the action out to second order in the metric perturbation, 
\begin{eqnarray*}
I[g_{\mu \mu }+h_{\mu \nu }] &=&\int R(g_{\mu \nu }+h_{\mu \nu }\text{, }%
g_{\mu \nu ,\alpha }+h_{\mu \nu ,\alpha }\text{,}\ldots )\sqrt{-g[g_{\mu \nu
}+h_{\mu \nu }]}d^{4}x \\
&=&\int R(g_{\mu \nu }\text{,}\ldots )\sqrt{-g}d^{4}x+\int \frac{\delta
\left( R\sqrt{-g}\right) }{\delta g_{\mu \nu }}h_{\mu \nu }d^{4}x \\
&&+\frac{1}{2}\int \left( \frac{\partial ^{2}\left( R\sqrt{-g}\right) }{%
\partial g_{\mu \nu }\partial g_{\alpha \beta }}h_{\mu \nu }h_{\alpha \beta
}+2\frac{\partial ^{2}\left( R\sqrt{-g}\right) }{\partial g_{\mu \nu
}\partial g_{\alpha \beta ,\gamma }}h_{\mu \nu }h_{\alpha \beta ,\gamma
}\right. \\
&&\left. +\frac{\partial ^{2}\left( R\sqrt{-g}\right) }{\partial g_{\mu \nu
,\tau }\partial g_{\alpha \beta ,\gamma }}h_{\mu \nu ,\tau }h_{\alpha \beta
,\gamma }+2\frac{\partial ^{2}\left( R\sqrt{-g}\right) }{\partial g_{\mu \nu
}\partial g_{\alpha \beta ,\gamma \tau }}h_{\mu \nu }h_{\alpha \beta ,\gamma
\tau }\right) d^{4}x \\
&&+O(3)\correction{}{.}
\end{eqnarray*}
The first term is the action for the background metric $g_{\mu \nu }$. The
second term vanishes (see \Eref{eq:var}), since we assume that the background
metric is a solution of Einstein vacuum equation itself, at least to lowest
order.

If we compare the above equation with \Eref{eq:total}, we can see
that the third term, complicated as it seems, is an effective ``matter''
Lagrangian for the gravitational field. Indeed, if one varies it with respect
to $h_{\mu \nu }$ holding $g_{\mu \nu }$ fixed (as we would do for  a
physical matter field on the background), then the complicated coefficients are
fixed and one can straightforwardly show that one gets exactly \emph{the
linear perturbation of the Einstein tensor itself}. Its vanishing is the
equation for the gravitational wave perturbation $h_{\mu \nu }$.
In this way we have shown that, for a \emph{small amplitude}
perturbation, the gravitational wave can be treated 
as a ``matter'' field with its own Lagrangian and field equations.

Given this Lagrangian, we should be able to calculate the effective
stress-energy tensor of the wave field by taking the variations of the
effective Lagrangian with respect to $g_{\mu \nu }$, holding the 
``matter'' field $h_{\mu \nu }$ fixed:
\begin{equation}
T^{\text{(GW)}\alpha \beta }\sqrt{-g}=2\frac{\partial L^{\text{(GW)}}[g_{\mu
\nu },h_{\mu \nu }]\sqrt{-g}}{\partial g_{\alpha \beta }}
\end{equation}
with 
\begin{eqnarray}
L^{\text{(GW)}}\sqrt{-g} &=&\frac{1}{32\pi }\left( \frac{\partial ^{2}\left(
R\sqrt{-g}\right) }{\partial g_{\mu \nu }\partial g_{\alpha \beta }}h_{\mu
\nu }h_{\alpha \beta }+2\frac{\partial ^{2}\left( R\sqrt{-g}\right) }{%
\partial g_{\mu \nu }\partial g_{\alpha \beta ,\gamma }}h_{\mu \nu
}h_{\alpha \beta ,\gamma }\right. \\
&&\left. +\frac{\partial ^{2}\left( R\sqrt{-g}\right) }{\partial g_{\mu \nu
,\tau }\partial g_{\alpha \beta ,\gamma }}h_{\mu \nu ,\tau }h_{\alpha \beta
,\gamma }+2\frac{\partial ^{2}\left( R\sqrt{-g}\right) }{\partial g_{\mu \nu
}\partial g_{\alpha \beta ,\gamma \tau }}h_{\mu \nu }h_{\alpha \beta ,\gamma
\tau }\right) \correction{}{.} \nonumber
\end{eqnarray}
This quantity is quadratic in the wave amplitude $h_{\mu \nu }$. It could be
simplified further by integrations by parts, such as by taking a derivative 
off $h_{\alpha \beta ,\gamma \tau }$. This would change the coefficients of
the other terms. We will not need to worry about finding the ``best'' form
for the expression (4.12), as we now show.

As in linearized theory, so also in the general case, the
quantity $h_{\mu \nu }$ behaves like a tensor with respect to background
coordinate transformations, and so does $T_{\mu \nu }^{\text{(GW)}}$. But
it is not gauge-invariant and so it is not a physical observable. 
However, since the integral of the action \emph{is} independent of coordinate
transformations that have compact support, so too is the integral of the
effective stress-energy tensor. In practical terms, this makes it possible
to localize the energy of a wave to within a region of about one wavelength
in size where the background curvature does not change significantly. In fact, if we restrict our gauge transformations to have a length scale
of a wavelength, and if we average (integrate) the stress-energy tensor of
the waves over such a region, then any gauge changes will be small surface terms.

By evaluating the effective stress-energy tensor on a smooth background
metric in a Lorentz gauge, and performing the averaging (denoted by symbol $\left\langle \cdots \right\rangle $
), one arrives at the
\textbf{Isaacson tensor}: 
\begin{equation}
T_{\alpha \beta }^{\text{(GW)}}=\frac{1}{32\pi }\left\langle h_{\mu \nu
;\alpha }h_{\hspace{0.24cm};\beta }^{\mu \nu }\right\rangle \correction{}{.}
\end{equation}
 
This is a convenient and compact form for the gravitational
stress-energy tensor. It localizes energy in short-wavelength 
gravitational waves to regions of the order of a wavelength. 
It is interesting to remind ourselves that our only
experimental evidence of gravitational waves today is the observation 
of the effect on a binary orbit of the loss of energy to the  
gravitational waves emitted by the system. So this energy formula, 
or equivalent ones, is central to our understanding of gravitational 
waves.

\section{Practical applications of the Isaacson energy}

If we are far from a source of graviational waves, we can treat the waves by
linearized theory. Then if we adopt the TT gauge and specialize the
stress-energy tensor of the radiation to a flat background, we get 
\begin{equation}\label{eq:ttflux}
T_{\alpha \beta }^{\text{(GW)}}=\frac{1}{32\pi }\left\langle h_{ij,\alpha }^{%
\text{TT}}h_{\hspace{0.24cm}\hspace{0.24cm},\beta }^{\text{TT}%.
ij}\right\rangle\correction{}{.}
\end{equation}
Since there are only two components, a wave travelling with frequencies $f$
(wave number $k=2\pi f$) and with a typical amplitude $h$ in both
polarizations carries an energy $F_{gw}$ equal to (see Ex.. 6 at the end of
this lecture) 
\begin{equation}
F_{gw}=\frac{\pi }{4}f^{2}h^{2}.
\end{equation}
Putting in the factors of $c$ and $G$ and scaling to reasonable values gives 
\begin{equation}
F_{gw}=3\text{ mW m}^{-2}\left[ \frac{h}{1\times 10^{-22}}\right] ^{2}\left[ 
\frac{f}{1\text{kHz}}\right] ^{2},
\end{equation}
which is a very large energy flux even for this weak a wave. It is twice
the energy flux of a full moon! Integrating over a sphere of radius $r$,
assuming a total duration of the event $\tau $, and solving for $h$, again
with appropriate normalizations, gives 
\begin{equation}
h=10^{-21}\left[ \frac{E_{gw}}{0.01\text{ M}_{\odot }\text{c}^{2}}\right] ^{%
\frac{1}{2}}\left[ \frac{r}{20\text{ Mpc}}\right] ^{-1}\left[ \frac{f}{1%
\text{ kHz}}\right] ^{-1}\left[ \frac{\tau }{1\text{ ms}}\right] ^{-\frac{1}{%
2}}\correction{}{.}
\end{equation}
This is the formula for the ``burst energy'', normalized to numbers
appropriate to a gravitational collapse occurring in the Virgo cluster. It
explains why physicists and astronomers regard the 10$^{-21}$ threshold as
so important. But this formula could also be applied to a binary system radiating
away its orbital gravitational binding energy over a long period of time $%
\tau $, for example.

\subsection{Curvature produced by waves}

W have assumed  that the background metric
satisfied the vacuum Einstein equations to linear order, but now it is 
possible to view the full
action principles as a principle for the background with a wave field $%
h_{\mu \nu }$ on it, and to let the wave energy affect the background
curvature.\cite{isa} This means that the background will actually solve, in
a self-consistent way, the equation 
\begin{equation}
G_{\alpha \beta }\left[ g_{\mu \nu }\right] =8\pi T_{\alpha \beta }^{\text{GW%
}}\left[ g_{\mu \nu }+h_{\mu \nu }\right]\correction{}{.}
\end{equation}
This does not contradict the vanishing of the first variation of the action,
which we needed to use above, because now we have an Einstein tensor that is
of quadratic order in $h_{\mu \nu }$, contributing a term of cubic order to
the first-variation of the action, which is of the same order as other terms
we have neglected.

\subsection{Cosmological background of radiation}

This self-consistent picture allows us to talk about, for example, a
cosmological gravitational wave background that contributes to the curvature
of the Universe. Since the energy density is the same as the flux (when $c=1$%
), we have 
\begin{equation}
\varrho _{\text{gw}}=\frac{\pi }{4}f^{2}h^{2},
\end{equation}
but now we must interpret $h$ in a statistical way. 
This will be treated in the contribution by Babusci et al., 
but basically it is done by replacing $h^{2}$ by a
statistical mean square amplitude per unit frequency (Fourier transform
power), so that the energy density per \emph{unit frequency} is
proportional to $f^{2}\left| \tilde{h}\right| ^{2}$. It is then
conventional to talk about the energy density per unit logarithm of
frequency, which means multiplying by $f$. The result, after being careful
about averaging over all directions of the waves and all independent
polarization components, is 
\begin{equation}
\frac{d\varrho _{\text{gw}}}{d\ln f}=4\pi ^{2}f^{3}\left| \bar{h}\left(
f\right) \right| ^{2}\correction{}{.}
\end{equation}
Finally, what is of the most interest is the energy density as a fraction of
the closure or critical cosmological density, given by the Hubble constant $%
H_{0}$ as $\varrho _{c}=3H_{0}^{2}/8\pi$. The resulting ratio is
the symbol $\Omega _{\text{gw}}(f)$ that we met in the previous \lecture: 
\begin{equation}
\Omega _{\text{gw}}(f)=\frac{32\pi ^{3}}{3H_{0}^{2}}f^{3}\left| \bar{h}%
\left( f\right) \right| ^{2}\correction{}{.}
\end{equation}

\subsection{Other approaches}

We finish this \lecture\ by observing that there is no unique approach to defining energy for
gravitational radiation or indeed for any solution of Einstein's equations.
Historically this has been one of the most difficult areas for physicists to
come to grips with. In the textbooks you will find discussions of 
pseudotensors, of energy measured at null infinity and at spacelike infinity, 
of Noether theorems and formulas for energy, and so on. None of these are 
worse than we have presented here, and in fact all of them are now known to 
be consistent with one another, if one does not ask them to do too much. 
In particular, if one wants only to localize the energy of a gravitational 
wave to a region of the size of a
wavelength, and if the waves have short wavelength compared to the background
curvature scale, then pseudotensors will give the same energy as the one 
we have defined here. Similarly, if one takes the energy flux defined 
here and evaluates it at null infinity, one gets the so-called Bondi 
flux, which was derived by H.\ Bondi in one of the pioneering steps in 
the understanding of gravitational radiation. Many of these issues are 
discussed in the Schutz-Sorkin paper referred to earlier.\cite{sork}

\section{Exercises}

\begin{enumerate}
\item[\textbf{5}]  \emph{In the notes above we give the general gauge
transformation } 
\[
h_{\mu \nu }\rightarrow h_{\mu \nu }-\xi _{\mu ;\nu }-\xi _{\nu ;\mu }.
\]
\emph{Use the formula for the derivation of Einstein's equations from an
action principle, } 
\[
\delta I=\frac{1}{16\pi }\int \frac{\delta \left( R\sqrt{-g}\right) }{\delta
g_{\mu \nu }}h_{\mu \nu }d^{4}x
\]
\emph{with } 
\[
\frac{\delta \left( R\sqrt{-g}\right) }{\delta g_{\mu \nu }}=-G^{\mu \nu }%
\sqrt{-g},
\]
\emph{but insert a pure gauge }$h_{\mu \nu }$\emph{. Argue that since this is
merely a coordinate transformation, the action should be invariant.
Integrate the variation of the action to prove the contacted Bianchi
identity } 
\[
G_{\quad ,\nu }^{\mu \nu }=0.
\]
\emph{This shows that the divergence-free property of }$G^{\mu \nu }$\emph{\
is closely related to the coordinate invariance of Einstein's
theory.\bigskip }

\item[\textbf{6}]  \emph{Suppose a plane wave, travelling in the $z$-direction in
linearized theory, has both polarization components }$h_{+}$\emph{\ and }$%
h_{\times }$\emph{. Show that its energy flux in the $z$-direction, }$T^{\text{%
(GW)}0z}$\emph{, is } 
\[
\left\langle T^{\text{(GW)}0z}\right\rangle =\frac{k^{2}}{32\pi }\left(
A_{+}^{2}+A_{\times }^{2}\right) ,
\]
\emph{where the angle brackets denote an average over one period of the wave.%
}\newpage 
\end{enumerate}

\chapter{Mass- and Current-Quadrupole Radiation}

In this \lecture\ we focus on the wave amplitude itself, and how 
it and the polarization depend on the motions in the source.
Consider an isolated source with a stress-energy tensor $T^{\alpha \beta }$.
As in \lecture~1, the Einstein equation is 
\begin{equation}\label{eq:linfield}
\left( -\frac{\partial ^{2}}{\partial t^{2}}+\nabla ^{2}\right) \overline{h}%
^{\alpha \beta }=-16\pi T^{\alpha \beta }
\end{equation}
($\overline{h}^{\alpha \beta }=h^{\alpha \beta }-\frac{1}{2}\eta ^{\alpha
\beta }h$ and $\overline{h}_{\quad ,\beta }^{\alpha \beta }=0$). Its general
solution is the following retarded integral for the field at a position $%
x^{i}$ and a time $t$ in terms of the source at a position $y^{i}$ and the
retarded time $t-R$:%
\begin{equation}
\overline{h}^{\alpha \beta }\left( x^{i},t\right) =4\int \frac{1}{R}%
T^{\alpha \beta }\left( t-R,y^{i}\right) d^{3}y,
\end{equation}
where we define 
\begin{equation}
R^{2}=\left( x^{i}-y^{i}\right) \left( x_{i}-y_{i}\right).
\end{equation}

\section{Expansion for the far field of a slow-motion source}

Let us suppose that the origin of coordinates is in or near the source, and
the field point $x^{i}$ is far away. Then we define $r^{2}=x^{i}x_{i}$ and
we have $r^{2}\gg y^{i}y_{i}$. We can therefore expand the term $R$ in the
dominator in terms of $y^{i}$. The lowest order is $r$, and all higher-order
terms are smaller than this by powers of $r^{-1}$. Therefore, they
contribute terms to the field that fall off faster than \correction{$\frac{1}{r}$}{$r^{-1}$}, and
they are negligible in the far zone. So we can simply replace $R$ by $r$ in
the dominator, and take it out of the integral.

The $R$ inside the time-argument of the source term is not so simple. If we
suppose that $T^{\alpha \beta }$ does not change very fast we can substitute 
$t-R$ by $t-r$ (the retarded time to the origin of coordinates) and expand 
\begin{equation}
t-R=t-r+n^{i}y_{i}+O\left( \frac{1}{r}\right) \text{,\qquad with }n^{i}=%
\frac{x^{i}}{r}\text{,\qquad }n^{i}n_{i}=1.
\end{equation}

The two conditions $r\gg y^{i}y_{i}$ and the slow-motion source, can be
expressed quantitatively as: 
\[
r\gg \bar{\lambda} 
\]
\[
R\ll \bar{\lambda} 
\]
where $\bar{\lambda}$ is the reduced wave length $\bar{\lambda}=%
\lambda/2\pi$ and $R$ is the size of source.

The terms of order $r^{-1}$ are negligible for the same reason as
above, but the first term in this expansion must be taken into account. It
depends on the direction to the field point, given by the unit vector $n^{i}$%
. We use this by making a Taylor expansion in time on the time-argument of
the source. The combined effect of these approximation is 
\begin{eqnarray}
\overline{h}^{\alpha \beta } &=&\frac{4}{r}\int \left[ T^{\alpha \beta
}\left( t^{\prime },y^{i}\right) +T_{\quad ,0}^{\alpha \beta }\left(
t^{\prime },y^{i}\right) n^{j}y_{j}+\frac{1}{2}T_{\quad ,00}^{\alpha \beta
}\left( t^{\prime },y^{i}\right) n^{j}n^{k}y_{j}y_{k}\right.  \nonumber \\
&&+\left. \frac{1}{6}T_{\quad ,000}^{\alpha \beta }\left( t^{\prime
},y^{i}\right) n^{j}n^{k}n^{l}y_{j}y_{k}y_{l}+\ldots \right] d^{3}y.
\end{eqnarray}
We will need all the terms of this Taylor expansion out to this order.

The integrals in expression (5.5) contain moments of the components of the
stress-energy. It is useful to give these names. Use $M$ for moments of the
density $T^{00}$, $P$ for moments of the momentum $T^{0i}$, and $S$ for the
moments of the stress $T^{ij}$. Here is our notation: 
\[
M\left( t^{\prime }\right) =\int T^{00}\left( t^{\prime },y^{i}\right) d^{3}y%
\text{,}\quad M^{j}\left( t^{\prime }\right) =\int T^{00}\left( t^{\prime
},y^{i}\right) y^{j}d^{3}y \text{,}
\]
\[
M^{jk}\left( t^{\prime }\right) =\int T^{00}\left( t^{\prime },y^{i}\right)
y^{j}y^{k}d^{3}y\text{,\quad }M^{jkl}\left( t^{\prime }\right) =\int
T^{00}\left( t^{\prime },y^{i}\right) y^{j}y^{k}y^{l}d^{3}y \text{,}
\]
\[
P^{l}\left( t^{\prime }\right) =\int T^{0l}\left( t^{\prime },y^{i}\right)
d^{3}y\text{,}\quad P^{lj}\left( t^{\prime }\right) =\int T^{0l}\left(
t^{\prime },y^{i}\right) y^{j}d^{3}y \text{,}
\]
\[
P^{ljk}\left( t^{\prime }\right) =\int T^{0l}\left( t^{\prime },y^{i}\right)
y^{j}y^{k}d^{3}y \text{,}
\]
\[
S^{lm}\left( t^{\prime }\right) =\int T^{lm}\left( t^{\prime },y^{i}\right)
d^{3}y\text{,}\quad S^{lmj}\left( t^{\prime }\right) =\int T^{lm}\left(
t^{\prime },y^{i}\right) y^{j}d^{3}y \text{.}
\]
These are the moments we will need.

Among these moments there are some identities that follow from the
conservation law in linearized theory, $T_{\hspace{0.12cm}\hspace{0.12cm}%
,\beta }^{\alpha \beta }=0$, which we use to replace time derivatives of
components of $T$ by divergences of other components and then integrate by
parts. The identities we will need are 
\begin{equation}\label{eq:identity1}
\dot{M}=0\text{,\quad }\dot{M}^{k}=P^{k}\text{,\quad }\dot{M}%
^{jk}=P^{jk}+P^{kj}\text{,\quad }\dot{M}^{jkl}=P^{jkl}+P^{klj}+P^{ljk}\text{,}
\end{equation}

\begin{equation}\label{eq:identity2}
\dot{P}^{j}=0\text{,\quad }\dot{P}^{jk}=S^{jk}\text{,\quad }\dot{P}%
^{jkl}=S^{jkl}+S^{jlk}\text{.}
\end{equation}
These can be applied recursively to show, for example, two further very
useful relations 
\begin{equation}\label{eq:identity3}
\frac{d^{2}M^{jk}}{dt^{2}}=2S^{jk}\text{,\quad }\frac{d^{3}M^{jkl}}{dt^{3}}=6%
\dot{S}^{^{(jkl)}}.
\end{equation}
where the round brackets on indices indicate full symmetrization.

Using this relations and notations it is not hard to show that 
\begin{eqnarray}
\overline{h}^{00}(t,x^{i})&=&\frac{4}{r}M+\frac{4}{r}P^{j}n_{j}+\frac{4}{r}%
S^{jk}(t^{\prime })n_{j}n_{k}+\frac{4}{r}\dot{S}^{jkl}(t^{\prime
})n_{j}n_{k}n_{l}+\ldots \label{eq:h00}\\
\overline{h}^{0j}(t,x^{i})&=&\frac{4}{r}P^{j}+\frac{4}{r}S^{jk}(t^{\prime
})n_{k}+\frac{4}{r}\dot{S}^{jkl}(t^{\prime })n_{k}n_{l}+\ldots 
\label{eq:h0j}\\
\overline{h}^{jk}(t,x^{i})&=&\frac{4}{r}S^{jk}(t^{\prime })+\frac{4}{r}\dot{S}%
^{jkl}(t^{\prime })n_{l}+\ldots . \label{eq:hjk}
\end{eqnarray}
In these three formulas there are different orders of time-derivatives, but 
in fact they are evaluated to the same final order in the slow-motion 
approximation. One can see that from the gauge condition $\overline{h}_{\hspace{0.12cm}%
\hspace{0.12cm},\beta }^{a\beta }=0$, which relates time-derivatives of 
some components to space-derivatives of others. 

In these expressions, one must
remember that the moments are evaluated at the retarded time $t^{\prime
}=t-r $ (except for those moments that are constant in time), and they are
multiplied by components of the unit vector to the field point $n^{j}=\left.
x^{j}\right/ r$.

\section{Application of TT gauge to the mass quadrupole field.}

In the expression for the amplitude that we derived so far, the final terms are
those that represent the current-quadrupole and mass-octupole radiation. The
terms before them represent the static parts of the field and the mass
quadrupole radiation. In this section we treat just these terms,
placing them into TT gauge. This will be simpler than treating it all at
once, and the procedure for the next terms will be a straightforward
generalization.

\subsection{The TT gauge transformations}

We are already in Lorentz gauge, and this can be checked by taking
derivatives of the expressions for the field that we have derived above. But
we are manifestly not in TT gauge. Making a gauge transformation consists of
choosing a vector field $\xi ^{\alpha }$ and modifying the metric by 
\begin{equation}
h_{\alpha \beta }\rightarrow h_{\alpha \beta }-\xi _{\alpha ,\beta }-\xi
_{\beta ,\alpha }.
\end{equation}
The corresponding expression for the potential $\overline{h}^{\alpha \beta }$
is 
\begin{equation}
\overline{h}^{\alpha \beta }\rightarrow \overline{h}^{\alpha \beta }+\xi
^{\alpha ,\beta }+\xi ^{\alpha ,\beta }-\eta ^{\alpha \beta }\xi _{\hspace{%
0.15cm},\mu }^{\mu }.
\end{equation}
For the different components this implies changes 
\begin{equation}
\delta \overline{h}^{00}=\xi ^{0,0}+\xi _{\,\hspace{0.15cm},j}^{j}
\end{equation}
\begin{equation}
\delta \overline{h}^{0j}=\xi ^{0,j}+\xi ^{j,0}
\end{equation}
\begin{equation}
\delta \overline{h}^{jk}=\xi ^{j,k}+\xi ^{k,j}-\delta ^{jk}\xi _{\hspace{%
0.15cm},\mu }^{\mu }
\end{equation}
where $\delta ^{jk}$ is the Kronecker delta (unit matrix). In practice, when
taking derivatives, the algebra is vastly simplified by the fact that we are
keeping only \correction{$\frac{1}{r}$}{$r^{-1}$} terms in the potentials. This means that spatial
derivatives do not act on \correction{$\frac{1}{r}$}{$r^{-1}$} but only on $t^{\prime }=t-r$. It
follows that $\partial t^{\prime }/\partial x^{j}=-n_{j}$, and $%
\partial h(t^{\prime })/\partial x^{j}=-\dot{h}(t^{\prime })n_{j}$.

It is not hard to show that the following vector field puts the metric into TT gauge to the order we are
working: 
\begin{eqnarray}
\xi ^{0}&=&\frac{1}{r}P_{\hspace{0.1cm}\hspace{0.12cm}k}^{k}+\frac{1}{r}%
P^{jk}n_{j}n_{k}+\frac{1}{r}S_{\hspace{0.12cm}lk}^{l}n^{k}+\frac{1}{r}%
S^{ijk}n_{i}n_{j}n_{k}\correction{}{,} \label{eq:gauge0} \\
\xi ^{i} &=&\frac{4}{r}M^{i}+\frac{4}{r}P^{ij}n_{j}-\frac{1}{r}P_{\hspace{%
0.12cm}k}^{k}n^{i}-\frac{1}{r}P^{jk}n_{j}n_{k}n^{i}+\frac{4}{r}%
S^{ijk}n_{j}n_{k} \correction{\\ \nonumber}{\nonumber \\}
&&\quad-\frac{1}{r}S_{\hspace{0.12cm}lk}^{l}n^{k}n^{i}-\frac{1}{r}%
S^{jlk}n_{j}n_{l}n_{k}n^{i}\correction{}{.}  \label{eq:gaugei}
\end{eqnarray}

\subsection{Quadrupole field in TT gauge}

The result of applying this gauge transformation to the original amplitudes
is 
\begin{eqnarray}
\overline{h}^{\text{TT}00}&=&\frac{4M}{r},\\
\overline{h}^{\text{TT}0i}&=&0,\\
\overline{h}^{\text{TT}ij}&=&\frac{4}{r}\left[ \perp ^{ik}\perp ^{jl}S_{lk}+%
\frac{1}{2}\perp ^{ij}\left( S_{kl}n^{k}n^{l}-S_{\hspace{0.12cm}%
k}^{k}\right) \right]
\end{eqnarray}
Remember that here we are not including $\dot{S}^{jkl}$, because it is a third order
effect.

The notation $\perp ^{ik}$ represents the projection operator perpendicular
to the direction $n^{i}$ to the field point. 
\begin{equation}
\perp ^{jk}=\delta ^{jk}-n^{j}n^{k}.
\end{equation}
It can be verified that this tensor is transverse to the direction $n^{i}$
and is a projection, in the sense that it projects to itself 
\begin{equation}
\perp ^{jk}n_{k}=0\text{,\qquad }\perp ^{jk}\perp _{k}^{\hspace{0.12cm}%
l}=\perp ^{jl}.
\end{equation}

The spherical component of the field is not totally eliminated in this gauge
transformation: the time-time component of the metric must contain the constant Newtonian field of the source. (In fact
we have succeeded in eliminating the dipole, or momentum part of the field, which is
also part of the non-wave solution. Our gauge transformation has incorporated a Lorentz
transformation that has put us into the rest frame of the source.) The
time-dependent part of the field is now purely spatial, transverse (because
everything is multiplied by $\perp $), and traceless (as can be verified by
explicit calculation).

The expression for the spatial part of the field actually does not depend on
the trace of $S_{jk}$, as can be seen by constructing the
trace-free part of the tensor, defined as: 
\begin{equation}\label{eq:tracefreequad}
\tilde{S}^{jk}=S^{jk}-\frac{1}{3}\delta ^{jk}S_{\hspace{0.12cm}l}^{l}.
\end{equation}
In fact, it is more conventional to use the mass moment here instead of the stress,
so we also define 
\begin{equation}
\tilde{M}^{jk}=M^{jk}-\frac{1}{3}\delta ^{jk}M_{l}^{l}\text{,\qquad }\tilde{S}%
^{jk}=\frac{1}{2}\frac{d^{2}\tilde{M}^{jk}}{dt^{2}}\correction{}{.}
\end{equation}
In terms of $\tilde{M}$ the far field is 
\begin{equation}\label{eq:quad}
\bar{h}^{\text{TT}ij}=\frac{2}{r}\left( \perp ^{ik}\perp ^{jl}\stackrel{%
\cdot \cdot }{\tilde{M}}_{kl}+\frac{1}{2}\perp ^{ij}\stackrel{\cdot \cdot }{%
\tilde{M}}_{kl}n^{l}n^{k}\right)\correction{}{.}
\end{equation}
This is the usual formula for the mass-quadrupole field. In textbooks the notation is somewhat
different than we have adopted here. In particular, our tensor $\tilde{M}$ is
what is called $\ibar$ in Misner, et al, (1973) and Schutz (1985). It is
the basis of most gravitational wave source estimates. We have derived it 
only in the context of linearized theory, but remarkably its form is 
identical if we go to the post-Newtonian approximation, where the 
gravitational waves are a perturbation of the Newtonian spacetime 
rather than of flat spacetime.

Given this powerful formula, it is important to try to interpret it 
and understand it as fully as possible. One obvious conclusion is 
that the dominant source of radiation, at least in the slow-motion limit, is
the second time-derivative of the second moment of the mass density $T^{00}$
(the mass-quadrupole moment). This is a very important difference between
gravitational waves and electromagnetism, in which the most important source
is the electric-dipole. In our case the mass-dipole term is not able
to radiate because it is constant, reflecting conservation of the linear 
momentum of the source. In electromagnetism, however, if the dipole
term is absent for some reason (all charges positive, for example) then 
the quadrupole term dominates and it looks very similar to \Eref{eq:quad}.

\subsection{Radiation patterns related to the motion of sources}

The projection operators in \Eref{eq:quad} show that the radiative 
field is transverse, as we expect. But the form of \Eref{eq:quad} hides 
two equally important messages, 
\begin{itemize}
\item the only motions that produce the 
radiation are the ones transverse to the line of sight; and 
\item the induced motions in a detector mirror the motions of the source 
projected onto the plane of the sky.
\end{itemize}
To see why these are true, 
we define the \emph{transverse traceless quadrupole tensor} 
\begin{equation}\label{eq:ttquad}
M\supr{TT}_{ij} = \perp\supr{k}\sub{i}\perp\supr{l}\sub{j}M_{kl} - 
\frac{1}{2}\perp_{ij}\perp^{kl}M_{kl}.
\end{equation}
(Notice that some of our definitions of tracelessness involve subtracting 
$1/3$ of the trace, as in \Eref{eq:tracefreequad}, and sometimes $1/2$ of 
the trace, as here in \Eref{eq:ttquad}. The appropriate factor is 
determined by the effective dimensionality (rank) of the tensor. Although 
we have 3 spatial dimensions, the projection tensor $\perp$ projects 
the mass quadrupole tensor onto a two-dimensional plane, where the trace 
involves only two components, not three.)

Now, if in \Eref{eq:quad} we replace $\tilde{M}_{ij}$ by its 
definition in terms of $M_{ij}$, and then collect terms appropriately, 
it is not hard to show that the equation simplifies to its most 
natural form:
\begin{equation}\label{eq:quadnatural}
\bar{h}^{\text{TT}ij}=\frac{2}{r}\stackrel{\cdot \cdot }{M}\supr{\text{TT}ij}.
\end{equation}
This could of course  have been derived directly by applying the TT operation to 
\Eref{eq:h00} to~\correction{\ref{eq:hjk}}{{\ref{eq:hjk}}}.
Since this equation involves only the TT-part of M, our first assertion above is proved.
According to this equation, in order to calculate the quadrupole 
radiation that a particular observer will receive, 
one need only compute the mass-quadrupole tensor's second 
time-derivative, project it onto the plane of the sky as seen by 
the observer looking toward the source, take away its trace, and 
rescale it by a factor of $2/r$. In particular, the TT-tensor that describes
the action of the wave (as in the polarization diagram in \Fref{fig:pol}) 
is a copy of the TT-tensor of the mass distribution. This proves our 
second assertion above.

Look again at \Fref{fig:pol}. Imagine a detector consisting of two 
free masses whose separation is being monitored. If the wave causes 
them to oscillate relative to one another along the $x$-axis (the $\oplus$ 
polarization), this means that the source motion contained a component 
that did the same thing. If the source is a binary, then 
the binary orbit projected onto the sky must involve motion of the 
stars back and forth along either the $x$- or the $y$-axis. 

It is possible from this to understand many aspects of quadrupole
radiation in a simple way. Consider a binary star system with a circular 
orbit. Seen by a distant observer in the orbital plane, the projected source motion
is linear, back and forth. The received polarization will be linear,  
the polarization ellipse aligned with the orbit. Seen by a distant 
observer along the axis of the orbit of the binary, the projected 
motion is circular, which is a superposition of two linear motions 
separated in phase by $90^o$. The received radiation will also 
have circular polarization. Because both linear polarizations are 
present, the amplitude of the wave emitted up the axis is twice that 
emitted in the plane. In this way we can completely determine the 
radiation pattern of a binary system.

Notice that, when viewed at an arbitrary angle to the 
axis, the radiation will be elliptically polarized, and the degree 
of ellipticity will directly measure the inclination of the orbital 
plane to the line of sight. This is a very special kind of information, 
which one cannot normally obtain from electromagnetic observations of 
binaries. It illustrates the complementarity of the two kinds of 
observing.

\section{Application of TT gauge to the current quadrupole field}

Now we turn to the problem of placing next-order terms of the wave field, 
the current-quadrupole and mass-octupole, into TT-gauge. Our interest 
here is to understand current-quadrupole radiation in the same 
physical way as we have just done for mass-quadrupole radiation. So 
we shall put the field into TT gauge and then see how to separate 
the current-quadrupole part from the mass-octupole, which we will
discard from the present discussion.

\subsection{The field at third order in slow-motion}

The next order terms in the non-TT metric bear a simple relationship to the
mass quadrupole terms (see equations~\ref{eq:h00}-\ref{eq:hjk}). In each of the metric components, just
replace $S^{jk}$ by $\dot{S}^{jkl}n_{l}$ to go from one order to the next.

This means that we can just skip to the end of the application of the gauge
transformations in equations~\ref{eq:gauge0} and \ref{eq:gaugei} and write the next order of the final field, only
using $S$ again, not $M$:
\begin{equation}
\bar{h}^{\text{TT}ij}=\frac{4}{r}\left[ \perp ^{ik}\perp ^{jl}\dot{S}%
_{lkm}n^{m}+\frac{1}{2}\perp ^{ij}\left( \dot{S}_{klm}n^{k}n^{l}n^{m}-\dot{S}%
_{\hspace{0.12cm}kl}^{k}n^{l}\right) \right],
\end{equation}
or more compactly 
\begin{equation}
\bar{h}^{\text{TT}ij}=\frac{4}{r}\left( \perp ^{ik}\perp ^{jl}\stackrel{%
\cdot }{\tilde{S}}_{klm}n^{m}+\frac{1}{2}\perp ^{ij}\stackrel{\cdot }{%
\tilde{S}}_{klm}n^{l}n^{k}n^{m}\right).
\end{equation}
The tilde above $S$ represents a trace-free operation \emph{on the first two
indices,} 
\[\stackrel{\cdot }{\tilde{S}}_{klm}=\dot{S}_{lkm}-%
\frac{1}{3}\delta _{kl}\dot{S}_{\hspace{0.12cm}im}^{i}.\]
These are the indices that come from the indices of $T^{jk}$, so the tensor 
is symmetric on these. By analogy with the quadrupole calculation, we 
can also define the TT part of $S_{ijk}$ by doing the TT projection on 
the first two indices,
\begin{equation}\label{eq:ttoct}
S\supr{TT}_{ijm} = \perp\supr{k}\sub{i}\perp\supr{l}\sub{j}S_{klm} - 
\frac{1}{2}\perp_{ij}\perp^{kl}S_{klm}.
\end{equation}
The TT projection of the equation for the metric is 
\begin{equation}\label{eq:octnatural}
h^{\text{TT}ij}=\frac{4}{r}\dot{}S\supr{\text{TT}ijk}n_k.
\end{equation}

\subsection{Separating the current-quadrupole from the mass-octupole}

The last equation is compact, but it does not have the ready interpretation
that we have at quadrupole order. This is because the moment of the 
stress, $S_{ijk}$, does not have such a clear physical interpretation. 
We see from \Eref{eq:identity1} to~\ref{eq:identity3} that $S_{ijk}$ is 
a complicated mixture of moments of momentum and density. To gain 
more physical insight into radiation at this order, we need to 
separate these different contributions. It is straightforward algebra 
to see that the following identity follows from the earlier ones:
\begin{equation}\label{eq:identity4}
\dot{S}^{ijk} = \frac{1}{6}\stackrel{\cdot\cdot\cdot}{M}{}^{ijk} + \frac{2}{3}
\stackrel{\cdot\cdot}{P}{}^{[jk]i} + \frac{2}{3}
\stackrel{\cdot\cdot}{P}{}^{[ik]j},
\end{equation}
where square brackets around indices mean antisymmetrization:
\[A^{[ik]}:= \frac{1}{2}\left(A^{ik} - A^{ki}\right).\]
This is a complete separation of the mass terms (in $M$) from the momentum 
terms (in $P$) because the only identities relating the momentum moments 
to the mass moments involve the symmetric part of $P^{ijk}$ on its first 
two indices, and this is absent from \Eref{eq:identity4}.

The first term in \Eref{eq:identity4} is the third moment of the 
density, and this is the source of the \emph{mass
octupole} field. It produces radiation through the third time-derivative. Since 
we are in a slow-motion approximation, this is smaller than the mass 
quadrupole radiation by typically a factor of $v/c$. Unless there were 
some very special symmetry conditions, one would not expect the mass 
octupole to be anything more than a small correction to the mass 
quadrupole. For this reason we will not treat it here.

The second and third terms in \Eref{eq:identity4} involve the second moment of the momentum, and together they are the source of  
the \emph{current quadrupole} field. It involves two time-derivatives, just 
as the mass-quadrupole does, but these are time-derivatives of the 
mometum moment, not the mass moment, so these terms produce a field that 
is also $v/c$ smaller than the typical mass quadrupole field. However, it 
requires less of an accident for the mass quadrupole to be absent and 
the current quadrupole present. It just requires motions that leave the 
density unchanged to lowest order. This happens in the $r$-modes. Therefore, 
the current-quadrupole deserves more attention, and we 
will work exclusively with these terms from now on.

The terms in \Eref{eq:identity4} that we need are the ones involving 
$\stackrel{\cdot\cdot}{P}\supr{ijk}$. These are atni-symmetrized on 
the first two indices, which involves effectively a vector product 
between the momentum density (first index) and one of the moment 
indices. This is essentially the angular momentum density. To make
the angular momentum explicit and to simplify the expression, we introduce 
the angular momentum and the first moment of the angular momentum density
\begin{eqnarray}
J^{i} := \epsilon^{ijk}P_{jk}, \label{eq:angmom} \\
J^{il} := \epsilon^{ijk}P\sub{jk}\supr{l}, \label{eq:angmommom}
\end{eqnarray}
where $\epsilon^{ijk}$ is the fully antisymmetric (Levi-Civita) symbol 
in 3 dimensions. It follows from this that 
\[P^{[jk]l} = \frac{1}{2}\epsilon^{jki}J\sub{i}\supr{l}.\]
These terms enter the TT projection of the field \Eref{eq:octnatural} with 
the last index of $S$ always contracted with the direction $n^i$ to the 
observer from the source. According to \Eref{eq:identity4}, this 
contraction always occurs on one of the antisytmmetrized indices, 
or if we use the form in the previous equation then we will always 
have a contraction of $n^i$ with $\epsilon^{ijk}$. This is a simple object, 
which we call 
\newcommand{\peps}{\sub{\perp}\epsilon}
\newcommand{\jdd}{\stackrel{\cdot\cdot}{J}}
\newcommand{\pj}{\sub{\perp}J}
\newcommand{\pjdd}{\sub{\perp}\stackrel{\cdot\cdot}{J}}
\begin{equation}\label{eq:epsperp}
\peps^{jk} := n_i\epsilon^{ijk}.
\end{equation}
This is just the two-dimensional Levi-Civita object in the plane 
perpendicular to $n^i$, which is the plane of the sky as seen by the 
observer. These quantities will be used in the current-quadrupole field, 
which contains projections on all the indices. Therefore, the only 
components of $J^{jk}$ that enter are those projected onto the sky, and
so it will simplify formulas to define the sky-projected moment of 
the angular momentum $\pj$
\begin{equation}\label{eq:pj}
\pj^{ij} := \perp\supr{i}\sub{l}\perp\supr{j}\sub{m}J^{lm}.
\end{equation}

Using this assembled notation, the current-quadrupole field is 
\begin{equation}\label{eq:currentquad}
h^{\text{TT}ij}=\frac{4}{3r}\left(
\peps^{ik}\;\pjdd\sub{k}\supr{j} + \peps^{jk}\;\pjdd\sub{k}\supr{i}
 + \perp^{ij}\peps^{km}\;\pjdd\sub{km}\right).
\end{equation}

This is similar in form and complexity to the mass-quadrupole field expression.
The interpretation of the contributions is direct. Only components of
the angular momentum in the plane of the sky contribute to the field. Similarly only
moments of this angular momentum transverse to the line of sight contribute. 
If one wants, say, the $xx$ component of the field, then the $\peps$ factor 
tells us it is determined by the $y$-component of momentum, ie the component 
perpendicular to the $x$-direction in the sky. In fact, it is much simpler
just to write out the actual components, assuming that the wave travels toward
the observer along the $z$-direction. Then we have 
\begin{eqnarray}
h^{\text{TT}xx}&=&\frac{4}{3r}(\jdd\supr{xy}+\jdd\supr{yx}), \\
h^{\text{TT}xy}&=&\frac{4}{3r}(\jdd\supr{yy}-\jdd\supr{xx}),
\end{eqnarray}
and the remaining components can be found from the usual symmetries of 
the TT-metric. I have dropped the prefix $\sub{\perp}$ on $J$ because in 
this coordinate system the given components are already transverse.

The simplicity of these expressions is striking. There are two basic 
cases where one gets current-quadrupole 
radiation.
\begin{itemize}
\item If there is an oscillating angular momentum distribution with 
a dipole moment along the angular momentum axis, as projected onto the sky, 
then in an appropriate coordinate system $\jdd\supr{xx}$ will be nonzero 
and we will have $\otimes$ radiation. To have a non-vanishing 
dipole moment,  the angular momentum density could, for example, be 
symmetrical under reflection through the origin along its axis, so that 
it points in opposite directions on opposite sides. 
\item If there is an oscillating angular momentum distribution with 
a dipole moment along an axis perpendicular to the angular momentum 
axis, as projected onto the sky, then in an appropriate coordiate system 
$\jdd\supr{xy}$ will be nonzero and we will have $\oplus$ radiation.
\end{itemize}

\subsection{A model system radiating current-quadrupole radiation}

\begin{figure}
\begin{center}
\caption{A simple current-quadrupole radiator. The left panel shows how 
the two wheels are connected with blade springs to a central axis. The wheels 
turn in opposite directions, each oscillating back and forth about its 
rest position. The right panel shows the side view of the system, 
and the arrows indicate 
the motion of the near side of the wheels at the time of viewing. The + signs 
indicate where the momentum of the mass of the wheel is toward the viewer 
and the - signs indicate where it is away from the viewer.}\label{fig:wheel}
\includegraphics[clip=true,width=0.45\textwidth]{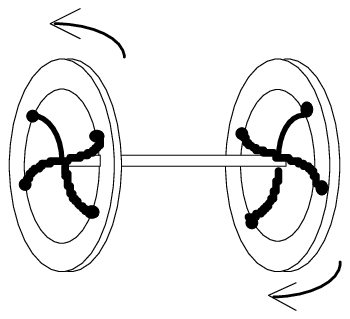}\hfill
\includegraphics[clip=true,width=0.45\textwidth]{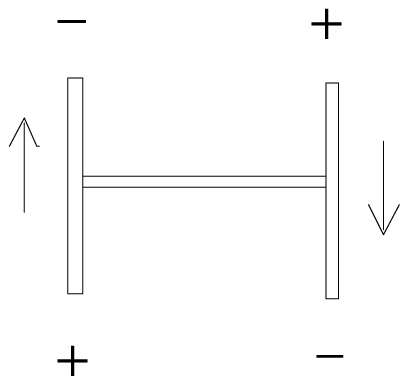}
\end{center}
\end{figure}

To see that the first of these two leads to physically sensible results, 
let us consider a simple model system that actually bears a close 
resemblance to the $r$-mode system. Imagine, as in the left panel 
of \Fref{fig:wheel}, two wheels connected 
by an axis, and the wheels are sprung on the axis in such a way 
that if a wheel is turned by some angle and then released, it will 
oscillate back and forth about the axis. 
Set the two wheels into oscillation with 
opposite phases, so that when one wheel rotates clockwise, the other 
rotates counterclockwise, as seen along the axis. 

Then when viewed along the axis, the angular momentum has no component 
transverse to the line of sight, so there is no radiation along the 
axis. This is sensible, because when projected onto the plane of the 
sky the two wheels are performing exactly opposite motions, so the 
net effect is that there is zero projected momentum density. 

When  viewed from a direction perpendicular to the axis, with the 
axis along the $x$-direction, then the angular momentum is transverse, 
and it has opposite direction for the two wheels. There is therefore 
an $x$-moment of the $x$-component of angular momentum, and the 
radiation field will have the $\otimes$ orientation.

To see that this has a physically sensible 
interpretation, look back again at the 
polarization diagram, \Fref{fig:pol}, and look at the bottom 
row of figures illustrating the $\otimes$ polarization. See what 
the particles on the $x$-axis are doing. They are moving up and 
down in the $y$-direction. What motions in the source could be 
producing this?

At first one might guess that it is the up-and-down motion of 
the mass in the wheels as they oscillate, because in fact the near 
side of each wheel does exactly what the test particles at the 
observer are doing. But this cannot be the explanation, because the 
far side of each wheel is doing the opposite, and when they both 
project onto the sky they cancel. What in fact gives the effect is 
that at the \emph{top} of the wheel the momentum density is first 
positive (towards the observer) and then negative, while at the 
\emph{bottom} of the wheel it is first negative and then positive. On 
the other wheel, the signs are reversed. 

\emph{Current quadrupole radiation is produced, at least in 
simple situations like the one we illustrate here, by the 
(second time-derivative \correction{derivative}{} of) the component 
of source momentum along the line of sight. If this is positive 
in the sense that it is towards the observer, then the momentum 
density acts as a positive gravitational ``charge''. If negative, 
then it is a negative ``charge''.} The wheels have an array of 
positive and negative spots that oscillates with time, and 
the test particles in the polarization diagram are drawn toward
the positive ones and pushed away from the negative ones.
Interestingly, in electromagnetism, magnetic dipole and magnetic quadrupole
radiation are also generated by the component of the electric current along
the line of sight.

This is a rather simple physical interpretation of some 
rather more complex equations. It is possible to re-write 
\Eref{eq:currentquad} to show explicitly the contribution 
of the line-of-sight momentum, but the expressions become 
even more complicated. Instead of dwelling on this, I will 
turn to the question of calculating the total energy radiated 
by the source.

\section{Energy radiated in gravitational Waves}

We have calculated the energy flux in \Eref{eq:ttflux}, and we now 
have the TT wave amplitudes. We need only integrate the flux over a distant 
sphere to get the total luminosity. We do this for the mass and current
quadrupoles in separate sections.

\subsection{Mass quadrupole radiation}

The mass quadrupole radiation field in \Eref{eq:quad} must be put into \correction{}{the }energy flux
formula, and the dependence on the direction $n^i$ can then 
be integrated over a sphere. It is
not a difficult calculation, but it does require some angular integrals over
over multiple products of the vector $n^{i}$, which depends on the angular direction on the
sphere. By symmetry, integrals of odd numbers of factors of $n^i$ vanish. 
For even numbers of factors, the result is essentially determined by 
the requirement that after integration the result must be fully 
symmetric under interchange of any two indices and it cannot have any 
special directions (so it must depend only on the Kronecker delta $\delta\supr{i}\sub{j}$\correction{}{)}. The identities we need are 
\begin{eqnarray}
\int n^{i}n^{j}d\Omega &=& \frac{4\pi }{3}\delta ^{ij},\label{eq:nn} \\
\int n^{i}n^{j}n^{k}n^{l}d\Omega &=&\frac{4\pi }{15}\left( \delta ^{ij}\delta
^{kl}+\delta ^{ik}\delta ^{jl}+\delta ^{il}\delta ^{jk}\right).\label{eq:nnnn}
\end{eqnarray}
Using these, one gets the following simple formula for the total luminosity
of \correction{from}{} mass-quadrupole radiation 
\begin{equation}
\fbox{$L_{gw}^{mass}$=$\frac{1}{5}\left\langle \stackrel{\cdot \cdot \cdot }{%
\tilde{M}^{jk}}\stackrel{\cdot \cdot \cdot }{\tilde{M}_{jk}}\right\rangle $}.
\end{equation}
Here we still preserve the angle brackets of  \Eref{eq:ttflux}, because
this formula only makes sense in general if we average in time over one cycle 
of the radiation.

\subsection{Current quadrupole radiation}

The energy radiated in the current quadrupole is nearly as simple to obtain
as the mass quadrupole formula. The extra factor of $n^{i}$ in the radiation
field makes the angular integrals longer, and  requires two further identities:
\begin{eqnarray}\label{eq:nnnnnn}
&&\int n^{i}n^{j}n^{k}n^{l}n^{p}n^{q}d\Omega = \frac{4\pi }{7}\delta
^{(ij}\delta ^{kl}\delta ^{pq)}, \\
&&\epsilon^{ijk}\epsilon^{i'j'k'}= \delta^{ii'}\delta^{jj'}\delta^{kk'} +
\delta^{ij'}\delta^{jk'}\delta^{ki'} + \delta^{ik'}\delta^{ji'}\delta^{kj'} \nonumber \\ && \qquad\quad - 
\delta^{ii'}\delta^{jk'}\delta^{kj'} - \delta^{ij'}\delta^{ji'}\delta^{kk'} - 
\delta^{ik'}\delta^{jj'}\delta^{ki'},\label{eq:epseps}
\end{eqnarray}
where the round brackets indicate full symmetrization on all indices. The 
expression is simplest if we define 
\[\tilde{J}^{jk}:=\epsilon^{jlm}\tilde{P}\sub{lm}\supr{k} + 
\epsilon^{klm}\tilde{P}\sub{lm}\supr{j}, \]
where
\[\tilde{P}^{kij} := P^{kij} - \frac{1}{3}\delta^{ij}P\supr{kl}\sub{l}.\]

The
result of the integration of the flux formula over a distant sphere is\cite{thorne}\cite{lind}, in our notation, 
\begin{equation}
L_{gw}^{current}=\frac{4}{5}\left\langle \stackrel{\cdot \cdot \cdot }{\tilde{J%
}^{jk}}\stackrel{\cdot \cdot \cdot }{\tilde{J}_{jk}}\right\rangle.
\end{equation}

\section{Radiation in the Newtonian limit}

The calculation so far has been within the assumptions of linearized
theory. Real sources are likely to have significant self-gravity. This
means, in particular, that there will be a significant component of the
source energy in gravitational potential energy, and this must be taken into
account.

In fact a more realistic equation than \Eref{eq:linfield} would be 
\begin{equation}\label{eq:fullfield}
\Box \bar{h}^{\alpha \beta } = -16\pi \left( T^{\alpha \beta } + 
t^{\alpha \beta,
}\right)
\end{equation}
where $t^{\alpha \beta }$ is the stress-energy \emph{pseudotensor} of gravitational
waves. This is hard to work with: \Eref{eq:fullfield}  is an implicit equation because $t^{\alpha
\beta }$ depends on $\bar{h}^{\alpha \beta }$.

Fortunately, the formulas that we have derived are more robust than they seem. It turns out that
the \emph{leading order} radiation field from a Newtonian source has the
same formula as in linearized theory. By leading order we mean the dominant
radiation. If there is mass-quadrupole radiation, then the mass-octupole
radiation from a Newtonian source will not be given by the formulas of the
linearized theory. On the other hand, 
current quadrupole and mass quadrupole radiation can 
co-exist, because they have
different symmetries, so the work we have done here is 
generally applicable.

More details on how one calculates radiation to 
higher order in the Newtonian limit will be given in the
lecture by Blanchet. This is particularly important for 
computing the radiation 
to be expected from coalescing binary systems, whose orbits become highly 
relativistic just before coalescence and which are therefore not well 
described by linearized theory.

\chapter{Source calculations}

Now that we have the formulas for the radiation from a system, we 
can use them for some simple examples.

\section{Radiation from a binary system}

The most numerous sources of gravitational waves are binary stars systems. 
In just half an orbital period, the non-spherical part of the  mass
distribution returns to its original configuration, so the angular
frequency of the emitted gravitational waves is twice the orbital angular
frequency.

We shall calculate here the mass quadrupole moment for two stars of masses $m_{1}$
and $m_{2}$, orbiting in the $x$-$y$ plane in a circular orbit with angular
velocity $\Omega $, governed by Newtonian dynamics. 
We take their total separation to be $R$, which 
means that the orbital radius of mass $m_1$ is $m_2R/(m_1+m_2)$ while 
that of mass $m_2$ is $m_1R/(m_1+m_2)$. We place the origin of coordinates 
at the center of mass of the system. Then for example 
the $xx$-component of $M^{ij}$ is 
\begin{eqnarray}
M_{xx} &=& m_1\left(\frac{m_2R\cos(\Omega t)}{m_1+m_2}\right)^2 + 
m_2\left(\frac{m_1R\cos(\Omega t)}{m_1+m_2}\right)^2 \nonumber \\
&=& \mu R^2\cos^2(\Omega t),
\end{eqnarray}
where $\mu:=m_1m_2/(m_1+m_2)$ is the reduced mass. By using a trigonometric 
identity and throwing away the part that does not depend on time (since 
we will use only time-derivatives of this expression) we have
\begin{equation}
M_{xx} = \frac{1}{2}\mu R^2\cos(2\Omega t).
\end{equation}
By the same methods, the other non-zero components are
\[M_{yy}=-\frac{1}{2}\mu R^{2}\cos(2\Omega t) \text{,\quad }
M_{xy}=\frac{1}{2}\mu R^{2}\sin(2\Omega t). \]
This shows that the radiation will come out at twice the orbital frequency.

In this case the trace-free moment $\tilde{M}^{ij}$ 
differs from $M^{ij}$ only by a
constant, so we can use these values for $M^{ij}$ to calculate the field and 
luminosity. 

As an example of calculating the field, let us compute $\bar{h}^{\text{TT}xx}$ 
as seen by an observer at a distance $r$ from the system 
along the $y$-axis, i.e.\ lying in the plane of the orbit. 
We first need the TT-part of the mass quadrupole moment, from \Eref{eq:ttquad}:
\[M^{\text{TT}xx} = M^{xx}-\frac{1}{2}\left(M^{xx} + M^{zz}\right).\]
But since $M^{zz}=0$, this is just $M^{xx}/2$. Then from \Eref{eq:quadnatural}
we find
\begin{equation}\label{eq:quadfield}
\bar{h}^{\text{TT}xx} = -2\frac{\mu}{r}(R\Omega)^2\cos\left[2\Omega (t-r)\right].
\end{equation}
Similarly, the 
result for the luminosity is 
\begin{equation}
L_{gw}=\frac{32}{5}\mu^{2}R^{4}\Omega^{6}\correction{}{.}
\end{equation}

The various factors in thse two equations are not independent, 
because the angular velocity is
determined by the masses and separations of the stars. 
When observing such a system,
we can not usually measure $R$ directly, but we can 
infer $\Omega $ from the
observed gravitational wave frequency, and we may 
often be able to make a guess at the masses. (Actually, we will see 
below that we can actually measure the important quantity about 
the masses.)  So we eliminate $R$ using the
Newtonian orbit equation 
\begin{equation}
R^{3}=\frac{m_1+m_2}{\Omega ^{2}}.
\end{equation}
If in addition we use the gravitational wave frequency $\Omega _{gw}=2\Omega 
$, we get 
\begin{eqnarray}
\bar{h}^{\text{TT}xx} &=& -2^{1/3}\frac{\mathcal{M}^{5/3}\Omega_{gw}^{2/3}}{r}\cos\left[\Omega_{gw}(t-r)\right],\\
L_{gw}&=&\frac{4}{5\cdot2^{1/3}}\left( \mathcal{M} \Omega_{gw}\right)^{\frac{10}{3}},
\end{eqnarray}
where we have introduced the symbol for the \emph{chirp mass} of the binary 
system:
\[\mathcal{M}:=\mu^{3/5}(m_1+m_2)^{2/5}.\]
Notice that both the field and the luminosity depend only on $\mathcal{M}$, not
on the individual masses in any other combination.

The power represented by $L_{gw}$ 
must be supplied by the orbital energy, $E=-m_1m_2/2R$. By
eliminating $R$ as before we find the equation
\[E = -\frac{1}{2^{5/3}}\mathcal{M}^{5/3}\Omega_{gw}^{2/3}.\]
This is remarkable because it too involves only the chirp mass $\mathcal{M}$. 
By 
setting the rate of change of $E$ equal to the (negative of the) luminosity, 
we find an equation for the rate of 
change of the gravitational wave frequency
\begin{equation}\label{eq:chirp}
\dot{\Omega}_{gw}=\frac{12\cdot 2^{1/3}}{5}\mathcal{M}^{5/3}\Omega_{gw}^{11/3}.
\end{equation}
As we mentioned in \lecture~3, since the frequency increases, 
the signal is said to ``chirp''. 

These results show that the chirp mass is the only mass
associated with the binary that can be deduced from observations of its 
gravitational radiation, at least if only the Newtonian orbit is important.
Moreover, if one can measure the field amplitude (e.g. $h^{\text{TT}xx}$) plus 
$\Omega_{gw}$ and $\dot{\Omega}_{gw}$,  one can 
deduce from these the value of $\mathcal{M}$ \emph{and} the distance $r$ 
to the system! \emph{A chirping binary with a circular orbit, 
observed in gravitational waves, 
is a \emph{standard candle:} one can infer its distance purely from the 
gravitational wave observations.} To do this one needs the full amplitude, 
not just its projection on a single detector, so one generally needs 
a network of detectors or a long-duration observation with a single 
detector to get enough information.

It is very unusual in astronomy to have standard candles, and they are 
highly prized. For example, one can in principle use this information to 
measure Hubble's constant.\cite{hubble}

\subsection{Corrections}
In the calculation above we made several simplifying assumptions. For example, 
how good is the assumption that the orbit is circular? The Hulse-Taylor 
binary is in a highly eccentric orbit, and this turns out to enhance 
its gravitational wave luminosity by more than a factor of 10, 
since the elliptical orbit brings the two stars much nearer to one 
another for a period of time than a circular orbit with the same period 
would do. So there are big corrections for this system. 

However, 
systems emitting at frequencies observable from ground-based 
interferometers are probably well-approximated by circular orbits, 
because they have arrived at their very close separation by 
gravitational-wave-driven in-spiral. This process removes eccentricity 
from the orbit faster than it shrinks the orbital radius, so by they
time they are observed they have insignificant eccentricity.

Another assumption is that the orbit is well described by Newtonian theory.
This is not a good assumption in most cases. 
Post-Newtonian orbit corrections will be very important in observations.
This is not because the stars eventually approach each other closely.
It is because they spend a long time at wide separations where the 
small post-Newtonian corrections accumulate systematically, eventually 
changing the phase of the orbit by an observable amount. So 
it is very important for observations that we match signals with 
a template containing high-order post-Newtonian corrections, as 
described in Blanchet's lecture. But even so, the information contained 
in the Newtonian part of the radiation is still there, so all our 
conclusions above remain important.

\section{The $r$-modes}

We consider rotating stars in Newtonian gravity and look at the effect 
that the emission of gravitational radiation has on their oscillations.
One might expect that the loss of energy to gravitational waves would 
damp out any perturbations, and indeed this is normally the case. However, 
it was a remarkable discovery of Chandrasekhar\cite{cha} that 
the opposite sometimes happens. 

A rotating star is idealized as an axially symmetric perfect-fluid system.
In the Newtonian theory the pulsations of a perturbed fluid can be described
by normal modes which are the solutions of perturbed Euler and 
gravitational field equations. If the star is stable, the eigenfrequencies 
$\sigma $ of the normal modes are real; if the star is unstable, 
there is at least one pair of complex-conjugate frequencies, one 
of which represents an exponentially growing mode and the other 
a decaying mode. (We take the convention that the time-dependence of 
a mode is $\exp(i\sigma t)$.)

In general relativity, the situation is in principle the same, except 
that there is a boundary condition on the perturbation equations 
that insists that gravitational waves far away be outgoing, i.e.\ that 
the star loses energy to gravitational waves. This condition forces all 
eigenfrequencies to be complex. The sign of the imaginary part of 
the frequency determines stability or instability. 

The loss of energy to gravitional radiation can destabilize a star 
that would otherwise (i.e.\ in Newtonian theory) be stable. This 
is because it opens a pathway to lower-energy configurations that 
might not be accessible to the Newtonian star. This normally happens 
because gravitational radiation also carries away angular momentum,
a quantity that is conserved in the Newtonian evolution of a perturbation.

The sign of the angular momentum lost by the star is a critical 
diagnostic for the instability. A wave that moves in the positive 
angular direction around a star will radiate positive angular momentum
to infinity. A wave that moves in the opposite direction, as seen 
by an observer at rest far away, will radiate negative angular momentum.
In a spherical star, both actions result in the damping of 
the perturbation because, for example, the postivive-going wave 
has intrinsically positive angular momentum, so when it radiates 
its angular momentum decreases and so its amplitude decreases. Similarly, 
the negative-going wave has negative angular momentum, so when it 
radiates negative angular momentum its amplitude decreases.

The situation can be different in a rotating star, as first pointed out 
by Friedman and Schutz.\cite{fri} The angular momentum carried by 
a wave depends on its pattern angular velocity \emph{relative to the star's 
angular velocity}, not relative to an observer far away. If a wave pattern 
travels backwards relative to the star, it represents a small effective 
slowing down of the star and therefore carries negative angular momentum. 
This can lead to an anomalous situation: if a wave travels backwards relative
to the star, but forwards relative to an inertial observer (because its
angular velocity relative to the star is smaller than the star's
angular velocity), then it will have negative angular momentum but it 
will radiate positive angular momentum. The result will be that its 
intrinsic angular momentum will get more negative, and its amplitude 
will grow. 

This is the  mechanism of the 
Chandrasekar-Friedman-Schutz (CFS) instability. In an ideal star, it 
is always possible to find pressure-driven waves of short enough 
wavelength around the axis of symmetry (high enough angular eigenvalue 
$m$) that satisfy this 
condition. But it turns out that even a small amount of viscosity 
can damp out the instability in such waves
, so it is not clear that pressure-driven 
waves will ever be significantly unstable in realistic stars. 

However, in 1997 Andersson\cite{and} pointed out that there 
was a class of modes called 
$r$-modes (Rossby modes) that no-one had previously investigated, 
and that were formally
unstable in all rotating stars.  Rossby waves are
well-known in oceanography, where they play an important role in energy 
transport around the
Earth's oceans. They are hard to detect, having long wavelengths and very
low density perturbations. They are mainly \emph{velocity perturbations} of the
oceans, whose restoring force is the Coriolis effect, 
and that is their character in neutron stars too. Because they 
have very small density perturbation, the gravitational radiation 
they emit is dominated by the current-quadrupole radiation.

For a slowly-rotating, nearly-spherical Newtonian star, the following 
velocity perturbation is characteristic of $r$-modes:
\begin{equation}\label{eq:rvel}
\delta v^{a}=\varsigma (r)\epsilon ^{abc}\nabla _{b}r\nabla _{c}Y_{lm},
\end{equation}
where $\varsigma(r)$ is some function of $r$ determined by the mode
equations. This velocity is a curl, so it is divergence-free; 
since it has no radial component, it does not change the density.
If the star is perfectly spherical, these perturbations are simply
a small rotation of some of the fluid, and it continues to rotate.
They have no oscillation, and have zero-frequency.

If we consider a star with a small rotational angular velocity $\Omega$, 
then the frequency $\sigma $ is no longer exactly 
zero and a Newtonian calculation to first order in $\Omega $ 
shows that there is a mode with \emph{pattern speed} 
$\omega_p=-\sigma/m$ equal to
\begin{equation}\label{eq:rangvel}
\omega_p = \Omega \left[ 1-\frac{2}{l\left( l+1\right) }\right].
\end{equation}
These
modes are now oscillating currents that move (approximately) along the
equipotential surfaces of the rotating star.

For $l\geqslant 2$, $\omega_p$ is positive but slower than the speed 
of the star, so  by the CFS 
mechanism these modes are unstable to the emission of gravitational radiation 
for an arbitrarily slowly rotating star. 

The velocity pattern given in \Eref{eq:rvel} for ($l=2$, $m=2$) is 
closely related to the wheel model we described for current-quadrupole 
radiation in  \Fref{fig:wheel}. Take two such wheels and orient their 
axels along the $x$- and $y$-axes, with the star rotating about the 
$z$-axis. Choose the sense of rotation so that the wheels at positive-$x$ 
and positive-$y$ are spinning in the opposite sense at any time, i.e.\ so
that their adjacent edges are always moving in the \emph{same} direction.
Then this relationship will be reproduced for all other adjacent pairs of 
wheels: adjacent edges move together. 

When seen from above the equatorial plane, the line-of-sight momenta of 
the wheels reinforce each other, and we get the same kind of pattern 
that we saw when looking at one wheel from the side. However, in this case
the pattern rotates with the angular velocity $2\Omega/3$ of \Eref{eq:rangvel}.
Since the pattern of line-of-sight momenta repeats itself every half 
rotation period, the gravitational waves are circularly polarized with 
frequency $4\Omega/3$. Seen along the $x$-axis, the wheel along the 
$x$-axis contributes nothing, but the other wheel contributes fully, 
so the radiation amplitude in this direction is half that going out 
the rotation axis. Seen along a line at $45^o$ to the  $x$-axis, 
the line-of-sight momenta of the wheels on the front part of the 
star cancel those at the back, so there is no radiation. Thus, along 
the equator there is a characteristic series of maxima and zeros, 
leading to a standard $m=2$ radiation pattern. This pattern also 
rotates around the star, but the radiation in the equator remains 
linearly polarized because there is only the $\otimes$ component, 
not the $\oplus$. Again, the radiation frequency is twice the pattern 
speed because the radiation goes through a complete cycle in half
a wave rotation period.

This discussion cannot go into the depth required to understand 
the $r$-modes fully. There are many issues of principle: what happens beyond linear 
order in $\Omega$; what happens if the star is described in relativity 
and not Newtonian gravity; what is the relation between $r$-modes 
and the so-called $g$-modes that can have similar frequencies; what happens 
when the amplitude grows large enough that the evolution is non-linear; 
what is the effect of magnetic fields on the evolution of the instability? The 
literature on $r$-modes is developing rapidly. We have included 
references where some of the most basic issues are discussed,\cite{and,lind,mor,aks,sch1}
but the interested student should consult the current literature carefully.

\subsection{Linear growth of the $r$-modes}

We have seen how the $r$-mode becomes unstable when coupled to 
gravitational radiation, and now we turn to the practical question: 
is it important. This will depend on the balance between the 
growth rate of the mode due to relativistic effects and the 
damping due to viscosity.

When coupled to gravitational radiation and viscosity, the mode 
has a complex frequency. If we define $\Im(\sigma):=1/\tau$, then 
$\tau$ is the characteristic damping time. When radiation and 
viscosity are treated as small effects, their contributions to the 
eigenfrequencies add, so we have that the total damping is given by
\begin{equation}
\frac{1}{\tau \left( \Omega \right) }=\frac{1}{\tau _{\text{GR}}}+\frac{1}{%
\tau _{\text{v}}}\text{,\qquad }\frac{1}{\tau _{\text{v}}}=\frac{1}{\tau _{%
\text{s}}}+\frac{1}{\tau _{\text{b}}},
\end{equation}
where \correction{$\frac{1}{\tau _{\text{GR}}}$}{$1/\tau _{\text{GR}}$}, \correction{$\frac{1}{\tau _{\text{v}}}$}{$1/\tau _{\text{v}}$} are the
contributions due to gravitational radiation emission and viscosity, and 
where the latter has been further 
divided between shear viscosity (\correction{$\frac{1}{\tau _{\text{s}}}$}{$1/\tau _{\text{s}}$}) and bulk viscosity (\correction{$%
\frac{1}{\tau _{\text{b}}}$}{$1/\tau _{\text{b}}$}). 

If we consider a ``typical'' neutron star with
a polytropic equation of state $p=k\rho ^{2}$ (for which $k$ has been 
chosen so that a 1.5 M%
$_{\odot }$ model has a radius $R=12.47$~km), and if we express the 
angular velocity in terms of the scale for the approximate maximum speed $\sqrt{\pi G\bar{%
\rho}}$ and the temperature in terms of $10^{9}$~K, then it can be shown 
that\cite{aks}
\begin{equation}\label{eq:tau}
\frac{1}{\tau}= {1\over \tau_{\text{gw}}} \left({{\rm 1 ms} \over
P}\right)^{p_{\text{gw}}} +  {1\over \tau_{\text{bv}}} \left({{\rm 1 ms} \over
P}\right)^{p_{\correction{{\text{bv}}}{{\text{bv}}}}} \left( { 10^9 {\rm K} \over T } \right)^6 + 
{1\over \tau_{\text{sv}}}\left( { T \over 10^9 {\rm K}  } \right)^2  \ ,
\end{equation}
where the scaling parameters 
$\tilde{\tau}_{\text{sv}}$, $\tilde{\tau}_{\text{bv}}$, 
$\tilde{\tau}_{\text{gw}}$ and the exponents $p_{\text{gw}}$ and $p_{\text{bv}}$ have to be calculated numerically. 
Some representative values relevant to the $r$-modes with $%
2\leqslant l\leqslant 6$ are in \Tref{tab:tau}\cite{aks}.
\begin{table}
\caption{Gravitational radiation and viscous time scales, in seconds. 
Negative values indicate instability, i.e. a growing rather than damping mode.}
\label{tab:tau}
\begin{center}
\begin{tabular}[t]{ccccccc}

$l$ & $m$ & $\tau_{gw}$(s) & $p_{gw}$ & $\tau_{bv}$ (s) & $p_{bv}$ & $\tau_{sv}$ (s)  \\  \hline 
&&&&&&\\
2 & 2 & -20.83 & 5.93 & $9.3\times 10^{10}$ & 1.77  & $2.25\times 10^8$ \\
3 & 3 & -316.1 & 7.98 & $1.89\times 10^{10}$ & 1.83 &  $3.53\times
10^7$ \\
\hline
\end{tabular}
\end{center}
\end{table}

The physics of the viscosity is interesting. It is clear from \Eref{eq:tau} 
that gravitational 
radiation becomes a stronger and stronger destabilizing influence as
the angular velocity of a star increases, but the viscosity is much 
more complicated. There are two contributions: shear and bulk. Shear 
viscosity comes mainly from electrons scattering off protons and 
other electrons. This effect falls with increasing temperature, just 
as does viscosity of every-day materials. So a cold, slowly rotating 
star will not have the instability, where a hotter star might. But 
at high temperatures, bulk viscosity becomes dominant. This effect 
arises in neutron stars from the nuclear physics.  Neutron-star matter 
always contains some protons and electrons. When it is compressed, 
some of these react to form neutrons, emitting a neutrino. When it 
is expanded, some of the neutrons beta-decay to protons and electrons, 
again emitting a neutrino. The emitted neutrino is not trapped in the 
star; within a short time, of the order of a second or less, it escapes.
This irreversible loss of energy each time the star is compressed 
creates a bulk viscosity. 
Now, bulk viscosity acts only due to the density perturbation, which 
is small in $r$-modes. So the effect of bulk viscosity only dominates 
at very high temperatures. 

\begin{figure}
\begin{center}
\caption{The balance of viscous and gravitational radiation effects in 
the $r$-modes is illustrated in a diagram of rotation speed, \correction{indiced 
by}{showing} the ratio of the maximum period $P_k$ to the rotation period $P$, 
versus the temperature of the star. The solid curve indicates the 
boundary between viscosity-dominated and radiation-dominated 
behavior: stars above the line are unstable. The dashed curves 
illustrate possible nonlinear evolution histories as a young 
neutron star cools.}\label{fig:rmode}
\includegraphics[clip=true,width=0.9\textwidth]{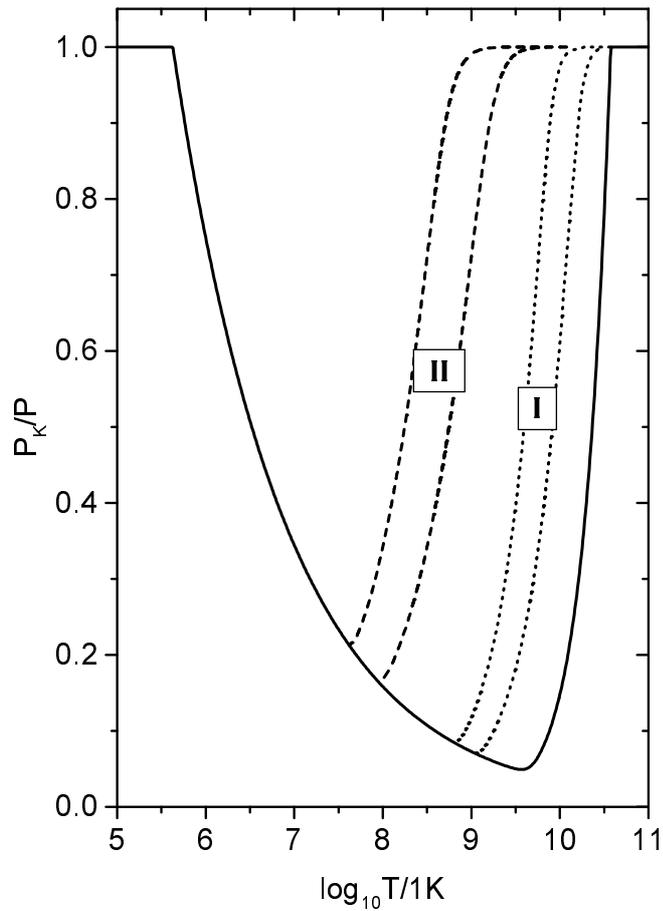}
\end{center}
\end{figure}

The balance of the viscous and gravitational effects is illustrated 
in \Fref{fig:rmode}\cite{aks}. This is indicative, but not definitive:
much more work is needed on the physics of viscosity and the 
structure of the modes at large values of $\Omega$ (small $P$).

\subsection{Nonlinear evolution of the star}
Our description so far is only a linear approximation.
To understand the full evolution of the $r$-modes we have to treat the
non-linear hydrodynamical effects that become important as the modes grow.
This could only be done with a numerical simulation, which some 
groups are now working on. But it is
possible to make simple estimates analytically.

We characterize the initial configuration with just two paramters: the
uniform angular velocity $\Omega $, and the amplitude $\alpha $ 
of the $r$-modes perturbation. The star is assumed to cool at the 
accepted cooling rate for neutron stars, independently of whether it 
is affected by the $r$-mode instability or not. The star is assumed 
to lose angular momentum to gravitational radiation at a rate given 
by the linear radiation field, with its large amplitude $\alpha$. 
This loss is taken to drive the star through a sequence of 
equilibrium states of lower and lower angular momentum. Details of this
approximation are in\cite{sch1}, here we report only the results.
The evolution turns out to have three phases:

\begin{enumerate}
\item[i)]  Initially the angular velocity $\Omega $ of the hot rapidly
rotating neutron stars is nearly constant, evolving on the viscous
time-scale $1/\tau _{\text{v}}$, while the amplitude $\alpha $ grows
esponentially on the gravitational radiation time-scale $1/\tau _{\text{GR}}$.

\item[ii)]  After a short time non-linear effects become important and
stop the growth of the amplitude $\alpha$. Most of the initial angular
momentum of the star is radiated away by gravitational 
radiation. The star spins down and
evolves to a point where the angular velocity $\Omega $ and the
temperature is sufficiently low that the $r$-mode is stable.

\item[iii)]  Finally gravitational radiation and viscosity damp out the 
$r$-mode and drive the star into its final equilibrium configuration.
\end{enumerate}

This may take about a year, a timescale governed by the cooling time 
of the star. During this year, the star would radiate away most 
of its angular momentum and rotational kinetic energy. This could 
be a substantial fraction of a solar mass in energy.

\subsection{Detection of $r$-mode radiation}

The large amount of energy radiated into the $r$-modes makes them 
attractive for detection, but detection will not be trivial. The 
$r$-mode event occurs at the rate of supernovae: some fraction (hopefully 
large) of all supernovae leave behind a rapidly spinning neutron star that 
spins down over a 1-year period. This means we should have sufficient 
sensitivity to reach to the Virgo Cluster (20 Mpc distance). Estimates \cite{sch1} suggest that 
a neutron star in the Virgo Cluster could be detected by second generation
of LIGO and VIRGO gravitational wave detectors with an amplitude
signal-to-noise of about 8, provided one can use matched filtering (exact 
template matching).

It will not be easy to use matched filtering, since one must follow 
all cycles of the signal as the star spins down, and we won't know this 
well because of many uncertainties: initial temperature, initial spin 
distribution, detailed physics of viscosity, and so on. But it would 
be helpful to have a parametrized model to take account of the 
uncertainties, so that we could look for a significant fit to one or 
more of the parameters.

In addition, it is likely that, if a significant proportion of all 
neutron stars went through the $r$-mode instability, then the 
Universe has been filled  by their radiation. There should be a 
background with an energy density $\Omega_{gw}$ that is a good 
fraction of the closure density. Its lower frequency limit should be 
around 200~Hz in the rest frame of the star. When we see radiation 
cosmologically, its lower frquency limit will indicate the epoch 
at which star formation began.

It is clear that the discovery of this new source of
gravitational waves will open several prospects for astronomy. Observations  
could be used as supernovae
detectors, revealing supernovae hidden in clouds of dust, identifying them 
about at year after they are formed. The existence of the radiation raises 
several prospects and questions about the physics of neutron stars, 
not least the interaction of magnetic fields with the instability.

\section{Conclusion}
These lectures have taken us through the basic theory of 
gravitational radiation and its applications in astrophysics, so far 
as we can understand and predict them now. In a few years, perhaps as 
little as 2, perhaps as many as 8, we will start to make observations 
of gravitational radiation from astrophysical sources. If gravitational 
wave astronomy follows other branches of observational astronomy, it 
will not be long before completely unexpected signals are seen, or 
unexpected features in long-predicted signals. To interpret these will 
require a joining a physical understanding of the relationship between 
gravitational 
radiation and its source to a wide knowledge of astronomical phenomena. I 
encourage the students who have attended these lectures, and others who 
may study them, to get themselves ready to contribute to 
this activity. It will be an exciting time!

\chapter*{References}

We have divided the references into two sections. The first gives generally 
useful references --- books, conference summaries, etc --- 
that interested students should go to for background 
and a more complete discussion of the theory. The second section contains 
specific references to the research literature. General references are 
indicated in the text by the author name plus year, as 
\emph{Misner, et al (1973).} The specific references are indicated by 
numbers, e.g. \emph{[1,3].}

\section{General references}

\small \textbf{General Relativity}. There are a number of good text books in 
GR. The following cover, at different levels of difficulty and 
completeness, linearized theory, gauges, and the definition of energy:

\begin{itemize}
\item[ ] \begin{thereferences}

\item[--] Ciufolini I., Wheeler, J.\ L., 1995 {\it Gravitation and Inertia} (Princeton University Press, Princeton).
\item[--] Landau L., Lifshitz E.\ M., 1962 {\it The Classical Theory of Fields} (Pergamon, New York).
\item[--] Misner, C.\ W., Thorne, K.\ S.,Wheeler, J.\ L., 1973 {\it Gravitation} (Freeman \& Co., San Francisco).
\item[--] Ruffini, R., Ohanian H.\ C., 1997 {\it Gravitazione e Spazio-Tempo} (Zanichelli, Bologna).
\item[--] Schutz, B.\ F., 1995 {\it A First Course in General Relativity} (Cambridge University Press, Cambridge).
\item[--] Wald, R.\ M., 1994?? {\it General Relativity} (Chicago University Press, Chicago).
\item[--] Weinberg S., 1972 {\it Gravitation and Cosmology} (Wiley \& Sons, New York).

\end{thereferences}
\end{itemize}

\small \textbf{Gravitational wave detectors}. Conference volumes on 
detector progress appear more than once per year these days. 
You can find progress reports on detectors on the web sites of 
the different groups, which you will find in the list of literature 
references below. The two references below are more tutorial, 
aimed at introducing the subject.

\begin{itemize}
\item[ ] \begin{thereferences}

\item[--] Blair, D. G., 1991 {\it The Detection of Gravitational Waves} (Cambridge University Press, England).
\item[--] Saulson, P. R., 1994 {\it Fundamentals of Interferometric Gravitational Wave Detectors} (World Scientific, Singapore).

\end{thereferences}
\end{itemize}

\small \textbf{Sources of gravitational waves}. Again, there have been a 
number of conference publications on this subject. The third and fourth 
 references are recent reviews of sources. The first two survey 
the problem of data analysis.

\begin{itemize}
\item[ ] \begin{thereferences}

\item[--] Schutz, B.\ F.,  ed., 1989 {\em Gravitational Wave Data Analysis} (Kluwer, Dordrecht).
\item[--] Schutz B. F., 1997 ``The Detection of Gravitational Waves'', in Marck, J. A., Lasota J. P., eds., {\it Relativistic Gravitation and Gravitational Radiation} (Cambridge University Press, Cambridge) p. 447--475.
\item[--] Thorne, K. S., 1987 ``Gravitational Radiation'', in Hawking, S. W., Israel, W., eds., {\it 300 Years of Gravitation} (Cambridge University Press, Cambridge), p. 330-458.
\item[--] Thorne, K. S., 1995 ``??'', in Kolb, E. W., Peccei, R., eds., {\it Proceedings of 1994 Summer Study on Particle and Nuclear Astrophysics and Cosmology}, (World Scientific, Singapore).

\end{thereferences}
\end{itemize}

\large \textbf{Text references}

\chapter*{Solutions to Exercises}
\section*{\lecture~1 Exercises}
\small
\textbf{Exercise~1\smallskip }

(a) Let us take the form  of the wave to be 
\[h^{\text{TT}jk}=\mathbf{e}^{jk}_\oplus h_+(t-\hat{\mathbf{n}}\cdot\hat{\mathbf{x}})\]
where $\e^{jk}_\oplus$ is the polarization tensor for the $\oplus$ polarization, 
and where $\hat{\mathbf{n}}$ is the unit vector in the direction of travel of 
the wave. We will let $h_+$ be an arbitrary function of its phase argument.

If the wave 
travels in the $x$-$z$ plane
at an angle $\theta $  to the $z$-direction, then the unit vector in our 
coordinates is 
\[\hat{\mathbf{n}}^i = (\sin \theta ,0,\cos \theta ).\]
We need to calculate the 
polarization tensor's components in the $x,y,z$ coordinates. We do this 
by rotating the $\oplus $ polarization tensor from its TT-form in coordinates 
parallel to the wavefront to its form in our coordinates. This requires  a simple rotation around the $y$-axis.
The transformation matrix is: 
\[
\Lambda\supr{j'}\sub{k}=\left( 
\begin{array}{ccc}
\cos \theta & 0 & \sin \theta \\ 
0 & 1 & 0 \\ 
-\sin \theta & 0 & \cos \theta
\end{array}
\right) .
\]
The polariation tensor in our coordinates (primed indices) becomes: 
\begin{eqnarray*}
\mathbf{e}^{j'k'} &=&\Lambda\supr{j'}\sub{l }\Lambda\supr{k'}\sub{m}\mathbf{e}^{lm} \\
 &=&\left( 
\begin{array}{ccc} 
\cos ^{2}\theta & 0 & -\sin \theta \cos \theta \\ 
0 & -1 & 0 \\ 
-\sin \theta \cos \theta & 0 & \sin ^{2}\theta
\end{array}
\right)
\end{eqnarray*}
Notice that the new polarization tensor is again traceless.

The gravitational wave will be, at an arbitrary time $t$ and position $(x,z)$ 
in our ($x,z$)-plane, 
\[h^{\text{TT}j'k'}=\mathbf{e}^{j'k' }h_+(t-x\sin \theta - z\cos\theta)\correction{]}{}. \]
For this problem we need the $xx$-component because the photon is propagating along this
direction, and we will always stay at $z=0$, so we have 
\[h^{\text{TT}xx}=\cos^{2}\theta h_+(t - x\sin\theta).\] 
We see that for this geometry 
the wave amplitude is reduced by a factor of $\cos^{2}\theta $.

Generalizing the argument in the text, the relation between time and 
position for the photon on its trip outwards along the $x$-direction is 
$t = t_0 + x$, where $t_0$ is the starting time. The analogous relation 
after the photon is reflected is $t = t_0 + L + (L-x)$, since 
in this case $x$ decreases in time from $L$ to 0.
If we put these into the equation for the linearized corrections 
to the return time, we get

\begin{eqnarray*}
t_{return} &=&t_0+2L+\frac{1}{2}\cos^2\theta\left\{ \int_{0}^{L}h_{+}[t_0+(1-\sin
\theta )x]dx\mathstrut \right.  \\
&&\left. \int_{0}^{L}h_{+}[t_0 + 2L - (1 + \sin\theta) x]dx\right\} .
\end{eqnarray*}
This expression must be differentiated with respect to $t_0$ to find the variation of 
the return time as a function of the start time. The key point is how 
to handle differentiation within the integrals. Consider, for example, the 
function $h_{+}[t_0+(1-\sin\theta )x]$. It is a function of a single argument,
\[\xi :=  t_0+(1-\sin\theta )x\]
so derivatives with respect to $t_0$ can be converted to derivatives 
with respect to $x$ as follows
\begin{eqnarray*}
\frac{dh_+}{dt_0} &=& \frac{dh_+}{d\xi}\frac{d\xi}{dt_0} = \frac{dh_+}{d\xi};\\
\frac{dh_+}{dx} &=& \frac{dh_+}{d\xi}\frac{d\xi}{dx} = (1-\sin\theta )\frac{dh_+}{d\xi}.
\end{eqnarray*}
It follows that 
\[\frac{dh_+}{dt_0} = \frac{dh_+}{dx} /  (1-\sin\theta ).\]
On the return trip the factor will be $-(1+\sin\theta)$. 
So when we differentiate we can convert the derivatives with respect to 
$t_0$ inside the integrals into derivatives with respect to $x$. Taking 
account of the factor $\cos^2\theta = (1-\sin\theta)(1+\sin\theta)$ in 
front of the integrals, the result is 
\begin{eqnarray*}
\frac{dt_{return}}{dt_0} &=&1+\frac{1}{2}(1+\sin\theta)\int_{0}^{L}\frac{dh_{+}}{dx}[t_0+(1-\sin
\theta )x]dx\mathstrut +  \\
&& \frac{1}{2}(1-\sin\theta)\int_{0}^{L}\frac{dh_{+}}{dx}[t_0 + 2L - (1 + \sin\theta) x]dx .
\end{eqnarray*}
The integrals can now be done, since they simply invert the differentiation
by $x$. Evaluating the integrands at the end points of the integrals gives
\Eref{eq:threeterm}:
\begin{eqnarray*}
\frac{dt_{return}}{dt_0}  &=& 1-\frac{1}{2}(1+\sin\theta)h_+(t_0)+ \sin\theta h_{+}[t_0+(1-\sin\theta )L] \\
+ && \frac{1}{2}(1-\sin\theta)h_+(t_0 + 2L).
\end{eqnarray*}

\medskip 

(b) If we Taylor-expand this equation in powers of $L$ about $L=0$, the 
leading term vanishes, and the first-order term is:
\begin{eqnarray*}
\frac{dt_{return}}{dt_0} &=& L\sin\theta(1-\sin\theta)\dot{h}_+(t_0) +  \\
&& \qquad L(1-\sin \theta)\dot{h}_+(t_0), \\
&=& L\cos^2\theta\dot{h}_+(t_0).
\end{eqnarray*}
This is just what was required. The factor of $\cos^2\theta$ comes, as we 
saw above, from the projection of the TT field on the $x$-coordinate direction.

\medskip 

(c) All the terms cancel and there is no effect on the return time. 

\bigskip

\textbf{Exercise 2}

This is part of the calculation in the previous example. All we need 
is the segment where the light travels from the distant end to the 
center:
\[
t_{out}=t_0+\frac{1}{2}\cos^2\theta\int_{0}^{L}h_+[t_0+L -(1+\sin \theta)x]dx
\]
and so $\frac{dt_{out}}{dt_0}$ is: 
\[
\frac{dt_{out}}{dt_0}=1+\frac{1}{2}(1-\sin \theta )\left[ h_{+}(t_0-\sin
\theta L)-h_{+}(t_0 + L)\right] 
\]
\bigskip 

\textbf{Exercise 3}

This question is frequently asked, but not by people who have done 
the calculation. The answer is that the two effects occur in different 
gauges, not in the same one. So they cannot cancel. The apparent 
speed of light changes in the TT-gauge, but then the positions of 
the ends remain fixed, so that the effect is all in the coordinate
speed. In a local Lorentz frame tied to one mass, the ends do move
back and forth, but then the speed of light is invariant.\bigskip 

\textbf{Exercise 4}

To the first order we have 
\[
R_{\alpha \beta \nu }^{\mu }=\Gamma _{\alpha \nu ,\beta }^{\mu }-\Gamma
_{\alpha \beta ,\nu }^{\mu } ,
\]
\begin{equation}
\Gamma _{\alpha \beta ,\nu }^{\mu }=\frac{1}{2}\eta ^{\mu \sigma }\left(
h_{\sigma \beta ,\alpha \nu }+h_{\sigma \alpha ,\beta \nu }-h_{\alpha \beta
,\sigma \nu }\right),  \tag{i}
\end{equation}
\begin{equation}
\Gamma _{\alpha \nu ,\beta }^{\mu }=\frac{1}{2}\eta ^{\mu \sigma }\left(
h_{\sigma \nu ,\alpha \beta }+h_{\alpha \sigma ,\nu \beta }-h_{\alpha \nu
,\sigma \beta }\right).  \tag{ii}
\end{equation}
The gauge transformation for a perturbation in linearized theory is 
\begin{equation}
h_{\alpha \beta }^{\prime }=h_{\alpha \beta }-\xi _{\alpha ,\beta }-\xi
_{\beta ,\alpha }.  \tag{iii}
\end{equation}
Substituting (iii) into (i) and (ii), we obtain 
\[
\text{(i)}=\frac{1}{2}\eta ^{\mu \sigma }\left( h_{\sigma \beta ,\alpha \nu
}^{\prime }+\xi _{\sigma ,\beta \alpha \nu }+h_{\sigma \alpha ,\beta \nu
}^{\prime }+\xi _{\sigma ,\alpha \beta \nu }-h_{\alpha \beta ,\sigma \nu
}^{\prime }\right) 
\]
\[
\text{(ii)}=\frac{1}{2}\eta ^{\mu \sigma }\left( h_{\sigma \nu ,\alpha \beta
}^{\prime }+\xi _{\sigma ,\nu \alpha \beta }+h_{\sigma \alpha ,\beta \nu
}^{\prime }+\xi _{\sigma ,\alpha \beta \nu }-h_{\alpha \nu ,\sigma \beta
}^{\prime }\right) .
\]
If  we find the difference between the two formulas above we get 
\[
\underline{R_{\alpha \beta \nu }^{\mu }}=\text{(ii)}-\text{(i)}=\Gamma
_{\alpha \nu ,\beta }^{\prime \mu }-\Gamma _{\alpha \beta ,\nu }^{\prime \mu
}=\underline{R_{\alpha \beta \nu }^{\prime \mu }} .
\]

\section*{\lecture~4 Exercises}
\small 

\textbf{Exercise 5}

The action principle is: 
\begin{equation}
\delta I=\int \frac{\delta \left( R\sqrt{-g}\right) }{\delta g_{\mu \nu }}%
h_{\mu \nu }d^{4}x=-\int G^{\mu \nu }\sqrt{-g}h_{\mu \nu }d^{4}x=0\correction{}{.}  \tag{i}
\end{equation}
If we perform a infinitesimal coordinate transformation $x^{\mu }\rightarrow
x^{\mu }+\xi ^{\mu }$ without otherwise varying the metric, then the action $I$ 
must not change:
\begin{eqnarray*}
0 &=&\delta I=\int G^{\mu \nu
}\left( \xi _{\mu ;\nu }+\xi _{\nu ;\mu }\right) \sqrt{-g}d^{4}x= \\
&&2\int G^{\mu \nu }\xi _{\mu ;\nu }d^{4}x\correction{}{.}
\end{eqnarray*}
This can be transformed in the
following way: 
\[
\delta I=\int \left( G^{\mu \nu }\xi _{\mu }\right) _{;\nu }\sqrt{-g}%
d^{4}x-\int \left( G_{\quad ;\nu }^{\mu \nu }\xi _{\mu }\right) \sqrt{-g}%
d^{4}x=0\correction{}{.}
\]
The first integral \correction{became}{is} a divergence and vanishes. The second, because of the
arbitrariness of $\xi _{\mu }$, gives the Bianchi's identities: 
\[
G_{\quad ;\nu }^{\mu \nu }=0\correction{}{.}
\]
\medskip 

\textbf{Exercise 6}

The two polarization components are $h_{+}^{xx}=-h_{+}^{yy}=\emph{A}%
_{+}e^{-i k \left( t-z\right) }$ and $h_{\times }^{xy}=\emph{A}_{\times
}e^{-i k \left( t-z\right) }$. The energy flux is \correction{}{the negative of}
\begin{eqnarray*}
\left\langle T_{0z}^{\text{(GW)}}\right\rangle  &=&\frac{1}{32\pi }%
\left\langle h_{\quad ,0}^{ij}h_{ij,z}\right\rangle =\correction{}{-}\frac{ k ^{2}}{%
16\pi }\left( \emph{A}_{+}^{2}+\emph{A}_{\times }^{2}\right) \left\langle
\sin ^{2} k \left( t-z\right) \right\rangle = \\
&&\correction{}{-}\frac{ k ^{2}}{32\pi }\left( \emph{A}_{+}^{2}+\emph{A}_{\times
}^{2}\right) \correction{}{.}
\end{eqnarray*}

\end{document}